\documentclass[
superscriptaddress,
reprint,
 amsmath,amssymb,
 aps,
 prd,
]{revtex4-1}

\usepackage{graphicx}
\usepackage{dcolumn}
\usepackage{bm}
\usepackage{amsmath}
\usepackage{amsfonts}
\usepackage{amssymb}
\usepackage{dsfont}
\usepackage{color}
\usepackage{xcolor}
\usepackage{siunitx}
\usepackage[colorlinks=true,
    linkcolor=blue,
    filecolor=blue,      
    urlcolor=blue,
    citecolor = blue]{hyperref}
\usepackage{braket}
\usepackage{mathtools}
\usepackage[mathlines]{lineno}
\usepackage[normalem]{ulem}
\usepackage{scalerel}
\usepackage{ bbold }
\usepackage{inputenc}
\usepackage[T1]{fontenc}

\graphicspath{{./figures/}}

\definecolor{mmagenta}{cmyk}{0,1,0,0.12}

\begin{document}

\title{Variational quantum simulation of U(1) lattice gauge theories with 
qudit systems} 

\date{\today}

\author{Pavel P. Popov}
\email{pavel.popov@icfo.eu}
\affiliation{ICFO - Institut de Ciencies Fotoniques, The Barcelona Institute of Science and Technology, Av. Carl Friedrich Gauss 3, 08860 Castelldefels (Barcelona), Spain}
\author{Michael Meth}
\affiliation{Universität Innsbruck, Institut für Experimentalphysik, Technikerstraße 25a, 6020 Innsbruck, Austria}
\author{Maciej Lewenstein}
\affiliation{ICFO - Institut de Ciencies Fotoniques, The Barcelona Institute of Science and Technology, Av. Carl Friedrich Gauss 3, 08860 Castelldefels (Barcelona), Spain}
\affiliation{ICREA, Pg. Lluis Companys 23, 08010 Barcelona, Spain}
\author{Philipp Hauke}
\affiliation{Pitaevskii BEC Center and Department of Physics, University  of  Trento,  Via Sommarive 14, I-38123 Trento, Italy}
\affiliation{INFN-TIFPA, Trento Institute for Fundamental Physics and Applications, Trento, Italy}
\author{Martin Ringbauer}
\affiliation{Universität Innsbruck, Institut für Experimentalphysik, Technikerstraße 25a, 6020 Innsbruck, Austria}
\author{Erez Zohar}
\affiliation{Racah Institute of Physics, The Hebrew University of Jerusalem, Givat Ram, Jerusalem 91904, Israel}
\author{Valentin Kasper}
\affiliation{ICFO - Institut de Ciencies Fotoniques, The Barcelona Institute of Science and Technology, Av. Carl Friedrich Gauss 3, 08860 Castelldefels (Barcelona), Spain}
\affiliation{Nord Quantique, 3000 boulevard de l'Université (P1-ACET), Sherbrooke J1K 2R1 QC, Canada}

\begin{abstract}
Lattice gauge theories are fundamental to various fields, including particle physics, condensed matter, and quantum information theory. Recent progress in the control of quantum systems allows for studying Abelian lattice gauge theories in table-top experiments. However, several challenges remain, such as implementing dynamical fermions in higher spatial dimensions and magnetic field terms. Here, we map D-dimensional U(1) Abelian lattice gauge theories onto qudit systems with local interactions for arbitrary D. We propose a variational quantum simulation scheme for the qudit system with a local Hamiltonian, that can be implemented on a universal qudit quantum device as the one developed in [Nat. Phys. 18, 1053–1057 (2022)]. We describe how to implement the variational imaginary-time evolution protocol for ground state preparation as well as the variational
real-time evolution protocol to simulate non-equilibrium physics on universal qudit quantum computers, supplemented with numerical simulations. Our proposal can serve as a way of simulating lattice gauge theories, particularly in higher spatial dimensions, with minimal resources, regarding both system sizes and gate count.
\end{abstract}
\maketitle

\section{Introduction}
Most quantum information processing platforms are based on qubits, the quantum generalization of classical bits. However, the underlying physical systems representing qubits frequently involve higher-dimensional Hilbert spaces that must be artificially restricted to two-level systems. Instead of limiting it, however, one can use the Hilbert space that the physical system provides for information processing. This leads to the multi-level analog of the qubit -- the qudit, which can be a powerful resource for quantum information processing~\cite{Wang2020}. The additional levels can enable alternative implementations of quantum algorithms~\cite{Lanyon2009} the implementation of optimal quantum measurements~\cite{Stricker2022}, as well as the native simulation of higher spin models or problems in quantum chemistry~\cite{MacDonell2020}. 
Moreover, the fundamentally different coherence~\cite{Ringbauer2018}, dissipation, and entanglement structure~\cite{Kraft2018} of qudit systems can be advantageous in terms of noise resilience~\cite{Cozzolino2019} or quantum error correction~\cite{Campbell2014}. These prospects of qudit systems and recent experimental progress make multi-level systems ideal for advanced quantum information processing.

So far, qudit experiments have been proposed~\cite{Bruss1998} and extensively used in quantum cryptography~\cite{Thew2004} for increased information capacity and improved resilience to perturbations~\cite{Cozzolino2019}. Beyond photons, almost all quantum technology platforms have demonstrated some degree of qudit control. 
More recently, superconducting systems~\cite{Morvan2021}, single photons~\cite{Chi2022} and trapped-ion experiments~\cite{Ringbauer2022} have demonstrated a universal set of gates for qudit quantum computing.
This rapid development of qudit hardware allows for the study of state-of-the-art quantum algorithms such as quantum simulation on these novel devices. Originally driven by the goal of developing a large-scale quantum computer, quantum simulation has been identified also as an attractive target for devices of the so-called noisy intermediate-scale quantum era \cite{Hauke2012_rev, Preskill2018, Bharti2022}. 
In particular, the quantum simulation of lattice gauge theories (LGTs) has made spectacular progress over the last decade~\cite{Wiese2013,Zohar2015,Dalmonte2016,Banuls2020,Aidelsburger2022,Zohar2022,Bauer2023}. 

LGTs are many-body systems with important applications in high-energy physics, condensed matter systems, and quantum information. In high energy physics, LGTs appear as the space-discretized description of the Standard Model of particle physics~\cite{Weinberg1996}. LGTs can also be found as an effective description in condensed matter physics and are important for quantum error correction, e.g., the toric code~\cite{Kitaev1997}. This versatility of LGTs makes them a central object of research for different communities. 
Quantum simulation protocols for LGTs have been proposed for numerous quantum platforms, from cold atoms~\cite{Tagliacozzo2013,Tagliacozzo2013_2,Zohar2013,Banerjee2013,Stannigel2014,Kasper2017,Zohar2017,Bender2018,Kasper2023,Gonzalez-Cuadra2022,Osborne2022, Zache2023}, through trapped ions~\cite{Hauke2013,Yang2016,Muschik2017,Paulson2021,Davoudi2021} to superconducting qubits~\cite{Mezzacapo2015,Klco2018,Atas2021,Ciavarella2021} and others~\cite{Klco2020,Mathis2020,Armon2021}. Indeed, first experimental implementations of LGT simulations are now a reality~\cite{Martinez2016,Schweizer2019,Kokail2019,Mil2020,Yang2020,Zhou2022,Nguyen2022,Mildenberger2022}.
Even though it is possible to quantum simulate LGTs in the laboratory, the experimental demonstrations remain constrained to specific scenarios. In particular, systems in one spatial dimension and some specific systems in two spatial dimensions with Abelian symmetry
have been experimentally realized. 
However, the implementation of Abelian LGTs beyond one spatial dimension remains challenging due to the presence of dynamical fermions and magnetic field terms (four-body interactions on a lattice) in the Hamiltonian of the theory~\cite{Zohar2022}.

Here, we propose a quantum simulation protocol for an Abelian U(1) LGT in (1+1) and in (2+1) space-time dimensions for qudit quantum processors based on trapped ions~\cite{Ringbauer2022}. Following~\cite{Zohar2018,Zohar2019}, we integrate out the fermionic fields and construct an LGT that can directly be mapped to a qudit system. The construction keeps the Hamiltonian of the theory local even in the case of higher spatial dimensions and evades the need of introducing Jordan--Wigner strings in order to encode the fermionic degrees of freedom in the qudits. In contrast to previous works that make use of the procedure of integrating out the fermions~\cite{Pardo2023,Irmejs2022}, we circumvent the need for the precise implementation of the resulting interactions in the Hamiltonian on the quantum device by proposing a hybrid quantum-classical variational simulation scheme for both ground-state preparation and quench dynamics. We also elaborate on which quantities have to be measured on the quantum device in order to execute the protocol and how to perform the measurement. We benchmark this approach by performing numerical simulations of the variational algorithm with the specific quantum circuits we propose for preparing the ground state of the LGT as well as simulating quench dynamics. As our results show, employing variational algorithms for static and dynamic properties of Abelian gauge theories is well within reach of current devices also for dimensions larger than (1+1). 

This paper is organized as follows. In Section~\ref{sec:lattice_gauge_theory_construction}, we briefly introduce the mathematical description of the target LGT and construct the corresponding qudit Hamiltonian. Then, in Section~\ref{sec:variational_quantum_simulation}, we give an outline of the variational algorithm we use. Here, we also explain how the relevant quantities for the execution of the algorithm can be measured in the experiment. Later, in Section~\ref{sec:implementation}, we present specific quantum circuits that need to be implemented on the quantum device and discuss how the simulation protocol would be implemented. In Section~\ref{sec:results}, we present the results of the numerical study and discuss the findings.

\section{U(1) LGT with fermionic matter }
\label{sec:lattice_gauge_theory_construction}
We consider a D-dimensional spatial lattice with sites $\mathbf{x} \in \mathbb{Z}^{\text{D}}$. The spinless fermionic matter is localized on the sites $\mathbf{x}$ where we denote the creation and annihilation operators by $\psi^{\dagger}_{\mathbf{x}}$ and $\psi_{\mathbf{x}}$, respectively. The gauge degrees of freedom are situated on the links between neighboring sites $\mathbf{x}$ and $\mathbf{y}$ and act on a $d$-dimensional Hilbert space with the basis $\ket{0}, \ldots, \ket{d-1}$. The electric field operator $E_{\mathbf{x},i}$ and the parallel transport (link) operator $U_{\mathbf{x},i}$, where the latter keeps track of the phase generated by the gauge field, are defined as
\begin{subequations}
\begin{align}
    E_{\mathbf{x},i} \ket{l} &= (l-d)\ket{l}  \, ,\\
    U_{\mathbf{x},i} \ket{l} &= u_{l}\ket{l+1}  \, ,
\end{align}
\end{subequations}
where the subscript $\mathbf{x},i$ denotes the link starting from the lattice site $\mathbf{x}$ and pointing in direction $i$, i.e., the link between sites $\mathbf{x}$ and $\mathbf{x}+\mathbf{e}_{i}$. For a $D$-dimensional spatial lattice, there are also $D$ distinct directions. We used the notation $u_l = \sqrt{d(d+1)-(l-d)(l-d+1)}$. 
Further, the fermionic degrees of freedom fulfill the algebra
\begin{align}
    \{ \psi_{\mathbf{x}}, \psi^{\dagger}_{\mathbf{y}} \} = \delta_{\mathbf{x},\mathbf{y}} \, ,\{ \psi_{\mathbf{x}}, \psi_{\mathbf{y}} \} = 0 , 
\end{align} 
and the electric field and the link operator fulfill 
\begin{align}
    [E_{\mathbf{x},i}, U_{\mathbf{y},j}]=\delta_{\mathbf{x},\mathbf{y}} \delta_{i,j} U_{\mathbf{y},j} \,.
\end{align}

The dynamics of the quantum many-body system are determined by the Hamiltonian
\begin{align}
    H = H_{\text{G}} + H_{\text{M}} + H_{\text{GM}}\,,
\label{eq:kogut-susskind_ham}
\end{align}
where the pure gauge part $H_G$ is given by
\begin{align}
    H_{G}&=\frac{g^{2}}{2} \sum_{\mathbf{x}, i} E^{2}_{\mathbf{x}, i} - \frac{1}{2 g^{2}} \sum_{\mathbf{p}} \left[ U_{\mathbf{p}}+U_{\mathbf{p}}^{\dagger} \right]\,,
\label{eq:gauge_part_Hamiltonian}
\end{align}
with the plaquette term
\begin{align}
    U_{\mathbf{p}} = U_{\mathbf{x}, 1} U_{\mathbf{x}+\mathbf{e}_{1}, 2} U_{\mathbf{x}+\mathbf{e}_{2}, 1}^{\dagger} U_{\mathbf{x}, 2}^{\dagger}\,
\label{eq:plaquette_2d}    
\end{align}
and the coupling constant $g$ is given by the charge of the electron. The first term in Eq.~\eqref{eq:gauge_part_Hamiltonian} has the meaning of the electric field energy, whereas the second term, which involves the plaquette terms, has the meaning of discretized magnetic field energy. The fermionic part $H_{M}$ of the Hamiltonian is given by
\begin{align}
    H_{M}&=M \sum_{\mathbf{x}}(-1)^{s_{\mathbf{x}}} n_{\mathbf{x}},
\end{align}
where we introduced the occupation number operator
\begin{align}
        n_{\mathbf{x}} = \psi^{\dagger}_{\mathbf{x}}\psi_{\mathbf{x}}
\end{align}
and the staggered charge
\begin{align}
    s_{\mathbf{x}} = \frac{1}{2}[1-(-1)^{\mathbf{x}}] \,.
\end{align}
$M$ denotes the fermionic rest mass. The gauge-matter interaction is 
\begin{align}
    H_{\mathrm{GM}}&=\frac{i}{2}\sum_{\mathbf{x}, i}\left[\psi^{\dagger}_{\mathbf{x}} U_{\mathbf{x}, i} \psi_{\mathbf{x}+\mathbf{e}_{i}} - {h.c.}\right] \,.
\end{align}
The system has local symmetries generated by the Gauss's law operator
\begin{align}
    G_{\mathbf{x}} = \sum_i\left[E_{\mathbf{x}, i}-E_{\mathbf{x}-\mathbf{e}_{i}, i}\right] - n_{\mathbf{x}} + s_{\mathbf{x}}.
\end{align}
Because of the fact that the Hamiltonian $H$ and the Gauss's law operator $G_{\mathbf{x}}$ commute, they have a common eigenvector system 
and we restrict the dynamics to the zero eigenvalues of each $G_{\mathbf{x}}$, i.e., 
\begin{align}
    G_{\mathbf{x}}\ket{\psi} = 0 \:\:\: \forall\mathbf{x}\,.
\end{align}
Precisely these local constraints on the physical states between the matter and the gauge field degrees of freedom will allow us later to formulate a unitarily equivalent system without the matter.
\subsection{U(1) LGT formulated as a qudit system}
\begin{figure}[t]
    \centering  \includegraphics[width=\columnwidth]{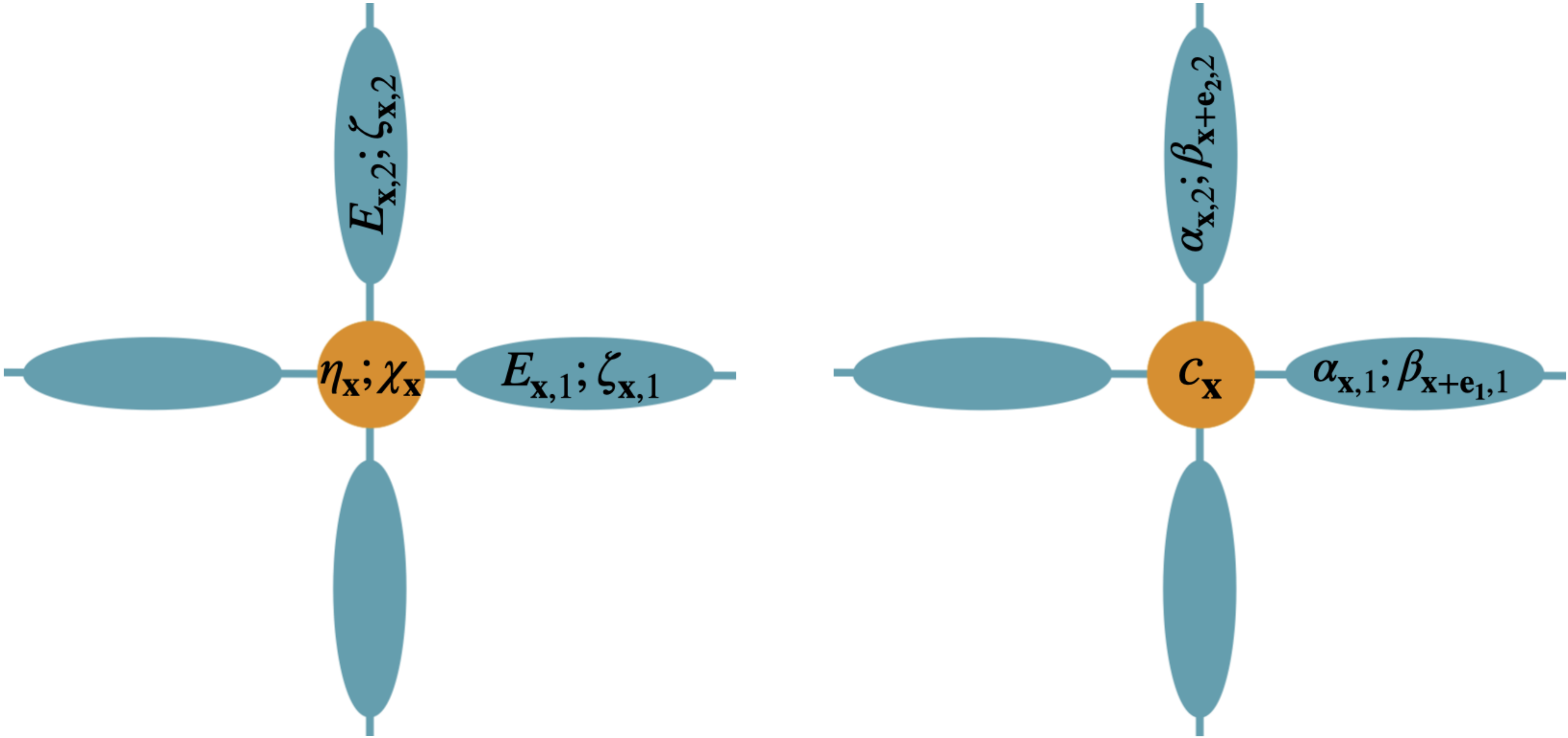}
        \caption{\textbf{Particle content on the lattice in the case of a two-dimensional spatial lattice.} \textbf{Left:} The hardcore bosons $\eta_{\mathbf{x}}$ and the fermions $\chi_{\mathbf{x}}$ reside on the lattice sites (orange circles), whereas the gauge fields $E_{\mathbf{x},i}$ and the fermions $\zeta_{\mathbf{x},i}$ reside on the links (blue ovals). \textbf{Right:} The Majorana modes $c_{\mathbf{x}}, \alpha_{\mathbf{x},i}$  and $\beta_{\mathbf{x},i}$ defined by Eq.~\eqref{eq:majorana_modes} and the corresponding sites/links on which they reside. Thanks to Gauss's law, this configuration permits one to replace the fermionic matter $\psi_{\mathbf{x}}$ residing in the original gauge theory on lattice sites, such that the final theory contains only the gauge fields $E_{\mathbf{x},i}$.}
    \label{fig:lattice_content}
\end{figure}
\begin{figure*}[t!]
    \centering
    \includegraphics[width=\textwidth]{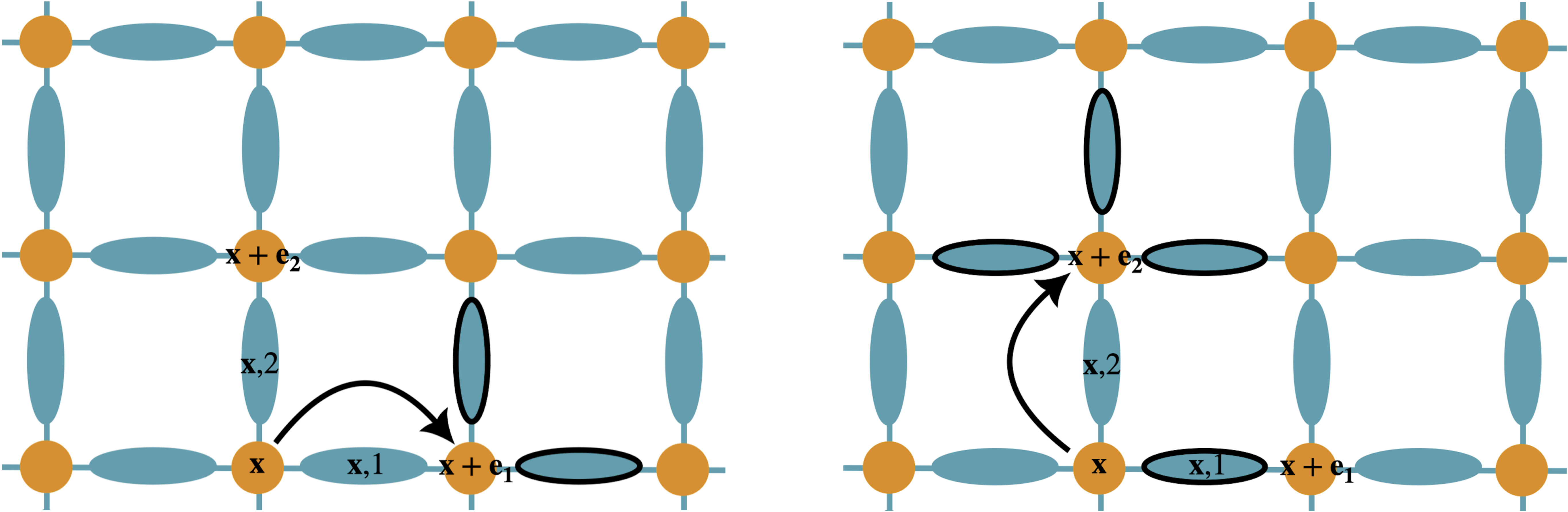}
    \caption{\textbf{
    Two-dimensional spatial lattice with matter fields on the sites (orange circles) and gauge fields on the links (blue ovals):} After performing the transformation from Eq.~\eqref{eq:V_n}, the hopping terms between two neighboring sites (indicated by an arrow) acquire prefactors that depend on the electric fields on the adjacent links. The involved electric fields depend on the hopping direction (horizontal or vertical) and result in direction-dependent phase factors as given in Eq.~\eqref{eq:prefactors}.} 
    \label{fig:lattice_hopping}
\end{figure*}
In this subsection, following two previous works~\cite{Zohar2018,Zohar2019}, we construct a gauge theory from a system of hardcore bosons, fermions, and gauge fields residing on a lattice. This construction introduces one fermionic and one hardcore bosonic degree of freedom per site, both constituting the matter of the LGT. The fermionic degree of freedom keeps track of the fermionic statistics and the hardcore bosonic degree of freedom allows to decouple the attachment of the gauge field to the matter. On top of that, on each link of the lattice, gauge field and fermionic degrees of freedom reside. Moreover, a gauge-invariant Hilbert space and a local Hamiltonian for these degrees of freedom are formulated, so that this Hamiltonian has the same matrix elements as the one of Eq.~\eqref{eq:kogut-susskind_ham}. The degrees of freedom of the constructed LGT are mapped onto a qudit system, while the locality of the interactions is preserved. This mapping allows for the efficient implementation of the LGT simulation in a qudit-based quantum device. In what follows, we present the details of this mapping. 

We consider a Hilbert space consisting of three kinds of particles:
 $\chi_{\mathbf{x}}$, $\eta_{\mathbf{x}}$, and $\zeta_{\mathbf{x},i}$ where $\chi_{\mathbf{x}}$ and $\zeta_{\mathbf{x},i}$ are fermionic operators and $\eta_{\mathbf{x}}$ is a hardcore boson operator (see the defining algebra below). While $\chi_{\mathbf{x}}$ and $\eta_{\mathbf{x}}$  live on the sites of the lattice, the fermion $\zeta_{\mathbf{x},i}$ lives on the links, see FIG.~\ref{fig:lattice_content}. Specifically, these operators fulfill the following 
 commutation and anti-commutation relations:
 \begin{subequations}
  \begin{align}
    \{\chi_{\mathbf{x}},\chi^{\dagger}_{\mathbf{y}}\}&= \delta_{\mathbf{x},\mathbf{y}}, \\ \{\zeta_{\mathbf{x},i},\zeta^{\dagger}_{\mathbf{y},j}\}&= \delta_{\mathbf{x},\mathbf{y}}\delta_{i,j}, \\
    [\eta_{\mathbf{x}},\eta^{\dagger}_{\mathbf{y}}]&= [\eta_{\mathbf{x}},\eta_{\mathbf{y}}]= 0\:\:\text{for}\:\: \mathbf{x}\neq\mathbf{y}\:\:\text{and}\\
    \{\eta_{\mathbf{x}},\eta^{\dagger}_{\mathbf{x}}\}&= 1,\:\:\{\eta_{\mathbf{x}},\eta_{\mathbf{x}}\}=0.    
\end{align}   
 \end{subequations}
For notational convenience, we introduce the following Majorana operators for $\chi_{\mathbf{x}}$ and $\zeta_{\mathbf{x},i}$
\begin{subequations}
    \begin{align}
    c_{\mathbf{x}} &= \chi_{\mathbf{x}} + \chi^{\dagger}_{\mathbf{x}},  \\
    \alpha_{\mathbf{x},i} &= \zeta_{\mathbf{x},i} + \zeta^{\dagger}_{\mathbf{x},i},\\
    \beta_{\mathbf{x},i} &= i(\zeta^{\dagger}_{\mathbf{x}-\mathbf{e_i},i}-\zeta_{\mathbf{x}-\mathbf{e_i},i}).
\end{align}
\label{eq:majorana_modes}
\end{subequations}
By having two modes per site ($\eta_{\mathbf{x}}$ and $c_{\mathbf{x}}$), and one fermionic mode on each link ($\zeta_{\mathbf{x},i}$), we have more degrees of freedom than in the original formulation of the LGT. Therefore, we restrict the operators to act on wave functions that fulfill
\begin{align}
    \chi_{\mathbf{x}} \ket{\psi} &= 0, \notag\\
    \zeta_{\mathbf{x},i} \ket{\psi} &= 0.
\label{eq:new_phys_H_space}
\end{align}
We construct the operator
\begin{align}
    \psi_{\mathbf{x}} = c_{\mathbf{x}} \eta_{\mathbf{x}} \, ,
\end{align}
which is fermionic as can be confirmed by direct calculation. By gauging these fermions, i.e., by coupling them to a gauge field residing on the links of the lattice, as explained above, we can define a Hamiltonian of a LGT like the one in Eq.~\eqref{eq:kogut-susskind_ham}, but acting on the Hilbert space defined by Eq.~\eqref{eq:new_phys_H_space}. This Hamiltonian reads

\begin{align}
 &H = \frac{g^{2}}{2} \sum_{\mathbf{x}, i} E^{2}_{\mathbf{x}, i} - \frac{1}{2 g^{2}} \sum_{\mathbf{p}} \left[ U_{\mathbf{p}}+U_{\mathbf{p}}^{\dagger} \right]\notag \\
+&M \sum_{\mathbf{x}}(-1)^{s_{\mathbf{x}}} \eta^{\dagger}_{\mathbf{x}} \eta_{\mathbf{x}}-\frac{i}{2}\sum_{\mathbf{x},i}\left(\eta^{\dagger}_{\mathbf{x}}c_{\mathbf{x}} U_{\mathbf{x},i}c_{\mathbf{x}+\mathbf{e_i}}\eta_{\mathbf{x}+\mathbf{e_i}}-\text{h.c.}\right)
 \end{align}
and, using the notation introduced above, Gauss's law is
\begin{align}
    G_{\mathbf{x}} = \sum_i\left[E_{\mathbf{x}, i}-E_{\mathbf{x}-\mathbf{e}_{i}, i}\right] - \eta^{\dagger}_{\mathbf{x}}\eta_{\mathbf{x}}+ s_{\mathbf{x}} .
\label{eq:gauss_law_hardcore}
\end{align}

In the following, we perform two unitary transformations to map the dynamics 
to a completely electric-field–dependent operator being a pure qudit system. As shown in Ref.~\cite{Zohar2018}, for specific gauge groups, one can define a unitary transformation that replaces the fermionic operators in the Hamiltonian with operators that act locally on the gauge fields. This unitary transformation is given by the expression
\begin{align}
    V_{\mathbf{x}} \equiv \prod_{i}(ic_{\mathbf{x}}\beta_{\mathbf{x},i})^{E_{\mathbf{x}-\mathbf{e_i},i}} \prod_{i}(ic_{\mathbf{x}}\alpha_{\mathbf{x},i})^{E_{\mathbf{x},i}}.
\label{eq:V_n}
\end{align}
Since the different parts of the product do not commute with each other, we have to explicitly specify the ordering. The choice we made by writing Eq.~\eqref{eq:V_n} coincides with the one in Ref.~\cite{Zohar2018}. Transforming the operators of the gauge theory degrees of freedom yields
\begin{subequations}
  \begin{align}
    V_{\mathbf{x}} U_{\mathbf{x,i}} V_{\mathbf{x}}^{\dagger} &= ip_{\mathbf{x},i}(E_{\cdot,\cdot})c_{\mathbf{x}} \alpha_{\mathbf{x},i} U_{\mathbf{x},i} \,, \\
    V_{\mathbf{x}+\mathbf{e_i}} U_{\mathbf{x,i}} V_{\mathbf{x}+\mathbf{e_i}}^{\dagger} &=i U_{\mathbf{x},i} c_{\mathbf{x}+\mathbf{e_i}} \beta_{\mathbf{x}+\mathbf{e_i},i} \tilde{p}_{\mathbf{x}+\mathbf{e_i},i}(E_{\cdot,\cdot})\,, \\
    V_{\mathbf{x}} c_{\mathbf{x}}V_{\mathbf{x}}^{\dagger} &=\exp[i \pi \sum\limits_i( E_{\mathbf{x},i}+E_{\mathbf{x}-\mathbf{e_i},i})] c_{\mathbf{x}}  \,,\\
     V_{\mathbf{x}} \eta_{\mathbf{x}} V_{\mathbf{x}}^{\dagger} &=\eta_{\mathbf{x}},   \, \\
     V_{\mathbf{x}} \zeta_{\mathbf{x},i} V_{\mathbf{x}}^{\dagger} &=\zeta_{\mathbf{x},i}.
\label{eq:trafo_for_x}
\end{align}   
 \end{subequations}
 We have used the following abbreviation for the prefactors of the transformed link operators:
 \begin{align}
     p_{\mathbf{x},i}(E_{\cdot,\cdot}) &= \exp[i\pi(\sum_{j>i}E_{\mathbf{x},j}+\sum_{j}E_{\mathbf{x}-\mathbf{e_j},j})],\notag\\
    \tilde{p}_{\mathbf{x}+\mathbf{e_i},i}(E_{\cdot,\cdot}) &=  \exp(i\pi\sum_{j>i}E_{\mathbf{x}+\mathbf{e_i}-\mathbf{e_j},j}).
 \end{align}
 Moreover, the property $[V_{\mathbf{x}},V_{\mathbf{y}}] = 0$ allows us to define a unitary transformation $\mathcal{V} = \prod_{\mathbf{x}} V_{\mathbf{x}}$. This global unitary transforms $H_{\text{GM}}$ as
\begin{align}
 &\mathcal{V}H_{\text{GM}}\mathcal{V}^{\dagger} =\notag\\ -&\frac{i}{2}\sum_{\mathbf{x},i}\left(f_{\mathbf{x},i}(E_{\cdot,\cdot})\eta^{\dagger}_{\mathbf{x}}\alpha_{\mathbf{x},i} U_{\mathbf{x},i}\beta_{\mathbf{x}+\mathbf{e_i},i}\eta_{\mathbf{x}+\mathbf{e_i}}-\text{h.c.}\right),
\end{align}
where $f_{\mathbf{x},i}(E_{\cdot,\cdot})$ is a prefactor that depends on the electric fields around the site $\mathbf{x}$ and is also dependent on the spatial direction $i$.
Specifically, in (2+1) dimensions, $i = 1,2$ for horizontal and for vertical hopping, respectively, and we have
\begin{align}
    &f_{\mathbf{x},1}(E_{\cdot,\cdot}) = (-1)^{E_{\mathbf{x}+\mathbf{e_1},1}+E_{\mathbf{x}+\mathbf{e_1},2}}\notag\\
  &f_{\mathbf{x},2}(E_{\cdot,\cdot}) = (-1)^{E_{\mathbf{x},1}+E_{\mathbf{x}-\mathbf{e_1}+\mathbf{e_2},1}+E_{\mathbf{x}+\mathbf{e_2},2}+E_{\mathbf{x}+\mathbf{e_2},1}}.
\label{eq:prefactors}
\end{align}
The electric fields involved in Eq.~\eqref{eq:prefactors} are highlighted in FIG.~\ref{fig:lattice_hopping}. 
In (1+1) dimensions, since there is only one direction, we omit the index $i$ and we have
\begin{align}
    f_{\mathbf{x}}(E_{\cdot,\cdot}) = (-1)^{E_{\mathbf{x+1}}}.
\end{align}
The outcome of this transformation is that the hopping term $H_{\text{GM}}$ does not include the Majorana modes $c_{\mathbf{x}}$ anymore. 

The plaquette term present in the Hamiltonian in higher spatial dimensions also transforms non-trivially:
\begin{align}
    &\mathcal{V}H_G\mathcal{V}^{\dagger} = \frac{g^{2}}{2} \sum_{\mathbf{x}, i} E^{2}_{\mathbf{x}, i} - \frac{1}{2 g^{2}} \sum_{\mathbf{p}} \left(\tilde{U}_{\mathbf{p}}+\tilde{U}_{\mathbf{p}}^{\dagger}\right)\,, \notag \\
\end{align}
with the transformed plaquette operator defined as

\begin{align}
    \tilde{U}_{\mathbf{p}} = &\exp[i\pi(E_{\mathbf{x}, 1}+E_{\mathbf{x}+\mathbf{e}_{1}, 2}+E_{\mathbf{x}+\mathbf{e}_{2}, 2}+E_{\mathbf{x}+\mathbf{e_2}-\mathbf{e_1}, 1})]\notag\\\times &U_{\mathbf{x}, 1} U_{\mathbf{x}+\mathbf{e}_{1}, 2}U^{\dagger}_{\mathbf{x}+\mathbf{e}_{2}, 1}U^{\dagger}_{\mathbf{x}, 2}.
\label{eq:plaquette_2d_tilde}  
\end{align}
After the transformation, the occupation numbers of the fermionic mode $\chi_{\mathbf{x}}$ in the physical space are the same as the occupation numbers of the hardcore bosons $\eta_{\mathbf{x}}$.
Furthermore, as easy to see from Eq.~\eqref{eq:trafo_for_x}, the modes $\zeta_{\mathbf{x},i}$ transform identically, therefore they stay in vacuum. The Hamiltonian of the system projected onto the subspace of absent fermions $\zeta_{\mathbf{x},i}$ reads 
\begin{align}
 H =& \frac{g^{2}}{2} \sum_{\mathbf{x}, i} E^{2}_{\mathbf{x}, i} - \frac{1}{2 g^{2}} \sum_{\mathbf{p}} \left[ \tilde{U}_{\mathbf{p}}+\tilde{U}_{\mathbf{p}}^{\dagger} \right]\notag \\
+&M \sum_{\mathbf{x}}(-1)^{s_{\mathbf{x}}} \eta^{\dagger}_{\mathbf{x}} \eta_{\mathbf{x}}\notag \\+&\frac{1}{2}\sum_{\mathbf{x},i}\left(f_{\mathbf{x},i}(E_{\cdot,\cdot})\eta^{\dagger}_{\mathbf{x}} U_{\mathbf{x},i}\eta_{\mathbf{x}+\mathbf{e_i}}+\text{h.c.}\right).
 \end{align}

The hardcore bosons on each site are furthermore related to the divergence of the gauge fields on the adjacent links due to the Gauss's law, Eq.~\eqref{eq:gauss_law_hardcore}. Precisely these local constraints allow for complete elimination of the matter from the Hamiltonian, as shown in Ref.~\cite{Zohar2019}. We can rewrite the Gauss's law also as
\begin{align}
g_{\mathbf{x}}|\psi\rangle=n_{\mathbf{x}}|\psi\rangle,
\label{eq:modifiedGaussLaw}
\end{align}
where we introduced 
\begin{align}
    g_{\mathbf{x}}= \sum_i\left[E_{\mathbf{x}, i}-E_{\mathbf{x}-\mathbf{e}_{i}, i}\right]+s_{\mathbf{x}} \,.
\end{align}
We introduce the operator
\begin{align}
    \mathcal{U}_{\mathbf{x}}=(\eta_{\mathbf{x}}+\eta^{\dagger}_{\mathbf{x}})^{g_{\mathbf{x}}} \,
\label{eq:unitary_remove_hardcore}
\end{align}
and the unitary transformation for the entire lattice
\begin{align}
\mathcal{U} = \prod_{\mathbf{x}} \mathcal{U}_{\mathbf{x}} \,.   
\end{align}
We denote physical states transformed by $\mathcal{U}$ by a tilde $|\tilde{\psi}\rangle$ and define them as 
\begin{align}
    |\tilde{\psi}\rangle=\mathcal{U}|\psi\rangle \,.
\end{align}
The matter transforms as
\begin{align}
    \mathcal{U}_{\mathbf{x}}  \eta^{\dagger}_{\mathbf{x}}\eta_{\mathbf{x}}   \mathcal{U}^{\dagger}_{\mathbf{x}} =(-1)^{g_{\mathbf{x}} } \eta^{\dagger}_{\mathbf{x}}\eta_{\mathbf{x}} +\frac{1}{2}(1-(-1)^{g_{\mathbf{x}}})   \,.
\end{align}

\begin{align}
    \eta^{\dagger}_{\mathbf{x}}\eta_{\mathbf{x}}|\tilde{\psi}\rangle=\left(1-(-1)^{g_{\mathbf{x}}}(1-2 g_{\mathbf{x}})\right)|\tilde{\psi}\rangle \,.
\end{align}
Since for a vector of the physical Hilbert space, the operator $g_{\mathbf{x}}$ has eigenvalues 0 or 1, we obtain
\begin{align}
\eta^{\dagger}_{\mathbf{x}}\eta_{\mathbf{x}}|\tilde{\psi}\rangle=0 \,.
\label{eq:physical_states}
\end{align}
This last equation signifies that all matter will be transformed to its vacuum state.

Next, we apply the unitary transformation on the Hamiltonian of the gauge theory. In order to write down the transformed Hamiltonian in a convenient form, we define on each vertex $\mathbf{x}$ 
the operators $P_{g,\mathbf{x}}$, which projects on the $g = 0$ or $g = 1$ eigenspace of the 
operator $g_{\mathbf{x}}$. Applying  Eq.~\eqref{eq:unitary_remove_hardcore} to the raising and the lowering operators in the interaction part of the Hamiltonian, we get precisely the projector $P_{1,\mathbf{x}}$. Note, that the ordering of the link operators and the projectors is now important since they do not commute. We end up with the following Hamiltonian in (2+1) dimensions

\begin{widetext}
\begin{align}
    H = &\frac{g^{2}}{2} \sum_{\mathbf{x}, i} E^{2}_{\mathbf{x}, i} - \frac{1}{2 g^{2}} \sum_{\mathbf{p}} \left[ \tilde{U}_{\mathbf{p}}+\tilde{U}_{\mathbf{p}}^{\dagger} \right]+2M\sum_{\mathbf{x},i}(-1)^{\mathbf{x}}E_{\mathbf{x}, i}\notag\\+&\frac{1}{2}\sum_{x}\left(P_{1,\boldsymbol{x}}(-1)^{E_{\boldsymbol{x},1}+E_{\boldsymbol{x+e_1-e_2},2}-s_{\boldsymbol{x+e_1}}}U_{\boldsymbol{x},1}P_{1,\boldsymbol{x+e_1}}-P_{1,\boldsymbol{x}}(-1)^{E_{\boldsymbol{x},1}+E_{\boldsymbol{x},2}-s_{\boldsymbol{x+e_2}}}U_{\boldsymbol{x},2}P_{1,\boldsymbol{x+e_2}}+\text{h.c.}\right).
\label{eq:final_Hamiltonian_2d}
\end{align}    
\end{widetext}
In (1+1) dimensions, the Hamiltonian simplifies to the form 
\begin{align}
    H &= \frac{g^2}{2}\sum_{\mathbf{x}} E_{\mathbf{x}}^2+2M \sum_{\mathbf{x}} (-1)^{{\mathbf{x}}} E_{\mathbf{x}} \notag \\
    &+\sum_{\mathbf{x}} \big[P_{1,\mathbf{x}}(-1)^{E_{\mathbf{x}+1}}U_{\mathbf{x}} P_{1,\mathbf{x}+1}+\text{h.c.}\big].
\label{eq:final_ham_1d}
\end{align}
Since we have decoupled the matter by a unitary transformation, we are left with a system of qudits that reside on the links of the lattice, with dynamics governed by a local Hamiltonian. This formulation allows for an efficient implementation on a qudit quantum device. 

In summary, we achieved the following: We applied two transformations on the LGT. The first transformation (Eq.~\eqref{eq:V_n}) removed the Majorana modes $c_{\mathbf{x}}$ from the Hamiltonian and replaced them effectively with electric-field--dependent prefactors. On the Hilbert space level, it coupled the modes $c_{\mathbf{x}}$ and $\eta_{\mathbf{x}}$ so that physical states only contain configurations with equal occupation numbers for both $c_{\mathbf{x}}$ and $\eta_{\mathbf{x}}$. This allowed us to keep the modes $\eta_{\mathbf{x}}$ only.

The second transformation (Eq.~\eqref{eq:unitary_remove_hardcore}), removed the hardcore bosons $\eta_{\mathbf{x}}$ by using the constraints between matter and gauge field degrees of freedom (Gauss's law). As indicated in Eq.~\eqref{eq:physical_states}, physical states after the transformation are the ones with all the matter fields in the vacuum state. Furthermore, the transformed Hamiltonian projects onto the subspace of physical states, making only the gauge fields have non-trivial dynamics. This allowed us to consider a system of gauge fields only. Combining both transformations, we arrived at local qudit Hamiltonian, given by Eq.~\eqref{eq:final_Hamiltonian_2d} or Eq.~\eqref{eq:final_ham_1d} in (2+1) or (1+1) dimensions, respectively.

Later in this work, for the numerical studies and the explicit circuits, we will consider a system of qutrits --- qudits with three internal levels ($d=3$). However, our approach is general and can be used (after modification of the quantum circuits) for general $d$-level qudits.

\section{Variational quantum simulation}
\label{sec:variational_quantum_simulation}
The simulation of quantum many-body systems stands as a top application of quantum computers. However, the task of engineering intricate Hamiltonians poses a significant challenge. In the realm of analog quantum simulation, the construction of the Hamiltonian itself proves to be the challenge, whereas, in digital quantum simulation, the primary difficulties lie in efficient digitization and fault tolerance with minimal resources.
To tackle these obstacles in both analog and digital quantum simulation, variational algorithms have emerged as particularly powerful and represent a potent tool for both classical and quantum systems. In this approach, a parametrized family of quantum states  is efficiently prepared through a variational Ansatz circuit, which is typically based on the physical properties and symmetries of the system under study. Here, the classical computer is tasked with optimizing a cost function, such as the energy of the candidate states for a given Hamiltonian, while the quantum computer is responsible for measuring the cost function on the system. Through an iterative process, this variational approach can provide an efficient representation of quantum many-body states, such as ground states of a given Hamiltonian. 


Recent technological advancements now permit the implementation and study of quantum variational algorithms. In the quantum domain, variational simulation finds frequent application in many-body physics, encompassing disciplines such as solid-state physics, high-energy physics, and chemistry~\cite{Tilly2022}. Notably, in the field of LGTs, there have been proposals and experimental implementations of simulation protocols based on the variational principle~\cite{Kokail2019,Ferguson2021,Atas2021,Paulson2021,Atas2023,Chan2023}. The principle has been used to create approximations to the ground state~\cite{Peruzzo2014,Kokail2019,McArdle2019, Atas2021}, thermal states~\cite{Verdon2019}, excited states~\cite{Higgott2019, Atas2021}, or the dynamics~\cite{Yuan2019, Barison2021} of a quantum many-body system.

\subsection{Parametrized quantum circuits}

We consider trial quantum states generated by a parametrized quantum circuit (PQC) of the form
\begin{align}
    |\psi(\boldsymbol{\theta})\rangle=\mathcal{U}(\boldsymbol{\theta})\ket{\psi_0}=U_{N}\left(\theta_{N}\right) \ldots U_{k}\left(\theta_{k}\right) \ldots U_{1}\left(\theta_{1}\right)\ket{\psi_0}\,,
    \label{eq:TrialState}
\end{align}
where $U_k(\theta_k)$ are unitary operations acting on the initial state $\ket{\psi_0}$. These unitaries are parameterized by a real number $\theta_k$ and which can be written as 
\begin{align}
    U_k(\theta_k) =  e^{-i \theta_k g_k} \,.
\label{eq:unitary_k}
\end{align}
The generators $g_k$ can be chosen in a hardware-efficient fashion according to the native operations that can be performed on the specific quantum device, where the variational algorithm should be executed. Here, we consider the gate set used in the trapped-ion qudit quantum computer of Ref.~\cite{Ringbauer2022}, where the native gates are two-level single-qudit rotations and entangling Mølmer--Sørensen (MS) gates. In addition, we will consider qudit controlled-rotation (CROT) gates, which can be constructed from the MS gate~\cite{Ringbauer2022}. For more details on the implementation, see Section~\ref{sec:implementation}.

In the following, we discuss the variational
principle for imaginary-time and real-time evolution, where the exact imaginary time evolution converges to the ground state of the quantum many-body
systems at infinite imaginary time and the real-time evolution describes the change of the quantum many-body dynamics in time. In addition, we explain a strategy to determine the imaginary and real-time evolution from a measurement of the quantum device. 
\subsection{Variational time evolution}
The equation describing the dynamics of a quantum state $\rho$ of a system with Hamiltonian $H$ is 
\begin{equation}
    \partial_t \rho(t) = \mathcal{L}[\rho(t)] \, ,
\end{equation}
where the super-operator $\mathcal{L}[\cdot]$ is given by
\begin{align}
      \mathcal{L}[\rho(t)] = -i[H,\rho(t)] 
\label{eq:real_time_usperop}
\end{align}
for Hamiltonian dynamics and 
\begin{align}
    \mathcal{L}[\rho(t)]=-\{H, \rho(t)\}+2\langle H\rangle \rho(t)
\label{eq:imag_time_superop}
\end{align}
for imaginary time evolution. The expectation value in Eq.~\eqref{eq:imag_time_superop} is taken in the state $\rho(t)$, leading to non-linear dynamics.

In the following, we parameterize the quantum state $\rho(t)$ by a set of variational parameters 
$\boldsymbol{\theta}(t)$, which are time-dependent on their own, and we can obtain the (approximate) 
time evolution by minimizing the so-called McLachlan distance,
\begin{equation}
   \Big\Vert \sum_{\mu}\frac{\partial \rho}{\partial\theta_{\mu}}\dot{\theta}_{\mu}-
    \mathcal{L}[\rho]\Big\Vert^2 \,, 
\end{equation}
with respect to $\boldsymbol{\theta}(t)$, i.e., 
trajectories of the variational parameters. The minimization leads to the equations of motion for the variational parameters
\begin{equation}
    \sum_{\nu}M_{\mu\nu}\dot{\theta}_{\nu}(t) = V_{\mu}\,.
\label{eq:eom_for_the_thetas}
\end{equation}
The real symmetric  matrix $M_{\mu\nu}$ represents the so-called Fubini-Study metric tensor~\cite{Wierichs2022} and is defined by
\begin{align}
M_{\mu\nu} = &\mathrm{Tr}\Bigg[\frac{\partial \rho}{\partial\theta_{\mu}}\frac{\partial \rho}{\partial\theta_{\nu}}\Bigg] \,
\label{eq:metric_tensor}
\end{align}
Note that $M_{\mu\nu}$ does not contain information about the time evolution operator. In turn, the vector
\begin{align}
V_{\mu} = \mathrm{Tr}\Bigg[\Bigg(\frac{\partial \rho}{\partial\theta_{\mu}}\mathcal{L}[\rho]\Bigg)\Bigg]
\label{eq:vector_V}
\end{align}
contains information about the time-evolution operator $\mathcal{L}[\cdot]$, which is different for real-time Eq.~\eqref{eq:real_time_usperop} and for imaginary-time evolution Eq.~\eqref{eq:imag_time_superop}.

In the following, we will assume that the variational state $\rho(\boldsymbol{\theta})$ is a pure state defined by a PQC:
\begin{align}
   \rho(\boldsymbol{\theta}) = \ket{\psi(\boldsymbol{\theta})}\bra{\psi(\boldsymbol{\theta})}. 
\label{eq:pure_state}
\end{align}
We will suppress the explicit dependence on $\boldsymbol{\theta}$ for notational convenience. Inserting the pure state from Eq.~\eqref{eq:pure_state} in the definition of the metric tensor Eq.~\eqref{eq:metric_tensor} leads to
\begin{align}
    M_{\mu\nu} = \text{Re}(\langle\partial_{\mu}\psi|\partial_{\nu}\psi\rangle) - \langle\partial_{\mu}\psi|\psi\rangle\langle\psi|\partial_{\nu}\psi\rangle.
\label{eq:Mmunu}
\end{align}
For imaginary-time evolution, the vector $V^{I}_{\mu}$, once we insert the pure state into Eq.~\eqref{eq:vector_V}, is given by 
\begin{align}
    V^{I}_{\mu} = \langle\partial_{\mu}\psi| H |\psi\rangle + \langle\psi| H |\partial_{\mu}\psi\rangle = \partial_{\mu}\langle H \rangle. 
\label{eq:vector_V_imag}
\end{align}
For real-time evolution, instead, we have
\begin{align}
    V^{R}_{\mu}= &i\left(\langle\psi| H |\partial_{\mu}\psi\rangle - \langle\partial_{\mu}\psi| H |\psi\rangle\right)\notag\\ + &i\left(\langle\partial_{\mu}\psi |\psi\rangle - \langle\psi |\partial_{\mu}\psi\rangle\right)\langle H \rangle.
\label{eq:vector_V_real}
\end{align}
The execution of the variational time evolution algorithm requires that the quantities $M_{\mu\nu}$ in Eq.~\eqref{eq:Mmunu}, $V^{I}_{\mu}$ in Eq.~\eqref{eq:vector_V_imag} and $V^{R}_{\mu}$ in Eq.~\eqref{eq:vector_V_real} are being measured on the quantum device. In the rest of this Section, we will give possible measurement schemes for these quantities.

\subsection{Direct measurement protocol for  $M_{\mu\nu}$ and $V_{\mu}$ in the experiment}
We proceed with the discussion on how to measure the quantities of Eq.~\eqref{eq:metric_tensor} and Eq.~\eqref{eq:vector_V} on the quantum device. First, we define the overlap functions
\begin{align}
    f_{\mu\nu}(a) &= \operatorname{Tr} [ \rho(\boldsymbol{\theta}+a\boldsymbol{e}_{\mu})\rho(\boldsymbol{\theta}+a\boldsymbol{e}_{\nu}) ]\,, \notag\\
    p_{\mu}(a) &= \text{Tr}[  \rho(\boldsymbol{\theta} + a\boldsymbol{e}_{\mu}) \rho(\boldsymbol{\theta})].
    \label{eq:DefOfG}
\end{align}
The components of the metric tensor are related to the second derivatives of these functions as
\begin{align}
    \partial^2_a f_{\mu\nu}(a) |_{a=0} =  2 M_{\mu\nu}  - \partial^2_a p_{\mu}(a) |_{a=0} - \partial^2_a p_{\nu}(a) |_{a=0}  \,.
\label{eq:connection_M_fp}
\end{align}
For the diagonal elements, since we consider pure states with the property
\begin{align}
 \text{Tr}[ \rho(\boldsymbol{\theta} + a \boldsymbol{e}_{\mu}) \rho(\boldsymbol{\theta} + a \boldsymbol{e}_{\mu})] = 1\,,
\end{align}
we obtain
\begin{align}
    M_{\mu\mu} &=  \partial^2_a p_{\mu}(a) |_{a=0}  \,.
\end{align}
Therefore, either measuring the derivatives of the functions in Eq.~\eqref{eq:DefOfG} directly on the quantum device or measuring the functions and then calculating the derivatives, will allow us to obtain the metric tensor. In Appendix~\ref{app:HowToMeasureM}, we explain how to extract the derivatives by measuring the functions, using the parameter shift rules \cite{Wierichs2022}.

The components of the vector $V_{\mu}$ for imaginary-time evolution are also derivatives of observables that can be measured on the quantum device. From Eq.~\eqref{eq:vector_V_imag}, it is evident that the first derivatives of the functions
\begin{align}
    v^I_{\mu}(a) &= \text{Tr}\{ \rho(\boldsymbol{\theta}+a\boldsymbol{e}_{\mu})H \}, 
\label{eq:function_for_V}
\end{align}
are directly connected to the components of $V^I_{\mu}$:
\begin{align}
    V^I_{\mu} = \partial_a v^I_{\mu}(a)|_{a=0}.
\label{eq:v_der}
\end{align}
In Appendix~\ref{app:HowToMeasureVImag}, we discuss how to extract $V^I$ from the values of the function $v^I_{\mu}(a)$.

For real-time evolution, in contrast, $V^R_{\mu}$ cannot be represented as a derivative of an observable. However, we can make use of the representation of the unitary operations involved in the PQC in Eq.~\eqref{eq:unitary_k}. Performing the derivatives with respect to the variational parameters in Eq.~\eqref{eq:vector_V_real}, we obtain
\begin{equation}
    V^{R}_{\mu} = \langle\{\Tilde{g}_{\mu},\Tilde{H}\}_c\rangle_0,
\label{eq:anti-commutator_V_real}
\end{equation}
where we defined
\begin{align}
    &\Tilde{g}_{\mu} = \mathcal{U}^{\dagger}(\boldsymbol{\theta}_{1:\mu})g_{\mu}\mathcal{U}(\boldsymbol{\theta}_{1:\mu})\notag\\
    &\Tilde{H} = \mathcal{U}(\boldsymbol{\theta})^{\dagger}H\mathcal{U}(\boldsymbol{\theta}),
\end{align}
using the notation
\begin{align}
    \mathcal{U}(\boldsymbol{\theta}_{i:j}) = U_{j}\left(\theta_{j}\right) \ldots U_{k}\left(\theta_{k}\right) \ldots U_{i}\left(\theta_{i}\right).
\end{align}
The subscript $c$ in Eq.~\eqref{eq:anti-commutator_V_real} means that we consider the connected (anti-)commutator.
One possible measuring procedure for such a quantity involves averaging over global random unitaries~\cite{Vermersch2019,Schuckert2020} 
\begin{align}
    \langle\{\Tilde{g}_{\mu},\Tilde{H}\}_c\rangle_0 = &\mathcal{N}_{H}(\mathcal{N}_{H}+1)\overline{\langle \Tilde{g}_{\mu} \rangle_u\langle \Tilde{H} \rangle_u[(\mathcal{N}_{H}+2)\langle \rho_{0} \rangle_u-1]}\notag\\
    -&\mathrm{Tr}(\Tilde{g}_{\mu})\langle\Tilde{H}\rangle_{\mu}-\mathrm{Tr}(\Tilde{H})\langle \Tilde{g}_{\mu}\rangle_{0}.
\end{align}
Here, $\overline{{\phantom{\hspace{0.3mm}}}\cdot{\phantom{\hspace{0.3mm}}}}$ denotes the averaging over random unitaries, drawn from the so-called circular unitary ensemble (see Ref.~\cite{Vermersch2019} for details), and $\mathcal{N}_H$ is the Hilbert space dimension. Hence, instead of having to deal with the unequal-time anti-commutator of Eq.~\eqref{eq:anti-commutator_V_real}, we just have to measure expectation values of observables ($\langle\cdot\rangle_u$) in states that result from the action of global random unitaries on the state to be measured and averaged over these unitaries. 

Here, we used a generic operator $H$ representing the whole Hamiltonian. However, the local structure of our particular Hamiltonian from Eq.~\eqref{eq:final_Hamiltonian_2d} or Eq.~\eqref{eq:final_ham_1d} is of great use, since we can represent the anti-commutator from 
Eq.~\eqref{eq:anti-commutator_V_real} as a sum of anti-commutators with each summand of the Hamiltonian.

One challenge related to the execution of this direct measurement protocol is the high number of repetitions that is needed for the reliable estimation of the matrix and vector elements $M_{\mu\nu}$ and $V_{\mu}$. This is due to the fact that these quantities contain derivatives of observables, the estimation of which requires the evaluation of the function at many nearby points. Remarkably, this difficulty can be circumvented by performing an alternative measurement procedure using an additional ancilla, as described below.

\subsection{Indirect measurement of $M_{\mu\nu}$ and $V_{\mu}$ through an ancilla}
\begin{figure}[t]
    \centering
    \includegraphics[width = 8.6cm]{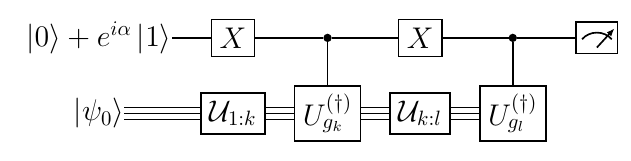}
    \caption{\textbf{Hadamard test for measuring unequal-time (anti-)commutators:} In order to measure the (anti-)~commutator, the ancillary qubit needs to be initialized in an equal superposition of $\ket{0}$ and $\ket{1}$ with a relative phase of $\alpha = 0$ ($\frac{\pi}{2}$). We obtain the desired correlation function by measuring the probability of the ancilla to be in the state $\ket{+}$.} 
    \label{fig:Hadamard_test}
\end{figure}
In this subsection, we give an alternative protocol for measuring the components of the matrix $M_{\mu\nu}$ and the vector $V_{\mu}$. This protocol is based on performing Hadamard tests on the quantum device in order to extract unequal-time (anti-)commutators. We have already shown in Eq.~\eqref{eq:anti-commutator_V_real}  that for real-time evolution, the components of $V^R_{\mu}$ can be written as unequal-time anti-commutators of observables. It is also possible for the components of $M_{\mu\nu}$ and $V^I_{\mu}$ to be written as unequal-time (anti-)commutators. A simple calculation shows that for pure states, we have
\begin{equation}
    M_{\mu\nu} = \langle\{\Tilde{g}_{\mu},\Tilde{g}_{\nu}\}_c\rangle_0
\end{equation}
and 
\begin{align}
V^I_{\mu} = i\langle[ \Tilde{g}_{\mu},\Tilde{H}] \rangle_0.
\label{eq:gradient_as_comm}
\end{align}
We now present a protocol for indirect measurement of such qudit quantities through an ancilla. In the following, we briefly outline the main idea of the protocol and show how to use it for measuring $M_{\mu\nu}$ and $V_{\mu}$. In a separate work~\cite{Popov2023}, we will present the algorithm in a more general  framework, show how one can use it to extract unequal-time correlation functions of observables in the experiment and compare it to other such algorithms.

We observe that we can write a general Hermitian matrix $H$ as a sum of two unitary matrices, i.e. $U_H + U^{\dagger}_H$.
Once we insert this expression in the unequal-time \break (anti-)commutator, we obtain a sum of four quantities. We measure the unequal-time (anti-)commutator of two generators $g_k$ and $g_l$ with the circuits shown in FIG.~\ref{fig:Hadamard_test}. In these circuits we have to couple an ancillary qubit to the qudit register that represents our system, using a controlled operation that corresponds to each of the unitaries $U_H$. These circuits are also known as the Hadamard test. In total, we can measure the (anti-)commutator by performing four Hadamard tests, one for each combination of unitaries. Further, we can measure anti-commutator or commutator, if we choose the initial phase $\alpha = 0$ or $\frac{\pi}{2}$, correspondingly.

\subsection{Measuring the Hamiltonian}
In this subsection, we rewrite the Hamiltonian in a form convenient for performing the measurement protocols from the previous subsection. For simplicity, we discuss the case of a (1+1)-dimensional theory and comment on the case of (2+1) dimensions. The first two terms in Eq.~\eqref{eq:final_ham_1d} are diagonal in the computational qudit basis, and thus can directly be measured by projectively reading out the qudit state. In addition, these terms are also local, so that the resulting controlled unitaries for the execution of the Hadamard test represent entangling operations between only the ancilla and the corresponding qudit.

The third term in Eq.~\eqref{eq:final_ham_1d} has to be rewritten as a tensor product of local observables in order to be measured projectively. In the case of qutrits, we have 
\begin{align}
    H_{\text{GM}} = P^1_{\mathbf{x}-1}\sigma^{1,2}_{X,\mathbf{x}}P^1_{\mathbf{x}+1}-P^{0}_{\mathbf{x}-1}\sigma^{0,1}_{X,\mathbf{x}}P^{0}_{\mathbf{x}+1}.
\end{align}
Here, we introduced the projector onto the $i$-th level of the qudit on site $\mathbf{x}$ as $P^i_{\mathbf{x}}$. The matrices $\sigma^{i,j}_{X,\mathbf{x}}$ are two-level Pauli-X matrices embedded in the qudit Hilbert space; these are defined as 
\begin{align}
    (\sigma^{i,j}_X)_{m,n} &= \begin{cases}
    1\,, \quad \text{if }  (m,n) = (i,j) \text{ or } (j,i)  \\
    0\,, \quad \text{otherwise}
    \end{cases}\\
    (\sigma^{i,j}_Y)_{m,n} &= \begin{cases}
    -i \,, \quad \text{if }  (m,n) = (i,j)  \\
    i \,, \quad \text{if }  (m,n) = (j,i)  \\    
    0\,, \quad \text{otherwise}
    \end{cases}\\
        (\sigma^{i,j}_Z)_{m,n} &= \begin{cases}
    1 \,, \quad \text{if }  (m,n) = (i,i)  \\
    -1 \,, \quad \text{if }  (m,n) = (j,j)  \\    
    0\,, \quad \text{otherwise.}
    \end{cases}\\
\notag
\label{eq:two_level_pauli_matrices}
\end{align}
Measuring this part of the Hamiltonian projectively would therefore involve a local change of bases. 

In order to implement the measurement procedure involving Hadamard tests for this part of the Hamiltonian in a quantum device, we need to rewrite the different terms as a sum of unitaries. For a Pauli matrix acting on a two-level subspace of a qutrit, the corresponding unitary decomposition is given by
\begin{align}
    \sigma^{1,2}_{X,\mathbf{x}} &= U^{1,2}_{X,\mathbf{x}} + \left(U^{1,2}_{X,\mathbf{x}}\right)^{\dagger}
\end{align}
where we defined
\begin{align}
    U^{1,2}_{X,\mathbf{x}} &= \frac{1}{2}(\sigma^{1,2}_{X,\mathbf{x}} + i\ket{0}_{\mathbf{x}}\bra{0}_{\mathbf{x}}).   
\end{align}
We construct the unitary decomposition of the product of all projectors
\begin{align}
P^1_{\mathbf{x}-1}P^1_{\mathbf{x}+1} = W^{P,1}_{\mathbf{x}} + \left(W^{P,1}_{\mathbf{x}}\right)^{\dagger}\,,
\end{align}
with 
\begin{align}
    W^{P,1}_{\mathbf{x}} = \frac{i}{2}\mathbb{1}+ \frac{1-i}{2}\left(\ket{1}_{\mathbf{x-1}}\otimes\ket{1}_{\mathbf{x+1}}\right)\left(\bra{1}_{\mathbf{x-1}}\otimes\bra{1}_{\mathbf{x+1}}\right).
\end{align}
The unitary decomposition of the first term in the Hamiltonian $H_{\text{GM}}$ is given by
\begin{align}
P^1_{\mathbf{x}-1}\sigma^{1,2}_{X,\mathbf{x}}P^1_{\mathbf{x}+1} = U^{1,2}_{X,\mathbf{x}}W^{P,1}_{\mathbf{x}} + U^{1,2}_{X,\mathbf{x}}\left(W^{P,1}_{\mathbf{x}}\right)^{\dagger} + \text{h.c.} \, .
\end{align}
The decomposition of the whole Hamiltonian as a sum of local  unitary operators has to be inserted in Eq.~\eqref{eq:anti-commutator_V_real} and Eq.~\eqref{eq:gradient_as_comm}. Furthermore, each term of the resulting expression can be measured by the circuits given in FIG.~\ref{fig:Hadamard_test}.

In the case of the Hamiltonian in (2+1) spatial dimensions, the number of summands of $H_{\text{GM}}$ is larger and we need to perform entangling gates that correspond to the few-body product of projectors. This means some few-body-controlled operations are involved. Recent proposals for the experimental implementation of such gates have been made~\cite{Katz2022}. Alternatively, we could write each projector as a sum of two unitaries; in this case, we would only have to perform two-body entangling operations at the expense of more Hadamard tests.

In summary, we have proposed two measurement protocols for the extraction of the geometric tensor $M_{\mu\nu}$ and of the vector $V_{\mu}$, which make the implementation of the variational time evolution on a qudit quantum device possible. 


\section{Towards implementation on qudit hardware}
\label{sec:implementation}
\begin{figure}[t!]
    \centering
    \includegraphics[width = \columnwidth]{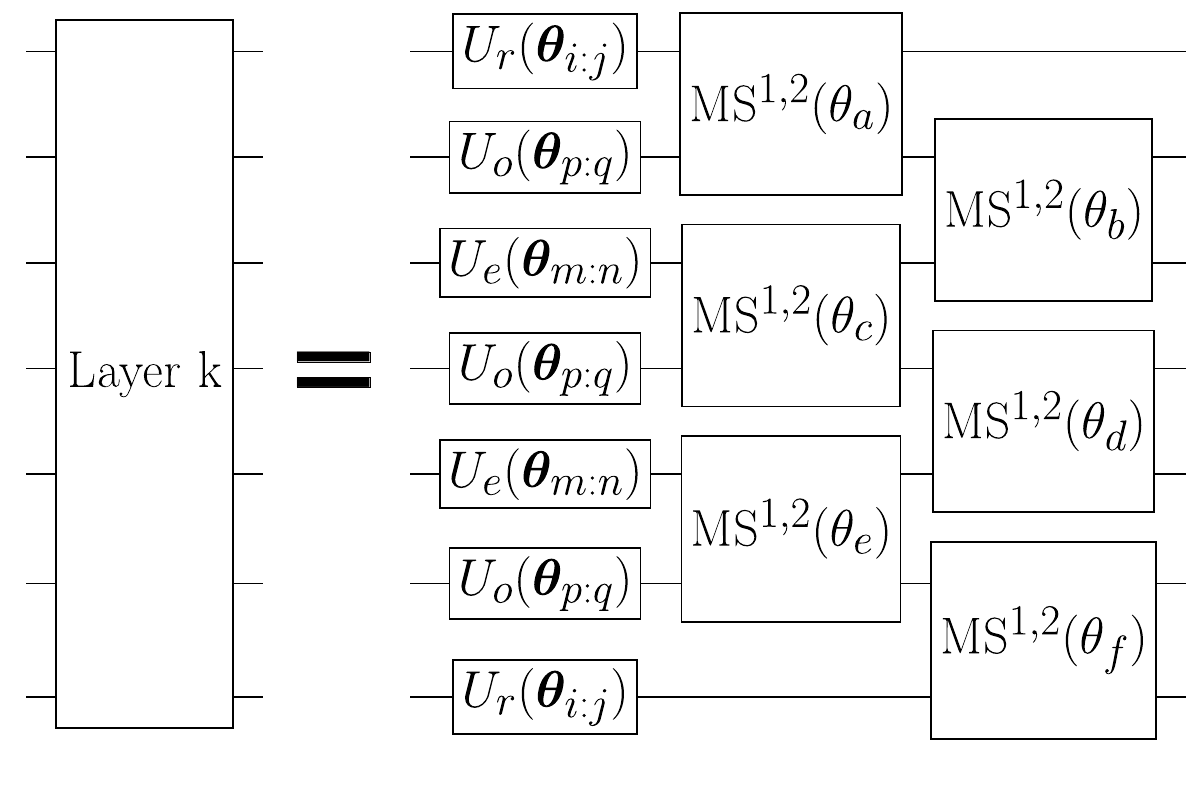}
    \caption{\textbf{Individual layers of the variational quantum circuit for seven qudits in (1+1) dimensions:} Each building block of the parametrized quantum circuit used for generating the trial states is built from a set of single qudit operations, given by two-level rotations, and entangling operations, given by the two-level Mølmer-Sørensen~\cite{Sorensen1999, Molmer1999} gate. }
    \label{fig:rte_quantum_circuit_1d}
\end{figure}

\begin{figure}[b]
    \centering
    \includegraphics[width = \columnwidth]{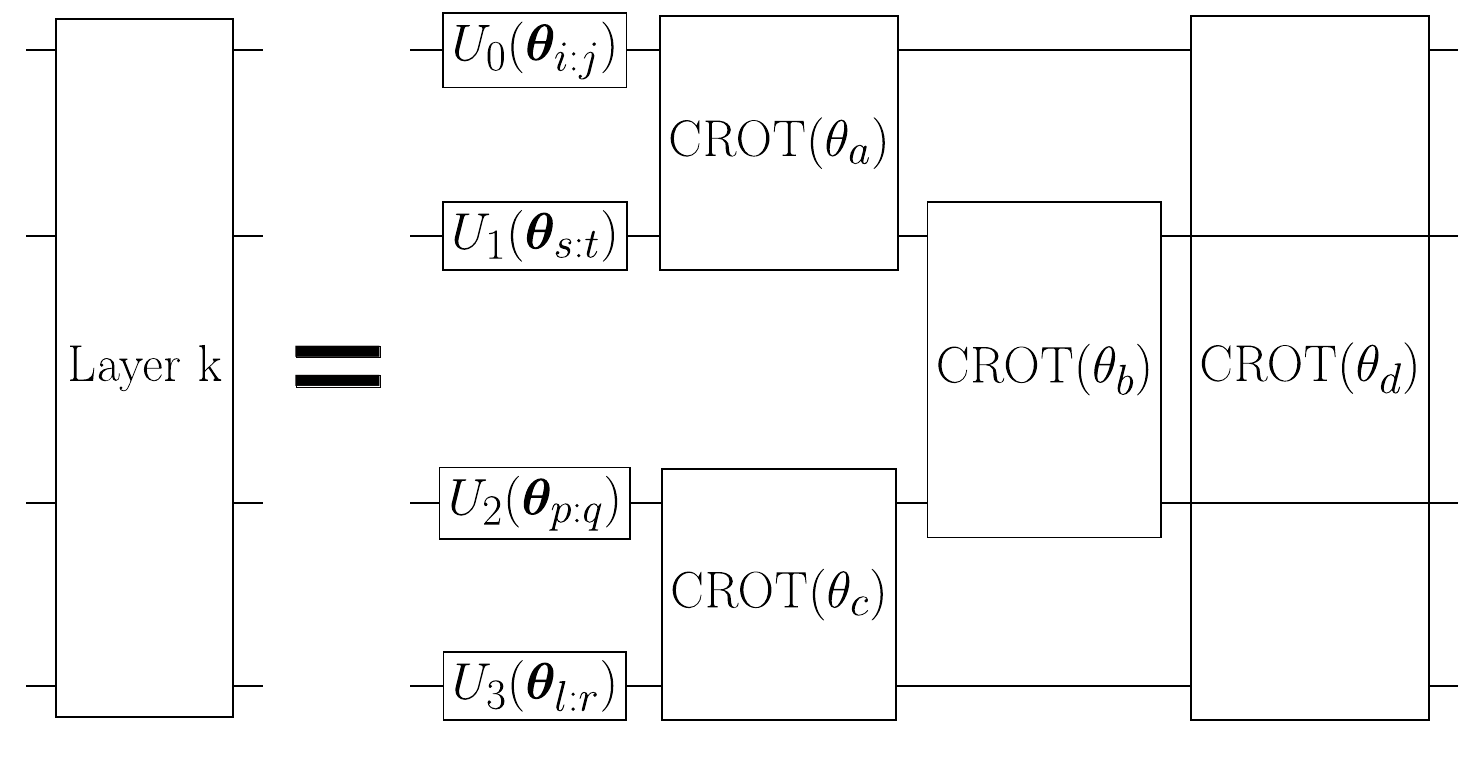}
    \caption{\textbf{Individual layers of the variational quantum circuit for one plaquette in (2+1) dimensions:} Each building block of the parametrized quantum circuit used for generating the trial states is built from a set of single-qudit operations, given by two-level rotations, and entangling operations, given by the CROT gate (see Eq.~\eqref{eq:crot_gate} for definition).}
    \label{fig:rte_quantum_circuit_2d}
\end{figure}
\begin{figure*}[t!]
    \centering
    \includegraphics[width=\textwidth]{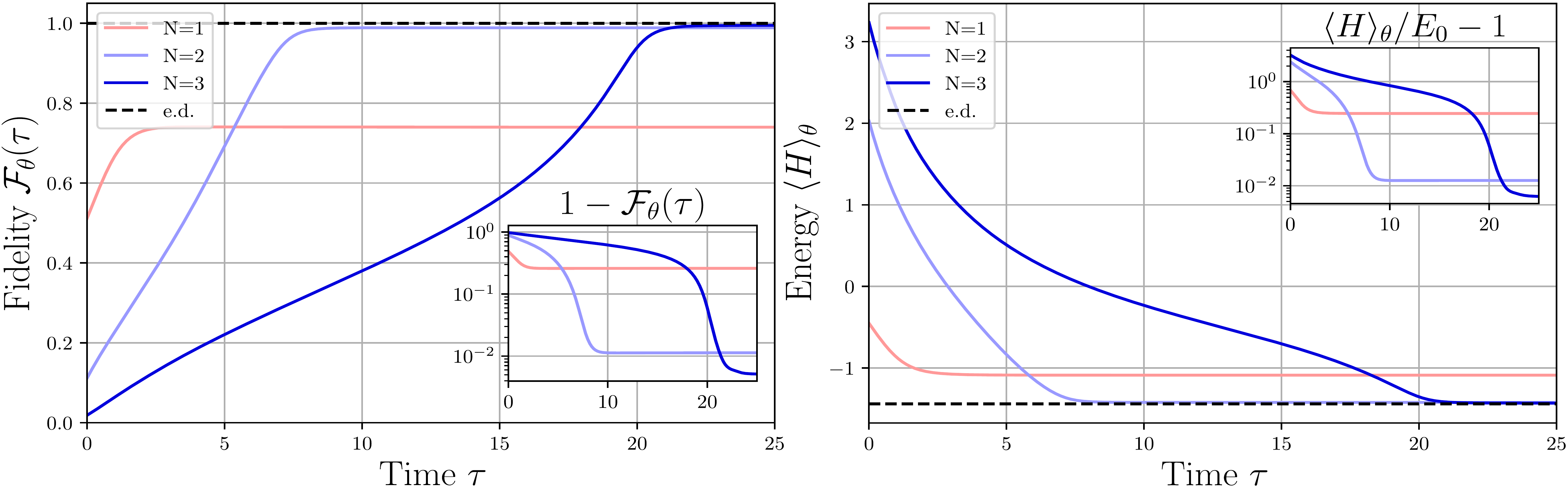}
    \caption{\textbf{Variational imaginary-time evolution of a system of seven qudits in (1+1) dimensions.} \textbf{Left:} Fidelity $\mathcal{F}_{\mathbf{\theta}}(\tau)$ of the variational state with respect to the exact ground state as a function of the imaginary time for different circuit depths ($N = 1,2$, and 3 layers). All of the results converge after a finite time. The highest reached fidelities are $\sim 74\%$, $\sim 99\%$, and $>99\%$ for $N = 1$, $2$, and $3$ layers, respectively. \textbf{Right:} Expectation value of the Hamiltonian of the system in the trial state as a function of the imaginary time. The error of the ground state energy in the final time of the simulation is below $1\%$ for $N = 3$ layers.} 
    \label{fig:imag_time_1d}
\end{figure*}

In this section, we explain how an implementation of the time evolution protocol would look on qudit quantum hardware. We briefly discuss what solution strategy for the equations of motion of the variational parameters we could employ and we present details about parametrized quantum circuits that can be implemented on the qudit device from Ref.~\cite{Ringbauer2022}.
\subsection{Solving the equation of motion for the variational parameters}
\begin{figure*}[t]
    \centering
    \includegraphics[width=\textwidth]{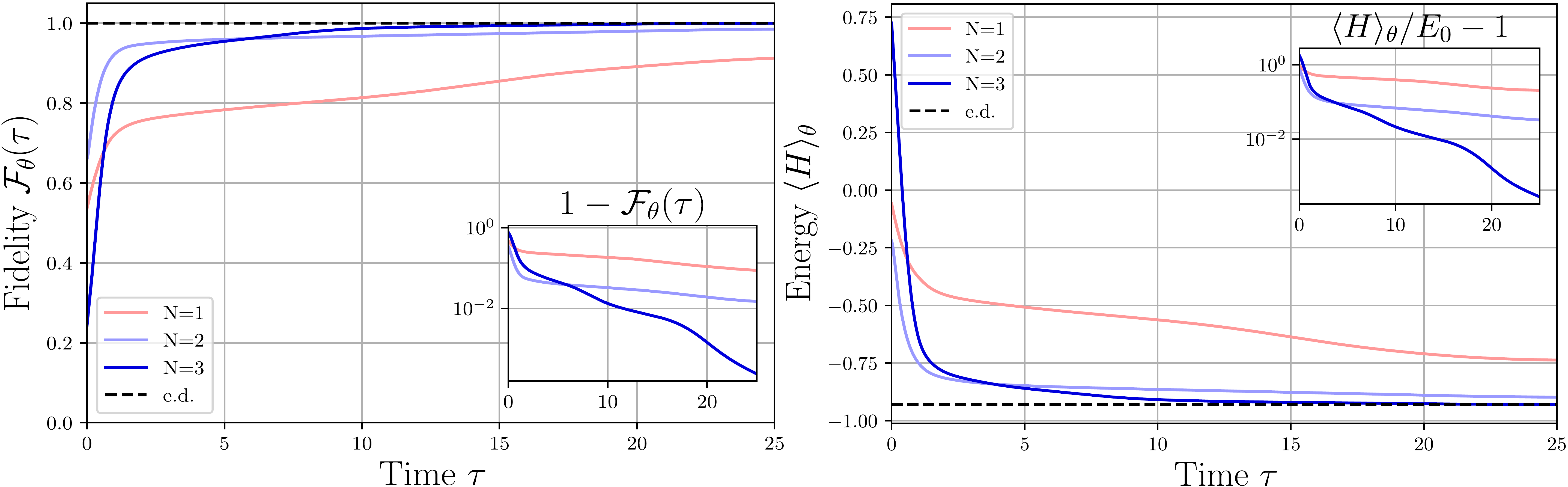}
    \caption{\textbf{Variational imaginary-time evolution of a system of one plaquette in (2+1) dimensions.} \textbf{Left:} Fidelity $\mathcal{F}_{\mathbf{\theta}}(\tau)$ of the variational state with respect to the exact ground state as a function of the imaginary time for different circuit depth ($N = 1,2$, and 3 layers). All of the results reach high fidelity after a finite time, around $\sim 91\%$, $\sim 99\%$, and $>99\%$ for $N = 1$, $2$, and $3$ layers, respectively. \textbf{Right:} Expectation value of the Hamiltonian of the system in the trial state as a function of the imaginary time. The error of the ground state energy in the final time of the simulation is below $1\%$ for $N = 3$ layers.} 
    \label{fig:imag_time_2d}
\end{figure*}

For both variational real- and imaginary-time evolution, solving Eq.~\eqref{eq:eom_for_the_thetas} would involve a feedback loop between the quantum device and a classical computer. In practice, one needs to solve a discretized version of Eq.~\eqref{eq:eom_for_the_thetas}. For example, a possible 
discretization is
\begin{align}
    \sum_{\nu}M_{\mu\nu}(t_n) [\theta_{\nu}(t_{n+1})- \theta_{\nu}(t_n)] = V_{\mu}(t_n) \Delta t\,,
\label{eq:discretized_eom}
\end{align}
with $t_n = n \Delta t$ and where $\Delta t$ is the time step. Here, we have used Euler discretization, but other discretizations (Runge-Kutta, etc.) for better numerical stability are also possible. The quantities $M_{\mu\nu}(t_n)$ and $V_{\mu}(t_n)$ have to be measured on the quantum device; then Eq.~\eqref{eq:discretized_eom} for the values of the parameters for the next time step $\theta_{\nu}(t_{n+1})$ has to be solved classically, and then the obtained values have to be given to the quantum device for the next iteration. The whole procedure has to be repeated until the end of the simulation.

\subsection{Ansatz for the variational quantum circuit}

One important aspect of our work is the choice of the parametrized quantum circuit. On the one hand, we need an expressive ansatz that allows for the parametrization of the gauge invariant Hilbert space. That is, we would ideally like to have a parametrized quantum circuit that can reproduce every physical state of the system we want to simulate, for some set of values for the variational parameters. On the other hand, we aim at near-term realizability of the simulation protocol on qudit hardware. This naturally restricts the types of quantum gates available for implementation and the number of entangling operations we can perform since each entangling operation introduces errors. Furthermore, high expressivity can restrict the efficiency of the classical optimization of the variational parameters. For these reasons, we have to look for the middle ground between expressivity and realizability. 

In the case of quench dynamics, the entanglement in the system may grow fast. This fact suggests that the expressivity of the quantum circuit for real-time evolution should be chosen high. Reversely, if the expressivity and in particular the entanglement that the ansatz allows, are not chosen correspondingly, the real-time evolution simulation may fail to reproduce the correct time evolved state already at early times. For ground states, in contrast, it is known for various systems that the entanglement structure follows an area law~\cite{Calabrese2004}, and a small number of entangling operations as well as a small number of single qudit operations should be sufficient to deliver the desired approximation with high precision.

The depth of the circuit, as well as the number of variational parameters, can be reduced by exploiting the symmetries of the system. For example, the (1+1)-dimensional Schwinger model has a CP symmetry that can be used to reduce the number of variational parameters by applying the same gates with the same variational parameters on various qudits in the system, as shown for qubits in Refs.~\cite{Kokail2019,Meth2022}.

For both variational imaginary- and real-time evolution, we employ a variational quantum circuit structured in layers. Each layer consists of a set of single qudit operations and entangling two-qudit gates, see FIG.~\ref{fig:rte_quantum_circuit_1d} and FIG.~\ref{fig:rte_quantum_circuit_2d}. Each gate in each layer of the circuit has a variational parameter given by the respective angle of rotation. We choose the set of single-qudit operations differently for imaginary- and for real-time evolution. In the case of imaginary-time evolution, we employ single qudit unitaries given by
\begin{align}
U_{e,o,r}(\boldsymbol{\theta}_{0:2}) = R^{0,1}_{X}(\theta_2)R^{0,2}_{X}(\theta_1)R^{1,2}_{Y}(\theta_0),
\end{align}
where the subscripts $e,o$, and $r$ stand for even, odd, and edge links, respectively and mean that for some set of qudits, we apply gates with the same variational parameter (in accordance with the symmetry of the system).
In the above unitary, the single qudit rotations are defined as
\begin{align}
	R^{i,j}(\theta,\varphi) = \exp\bigg(-\frac{i\theta}{2}\sigma^{i,j}_{\varphi}\bigg)\,,
\end{align}
\noindent
with $\sigma^{i,j}_{\varphi} = \cos(\varphi)\sigma^{i,j}_X + \sin(\varphi)\sigma^{i,j}_Y $.

\noindent For real-time evolution, we choose the single-qudit unitary operations more expressively as
\begin{align}
U_{e,o}(\boldsymbol{\theta}_{0:7}) = &R^{0,1}_{X}(\theta_7)R^{0,1}_{Y}(\theta_6)R^{0,2}_{X}(\theta_5)R^{0,2}_{Y}(\theta_4)\notag\\
\times&R^{1,2}_{X}(\theta_3)R^{1,2}_{Y}(\theta_2)R^{0,1}_{Z}(\theta_1)R^{0,2}_{Z}(\theta_0),\notag\\
U_{r}(\boldsymbol{\theta}_{0:7}) = &R^{1,2}_{X}(\theta_3)R^{1,2}_{Y}(\theta_2)R^{0,1}_{Z}(\theta_1)R^{0,2}_{Z}(\theta_0)\,.
\end{align}

In each layer, after a set of single-qudit unitaries, we apply a set of two-qudit entangling operations. For most of cases, these entangling gates will be Mølmer-Sørensen gates
\begin{align}
    \text{MS}^{i,j}(\theta) = \exp\left(-\frac{i\theta}{4}(\sigma^{i,j}_X\otimes\mathbb{1}+\mathbb{1}\otimes\sigma^{i,j}_X)^2\right).
\end{align}
Note that the set of single-qudit unitaries used for the simulation of real-time evolution, together with the entangling Mølmer-Sørensen gate, constitute a universal set of operations for the qudit device. 

In the case of real-time evolution of a plaquette in (2+1) dimensions, we use CROT entangling gates, which for qutrits is defined as 
\begin{align}
    \text{CROT}(\theta) = &(\mathbb{1}-\ket{2}\bra{2})\otimes\mathbb{1} \notag\\+&\ket{2}\bra{2}\otimes\exp\left(-\frac{i\theta}{2}\sigma^{1,2}_X\right).
\label{eq:crot_gate}
\end{align}
Due to the connectivity in the trapped-ion quantum device of Ref.~\cite{Ringbauer2022}, entangling MS and CROT gates can be implemented between any pair of qudits, thus not restricting the setup to nearest-neighbour entangling gates. 

\section{Results}
\label{sec:results}
To benchmark the experimental realizability, we perform numerical simulations for the real- and imaginary-time evolution of an Abelian LGT in (1+1) and (2+1) dimensions. As proof of principle, we set the coupling constant $g = 1$ and the fermion mass to $M = 0.1$ and work with a number of qudits and gate sets feasible with current hardware~\cite{Ringbauer2022}. In terms of entangling gate count, for simulating a system with $L$ links in (1+1) dimensions, we construct circuits that use $L-1$ entangling gates per layer of the variational circuit, or $L$ gates for the (2+1)-dimensional case. Hence, for $N$ layers, we need at most $LN$ qudit entangling gates; see Appendix~\ref{app:rte_7qudits}.
\subsection{Variational imaginary-time evolution}
\begin{figure*}[htp]
    \centering
    \includegraphics[width=\textwidth]{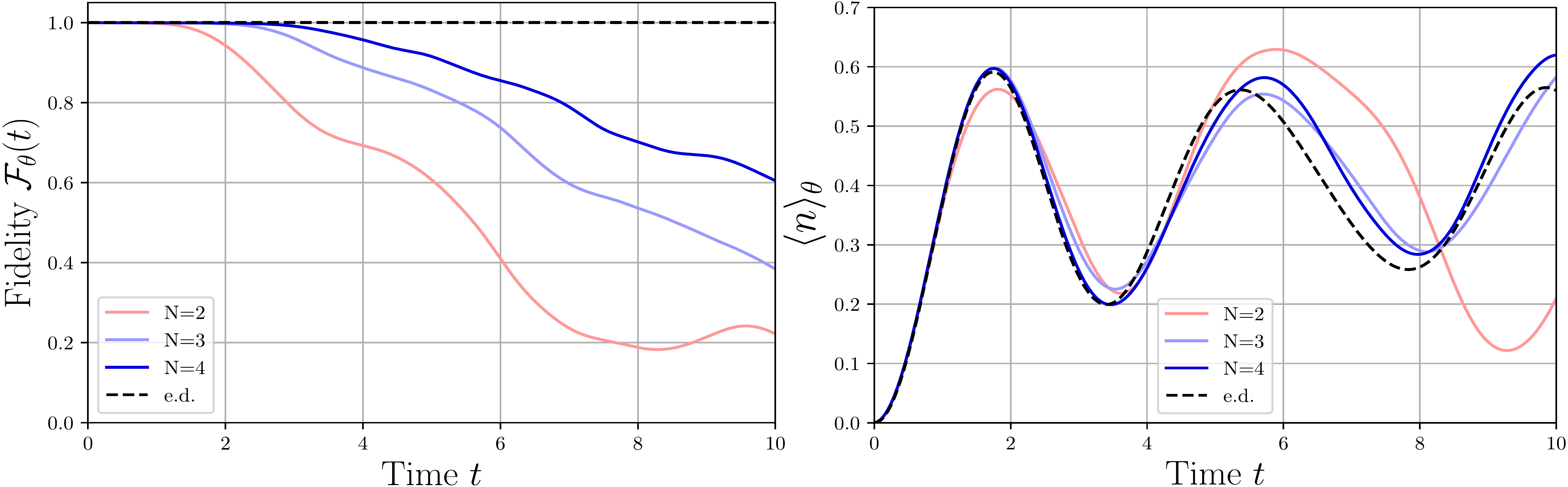}
    \caption{\textbf{Variational real-time evolution of a system of five qudits in (1+1) dimensions.} \textbf{Left:} Fidelity of the variational state with respect to the exactly time-evolved state (see Eq.~\eqref{eq:fidelity_rte}). The fidelity decreases with time as the correct time evolution of the initial state becomes harder and harder to approximate with fixed number of layers. \textbf{Right:} Real-time dynamics after a quench of the gauge invariant fermionic numbers in the middle of the one-dimensional system. The dashed line corresponds to the exact time evolution of the fermion numbers, and the colorful lines correspond to the approximate time evolution with different number of layers $N$.} 
    \label{fig:rte_1d_joint}
\end{figure*}

\begin{figure}
    \centering
    \includegraphics[width = \columnwidth]{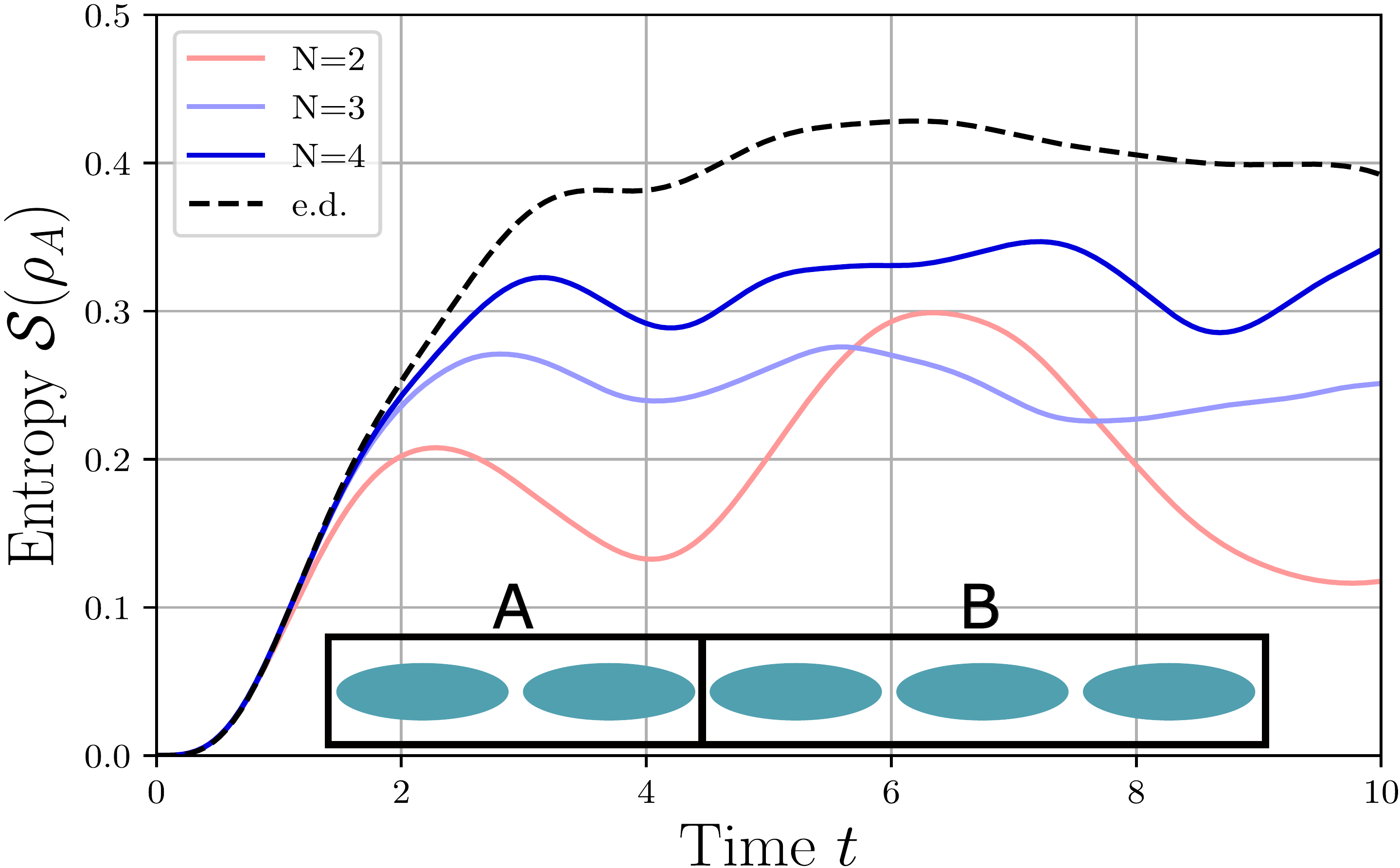}
    \caption{\textbf{Evolution of the entanglement entropy of subsystem A after a quench in a system of five qudits in (1+1) dimensions:} After initial linear growth, the entropy $\mathcal{S}(\rho_A)$ saturates to a finite constant up to fluctuations due to finite size effects. As we increase the number of layers $N$ in the variational circuits, we observe better and better approximations to this behavior of the entropy in the trial state.}
    \label{fig:rte_1d_entropy}
\end{figure}

For simulating variational imaginary-time evolution, we employ a quantum circuit with variable number of layers $N$, where each individual layer looks like the one in FIG.~\ref{fig:rte_quantum_circuit_1d} for a system in (1+1) dimensions and FIG.~\ref{fig:rte_quantum_circuit_2d} for a system in (2+1) dimensions. The variational circuit is applied on an initial state that is easy to implement, for example, the product state of all qudits initialized in the state $\ket{1}$, $\ket{\psi_0} = \bigotimes_i\ket{1}_i$. For performing the imaginary-time evolution, we are interested in initializing random states captured by the parametrized circuit. To prepare such random states, we have to choose a random set of variational parameters and apply the corresponding circuit on the state $\ket{\psi_0}$.

By numerically solving the discretized equation of motion for the variational parameters during imaginary-time evolution, we obtain a set of values for the parameters for each time instance. For each such set of parameters, we calculate the fidelity of the variational state with the exact ground state 
\begin{align}
    \mathcal{F}_{\mathbf{\theta}}(\tau) = |\langle\psi(\mathbf{\theta}(\tau))|\psi_{\text{ground}}\rangle|^2
\end{align}
and the energy of the variational state $\langle H \rangle_{\mathbf{\theta}}$.

In FIG.~\ref{fig:imag_time_1d} and FIG.~\ref{fig:imag_time_2d}, these two quantities are plotted as a function of the imaginary time for a system of 7 qudits in (1+1) dimensions and for a plaquette in (2+1) dimensions, respectively. The different lines correspond to variational circuits with different number of layers $N = 1,2$, and $3$. We see that for all circuits, the fidelity and the energy converge. In addition, the quantities converge to the exact ones with the infidelity of less than $1\%$ for three layers. Correspondingly, we can obtain the correct energy with a relative error below $1\%$.

\subsection{Variational real-time evolution}
\label{subsection:rte}
\begin{figure*}[htp]
    \centering
    \includegraphics[width=\textwidth]{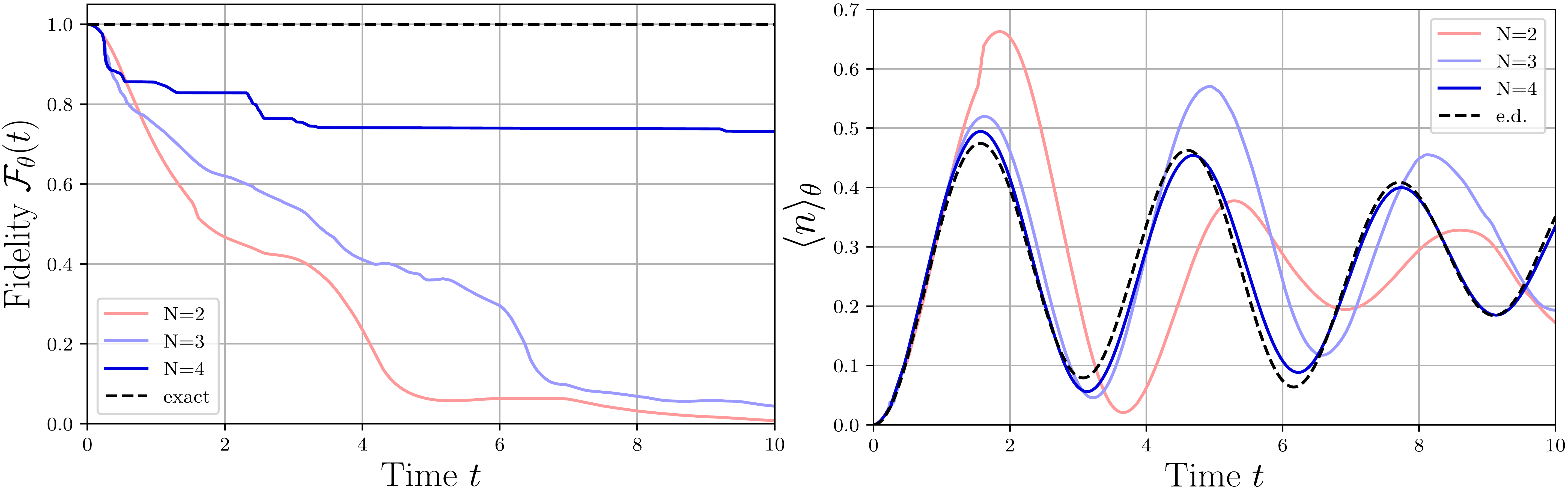}
    \caption{\textbf{Variational real-time evolution of a system of one plaquette in (2+1) dimensions.} \textbf{Left:} Fidelity of the variational state with respect to the exactly time-evolved state (see Eq.~\eqref{eq:fidelity_rte}). The fidelity for $N = 2$ and $3$ layers decreases with time, while the decrease for $N = 4$ is much slower. This suggests that for a system of 1 plaquette, there might not be a need to gradually increase the circuit depth with time to approximate the time evolution reliably. \textbf{Right:} Real-time dynamics after a quench of the gauge invariant fermionic numbers in the middle of the one-dimensional system. The dashed line corresponds to the exact time evolution of the fermion numbers, and the colored lines correspond to the approximate time evolution with a different number of layers $N$.} 
    \label{fig:rte_2d_joint}
\end{figure*}

\begin{figure}
    \centering
    \includegraphics[width = \columnwidth]{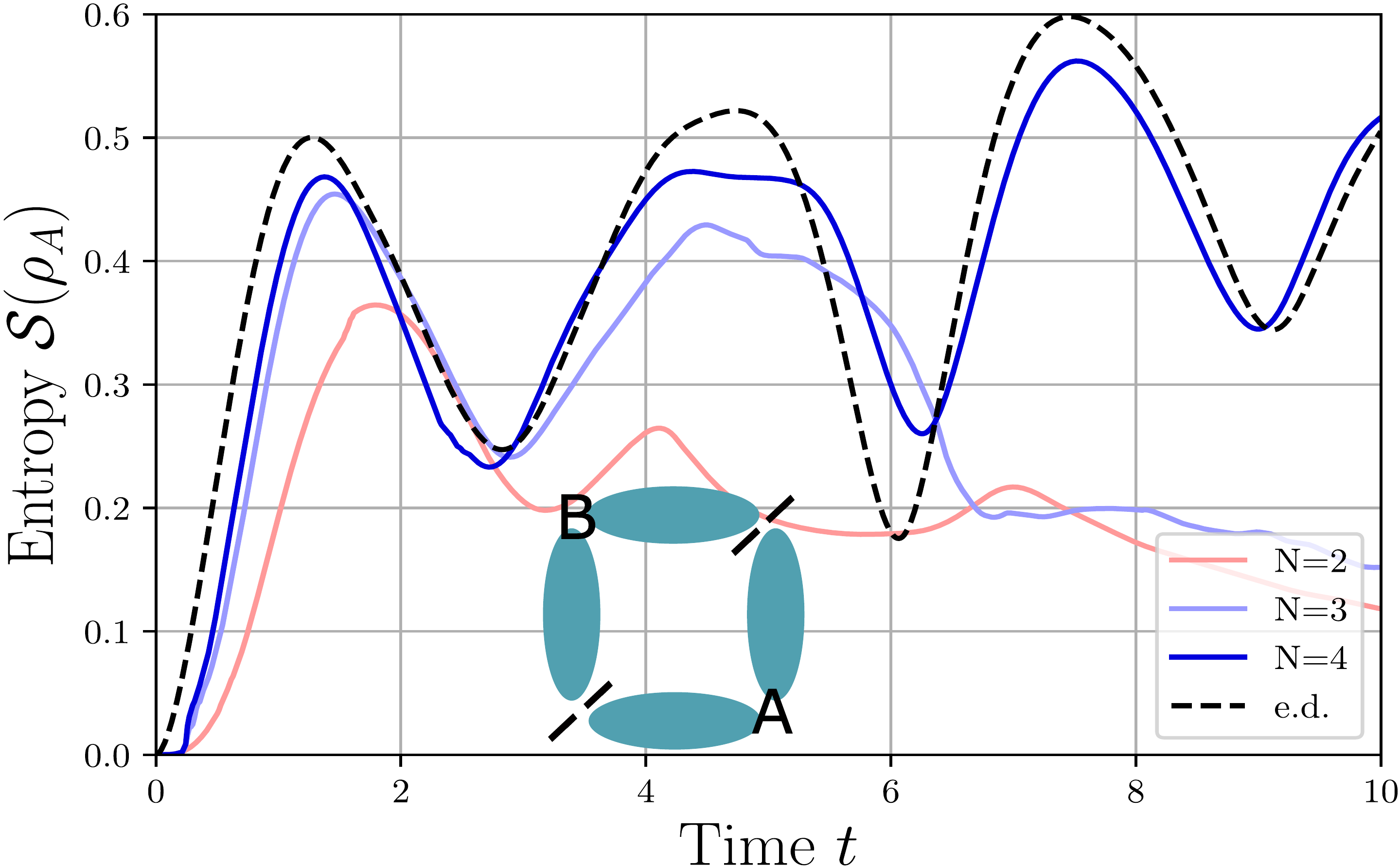}
    \caption{\textbf{Evolution of the entanglement entropy of subsystem A after a quench in a system of one plaquette in (2+1) dimensions:}
    We observe that the growth of entropy at very early times is not correct; however, there is a qualitative agreement with the exact diagonalization. This growth of entropy  leads to small deviations from the exact result for the fermionic occupation numbers; however, these deviations do not grow with time.}
\label{fig:rte_2d_entropy}
\end{figure}

With the variational time-evolution approach, we can also approximate quench dynamics. Initially, a system of 5 qudits is prepared in the product state $\ket{\psi_0} = \bigotimes_i\ket{1}_i$, which corresponds to the ground state of the Hamiltonian in Eq.~\eqref{eq:kogut-susskind_ham} for infinite fermion rest mass $M\rightarrow\infty$. Then, we apply the time evolution operator with respect to the Hamiltonian in Eq.~\eqref{eq:kogut-susskind_ham}. In FIG.~\ref{fig:rte_1d_joint}, the fidelity of the trial state with respect to the ideally time-evolved state,
\begin{align}
    \mathcal{F}_{\mathbf{\theta}}(t) = |\langle\psi(\mathbf{\theta}(t))|\psi(t)\rangle|^2,
\label{eq:fidelity_rte}
\end{align}
is shown as a function of time. The fidelity can serve as a strict quantifier of how reliable the variational approach reproduces the time evolution of the state of the system. In FIG.~\ref{fig:rte_1d_joint} also the evolution in time of a gauge invariant observable---the fermion number $\langle n \rangle_{\boldsymbol{\theta}}$---is given for different circuit depths (number of layers $N = 2,3$, and $4$). The dashed black line represents the exact time evolution of the fermion numbers under the Hamiltonian of the gauge theory. We observe good qualitative agreement between the variationally evolved and the exactly evolved fermion numbers in the case of $N = 3,4$ (blue lines). Furthermore, for 4 layers, we also obtain quantitatively high fidelity ($>80\%$) for times up to two oscillation periods of the fermion numbers.

A quantity that can give us insight about why the variational time evolution starts to deviate after some time is the entanglement (von Neumann) entropy. We introduce a cut in the system of qudits (for example, in the middle of the 1 dimensional chain) and denote the two resulting subsystems as A and B. Then, the bipartite entanglement entropy of subsystem A reads
\begin{align}
    \mathcal{S}(\rho_A) = -\text{Tr}_A\left[\rho_A\ln\rho_A\right],
\end{align}
where the reduced density matrix of subsystem A is 
\begin{align}
    \rho_A = \text{Tr}_B \ket{\psi}\bra{\psi}.
\end{align}
In FIG.~\ref{fig:rte_1d_entropy}, we plot this entropy as a function of time. As we expect, initially, the entropy grows linearly with time. After that, it saturates and becomes constant, up to some fluctuations due to finite-size effects. We observe a correlation between the ability of the quantum circuit to capture this entropy growth up to the saturated value and its ability to reproduce the correct behavior of the time-evolved system. Therefore we need a circuit capable of generating sufficient entanglement for the correct simulation of the time evolution. This capability grows with the number of entangling operations in the circuit. However, we need to restrict this number to keep the experimental error low.

We also simulate the variational real-time evolution for a system of 1 plaquette in (2+1) dimensions. In FIG.~\ref{fig:rte_2d_joint}, the gauge-invariant fermion occupation numbers $\langle n\rangle_{\theta}$ in the lower left corner of the plaquette is shown as a function of time. Also, the fidelity with respect to the exactly time-evolved state is shown as a function of time. Quantitatively, the fidelity drops below $80\%$ for early times, and the time evolution of the fermion occupation numbers slightly deviates from the exact one. However, qualitatively, the time evolution of $\langle n\rangle_{\theta}$ is captured very well. As we see from the fidelity plot, even for later times, the fidelity remains constant. Here again, we investigate the bipartite entanglement entropy (FIG.~\ref{fig:rte_2d_entropy}) and establish the correlation between quantitative results and entropy growth at early times. In Appendix~\ref{app:four_body_gate}, we show how to modify the variational circuit to give very good qualitative results throughout the simulation. The new element is an entangling gate motivated by the plaquette term present in the Hamiltonian of the gauge theory in (2+1) dimensions. Though currently not state of the art, there are considerable efforts towards constructing native multipartite gates in trapped-ion quantum computers~\cite{Katz2022, Andrade2022}.

\section{Conclusion }
The core of the presented approach to the quantum
simulation of Abelian LGTs is based
on three ingredients: the hardware-efficient implementation of the LGT by integrating of redundant degrees of freedom, while preserving the locality of the Hamiltonian; the mapping of the resulting LGT on a qudit model that can be directly implemented on qudit quantum hardware; and the application of a variational quantum simulation protocol for ground state preparation as well as quench dynamics.

In particular, we showed how to encode a U(1) LGT with dynamical fermions in arbitrary dimensions in a qudit system that can be implemented on already existing trapped-ion quantum computers~\cite{Ringbauer2022}. In a proof-of-principle numerical simulation, we showed the power of the variational time evolution algorithm in finding the ground state of this LGT in (1+1) and in (2+1) dimensions. Using only gates available in present-day experimental platforms and a small number of entangling operations, we successfully found a good approximation of the ground state of the LGT for various system sizes, including a system in (2+1) dimensions.

Furthermore, we also showed that we can variationally simulate the real-time evolution of an LGT and provided a circuit that is implementable on present-day devices. Gauge invariant quantities such as fermion occupation numbers showed several periods of oscillating behavior, a clear improvement with respect to previous studies~\cite{Martinez2016, Nguyen2022}. We also simulated the real-time dynamics of an LGT in (2+1) dimension and showed an improvement over various oscillation periods by extending the circuit with a four-body entangling gate.

So far, we were primarily concerned with the case of Abelian LGTs. In the future, our methods can be extended to gauge theories with a non-Abelian symmetry, such as SU(2) and SU(3), relevant for the Standard model of particle physics, but also some discrete non-Abelian gauge groups such as the dihedral groups $D_n$. The gauge field Hilbert space of the latter is naturally finite-dimensional and thus can be implemented without truncation in a qudit quantum device. By combining the resource-friendly variational approach with a qudit-based quantum computation strategy, we anticipate significant improvement in the quantum simulation of non-Abelian gauge theories in the near future. 

\section{Acknowledgements}
We thank J.~Berges, R.~Blatt, E.~Demler, A.~Garcia Sala, D.~González-Cuadra, C.~Muschik, G.~Müller-Rigat T.V.~Zache, and P.~Zoller for
fruitful discussions.\!
ICFO group acknowledges support from: ERC AdG NOQIA; Ministerio de Ciencia y Innovation Agencia Estatal de Investigaciones (PGC2018-097027-B-I00/10.13039/501100011033,  CEX2019-000910-S/10.13039/501100011033, Plan National FIDEUA PID2019-106901GB-I00, FPI, QUANTERA MAQS PCI2019-111828-2, QUANTERA DYNAMITE PCI2022-132919,  Proyectos de I+D+I “Retos Colaboración” QUSPIN RTC2019-007196-7); MICIIN with funding from European Union NextGenerationEU(PRTR-C17.I1) and by Generalitat de Catalunya;  Fundació Cellex; Fundació Mir-Puig; Generalitat de Catalunya (European Social Fund FEDER and CERCA program, AGAUR Grant No. 2021 SGR 01452, QuantumCAT \ U16-011424, co-funded by ERDF Operational Program of Catalonia 2014-2020); Barcelona Supercomputing Center MareNostrum (FI-2022-1-0042); EU (PASQuanS2.1, 101113690); EU Horizon 2020 FET-OPEN OPTOlogic (Grant No 899794); EU Horizon Europe Program (Grant Agreement 101080086 — NeQST), National Science Centre, Poland (Symfonia Grant No. 2016/20/W/ST4/00314); ICFO Internal “QuantumGaudi” project; European Union’s Horizon 2020 research and innovation program under the Marie-Skłodowska-Curie grant agreement No 101029393 (STREDCH) and No 847648  (“La Caixa” Junior Leaders fellowships ID100010434: LCF/BQ/PI19/11690013, LCF/BQ/PI20/11760031,  LCF/BQ/PR20/11770012, LCF/BQ/PR21/11840013). 

E.Z. acknowledges the support of the ISRAEL SCIENCE FOUNDATION
(grant No. 523/20).

P.H.\ acknowledges support from Provincia Autonoma di Trento, by Q@TN, the joint lab between University of Trento, FBK-Fondazione Bruno Kessler, INFN-National Institute for Nuclear Physics, and CNR-National Research Council, and from ICSC – Centro Nazionale di Ricerca in HPC, Big Data and Quantum Computing, funded by the European Union under NextGenerationEU. 
This project has received funding from the ERC under the European Union’s Horizon 2020 research and innovation programme (GA No 804305).
This project was funded within the QuantERA II Programme that has received funding from the European Union’s Horizon 2020 research and innovation programme under GA No 101017733.
Funded by the European Union under Horizon Europe Programme - Grant Agreement 101080086 — NeQST. 

Funded by the European Union (ERC, QUDITS, 101039522) and by the EU-QUANTERA project TNiSQ (N-6001).

Views and opinions expressed are, however, those of the author(s) only and do not necessarily reflect those of the European Union, European Commission, European Climate, Infrastructure and Environment Executive Agency (CINEA), nor any other granting authority. Neither the European Union nor any granting authority can be held responsible for them. 

\appendix
\section{Parameter shift rules for the components of the metric tensor}
\label{app:HowToMeasureM}
\begin{figure*}[t]
    \centering
    \includegraphics[width=\textwidth]{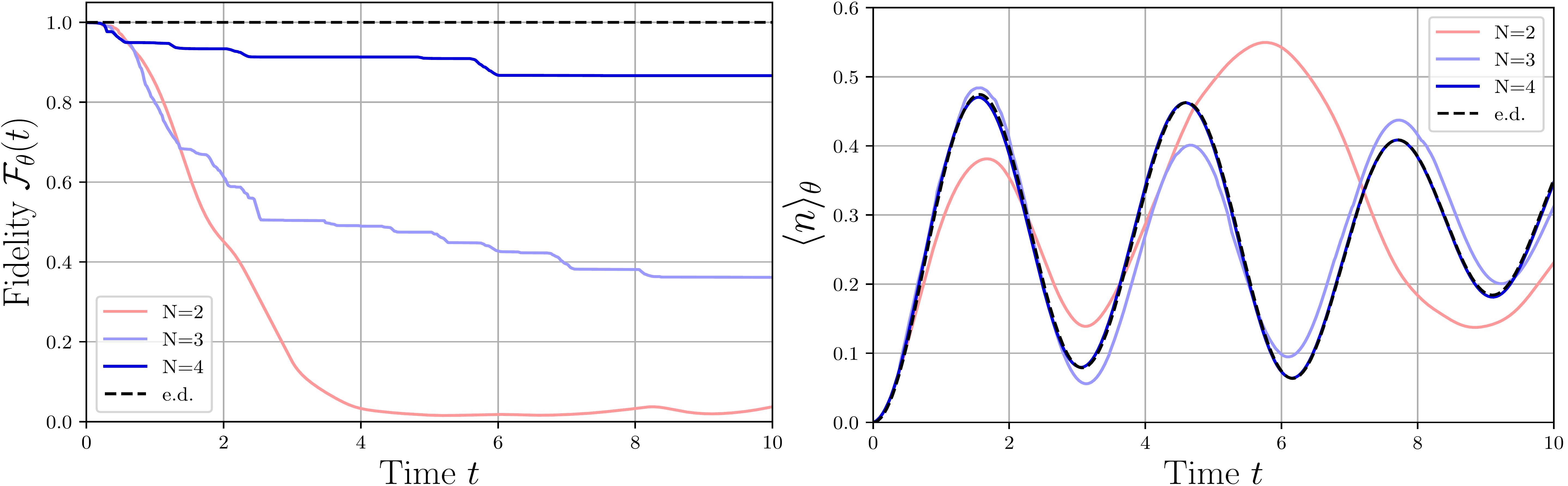}
    \caption{\textbf{Variational real-time evolution of a system of 1 plaquette in (2+1) dimensions with a circuit that includes a four-body gate.} \textbf{Left:} Fidelity of the variational state with respect to the exactly time-evolved state (see Eq.~\eqref{eq:fidelity_rte}). For $N=4$, the fidelity does not drop below $80\%$ for the whole simulation period. \textbf{Right:} Real-time dynamics after a quench of the gauge invariant fermionic numbers in the bottom left corner site of the plaquette. We see virtually perfect agreement between the exactly diagonalized (dashed line) result and the result from the variational simulation for $N = 4$ layers.} 
    \label{fig:rte_1plaquette_plaquette_joint}
\end{figure*}
The problem we faced in the main text is how to compute the derivatives of a function given we can measure the function itself. We start by writing the function  
$f_{\mu\nu}(a)$ in Eq.~\eqref{eq:DefOfG} in an analytical form. We then perform
the derivative with respect to $a$ analytically, motivated
by the so-called parameter shift-rules, see for example~\cite{Wierichs2022}. 

In order to derive an analytical expression for the derivatives, we focus on $f_{\mu\nu}(a)$ for pure states.
If we assume trial states of the form as in Eq.~\eqref{eq:TrialState}, after we insert this trial state into the definition of the function $f_{\mu\nu}(a)$ from Eq.~\eqref{eq:DefOfG}, we obtain the expression
\begin{align}
    f_{\mu\nu}(a) = |\braket{\psi_{\mu}| W_{\mu}(a) V U_{\nu}(a) |\psi_{\nu}}|^2 \,,
    \label{eq:SimplifiedFormOfF}
\end{align}
where the states $\ket{\psi_{\mu}}$ is defined as
\begin{align}
    \ket{\psi_{\mu}} = U_{\mu-1}(\theta_{\mu-1})\dots U_{1}(\theta_{1})\ket{\psi_0}
\end{align}
and $W_{\mu}(a)$, $U_{\nu}(a)$ and $V$ are unitary matrices and
\begin{align}
    W_{\mu}(a) &= e^{ia \, w_{\mu} } \, , \\
    U_{\nu}(a) &= e^{ia \, u_{\nu} } \, ,
\end{align}
with Hermitian operators $w_{\mu}$ and $u_{\nu}$. Because of Eq.~\eqref{eq:TrialState}, the eigenstates of
$W$ and $U$ can be written as
\begin{subequations}
\begin{align}
    W_{\mu}(a) \ket{w_{\mu,m}} &= e^{ia w_{\mu,m} } \ket{w_{\mu,m}} \, , \\
    U_{\nu}(a) \ket{u_{\nu,n}} &= e^{ia u_{\nu,n}} \ket{u_{\nu,n}} \, ,
\end{align} \label{eq:Eigenvalues}
\end{subequations}
\noindent where the labels $m$ and $n$ enumerate  
possible degeneracies. 
Inserting the resolution of the identity in Eq.~\eqref{eq:SimplifiedFormOfF}, we obtain 
\begin{align}
    f_{\mu\nu}(a) = \sum_{m,n,p,q} f^{\mu\nu}_{m,n,p,q} 
    e^{ia(w_{\mu,m} + u_{\nu,n} - w_{\mu,p} - u_{\nu,q})}\,,
\end{align}
with $f^{\mu\nu}_{m,n,p,q} $ being (up to a linear transformation) the Fourier coefficients of $f_{\mu\nu}(a)$.
Considering all possible combinations of $m$, $n$, $p$, and 
$q$ there are $R_{f,\mu\nu}$ different values of 
$w_{\mu,m} + u_{\nu,n} - w_{\mu,p} - u_{\nu,q}$. We denote these values by $\omega_{r}$ with 
$r \in \{1, \ldots, R_{f,\mu\nu}\}$. Summarizing the 
eigenvalues in this form allows us to
rewrite the above expression for $f_{\mu\nu}(a)$ as
\begin{align}
    f_{\mu\nu}(a) = \sum_{r=1}^{R_{f,\mu\nu}} F_{\mu\nu,r} e^{i a \omega_r} \,. \label{eq:AnalyticalForm}
\end{align}
We use this analytical form to determine the coefficients $F^{\mu\nu}_r$. 
To this end, we evaluate the function $f_{\mu \nu}(a)$
at $R_{f,\mu\nu}$ different values of $a_l$ with $l \in \{1, \ldots, R_{f,\mu\nu} \}$ such that we obtain $R_{f,\mu\nu}$
different linear equations
\begin{align}
    f_{\mu\nu}(a_l) = \sum_{r=1}^{R_{f,\mu\nu}} e^{i a_l \omega_r} F_{\mu\nu,r}  \, \text{for } l \in \{1, ..., {R_{f,\mu\nu}} \} \,
\end{align}
with the unknown variables $F_{\mu\nu,r}$.
The above system of equations has a solution if the 
matrix $\mathcal{A}_{lr} = e^{i a_l \omega_r}$ has full rank.
If that is not the case, 
we change the evaluation points $a_l$ until the matrix
$\mathcal{A}_{lr}$ achieves full rank. Having the coefficients of the Fourier sum of the function $f_{\mu\nu}(a)$, we can calculate its derivatives by deriving this Fourier sum. These  derivatives, on the other hand, direclty give us the matrix elements of $M_{\mu\nu}$, as shown in Eq.~\eqref{eq:connection_M_fp}.

In order to obtain the diagonal elements $M_{\mu\mu}$, we 
determine the function $p_{\mu}(a)$ and use Eq.~\eqref{eq:DefOfG}. For this, we rewrite $p(a)$ as
\begin{align}
    p_{\mu}(a) = |\braket{\psi_{\mu}| P_{\mu}(a)|\psi_{\mu}}|^2 \,,
    \label{eq:SimplifiedFormOfP}
\end{align}
where $T$ is a unitary matrix and $P_{\mu}(a)$ is of the form
\begin{align}
    P_{\mu}(a) &= e^{ia \, p_{\mu} } \, .
\end{align}
This is a unitary matrix with the eigenvectors and eigenvalues given by 
\begin{align}
    P_{\mu}(a) \ket{p_{\mu,m}} &= e^{ia p_{\mu,m} } \ket{p_{\mu,m}} \,,
    \label{eq:EigenvaluesP}
\end{align} 
where the label $m$ enumerates the eigenvectors. 
Inserting the resolution of the identity in Eq.~\eqref{eq:SimplifiedFormOfP}, we obtain 
\begin{align}
    p_{\mu}(a) = \sum_{k,l} p^{\mu}_{k,l} 
    e^{ia(p_{\mu,k} - p_{\mu,l})} \,.
\end{align}
Considering all possible combinations of $k$ and $l$, there are $R_{\mu,p}$ different values of 
$p_{\mu,k} - p_{\mu,l}$.
We denote these $R_{\mu,p}$ different values by $\lambda_{\mu, r}$ with
$r \in \{1, \ldots, R_{\mu,p}\}$, giving 
\begin{align}
    p_{\mu}(a) = \sum_{r} p_{\mu,r} e^{i a \lambda_{\mu,r}}  \,. \label{eq:AnalyticalFormOfP}
\end{align}
We use the above analytical form of $p_{\mu}(a)$ 
to determine the coefficients $P_{\mu,r_p}$. 
Tho this end, we evaluate the function $p_{\mu}(a)$
at $R_{\mu,p}$ different values of $a$, which we denote $a_{\mu,r_p}$ with $r_p \in \{1, \ldots, R_p \}$, such that we obtain a
system of linear equations
\begin{align}
    p_{\mu}(a_{r_p}) = \sum_{r} p_{\mu,r} e^{i a_{\mu,r_p} \lambda_{\mu,r}} \,, \label{eq:AnalyticalFormPSystem}
\end{align}
with the unknown variable $p_{\mu,r}$.
Having the derivatives of $p_{\mu}$ and $f_{\mu\nu}$, we can calculate the coefficients of the metric tensor, using Eq.~\eqref{eq:connection_M_fp}.
\begin{figure}
    \centering
    \includegraphics[width = \columnwidth]{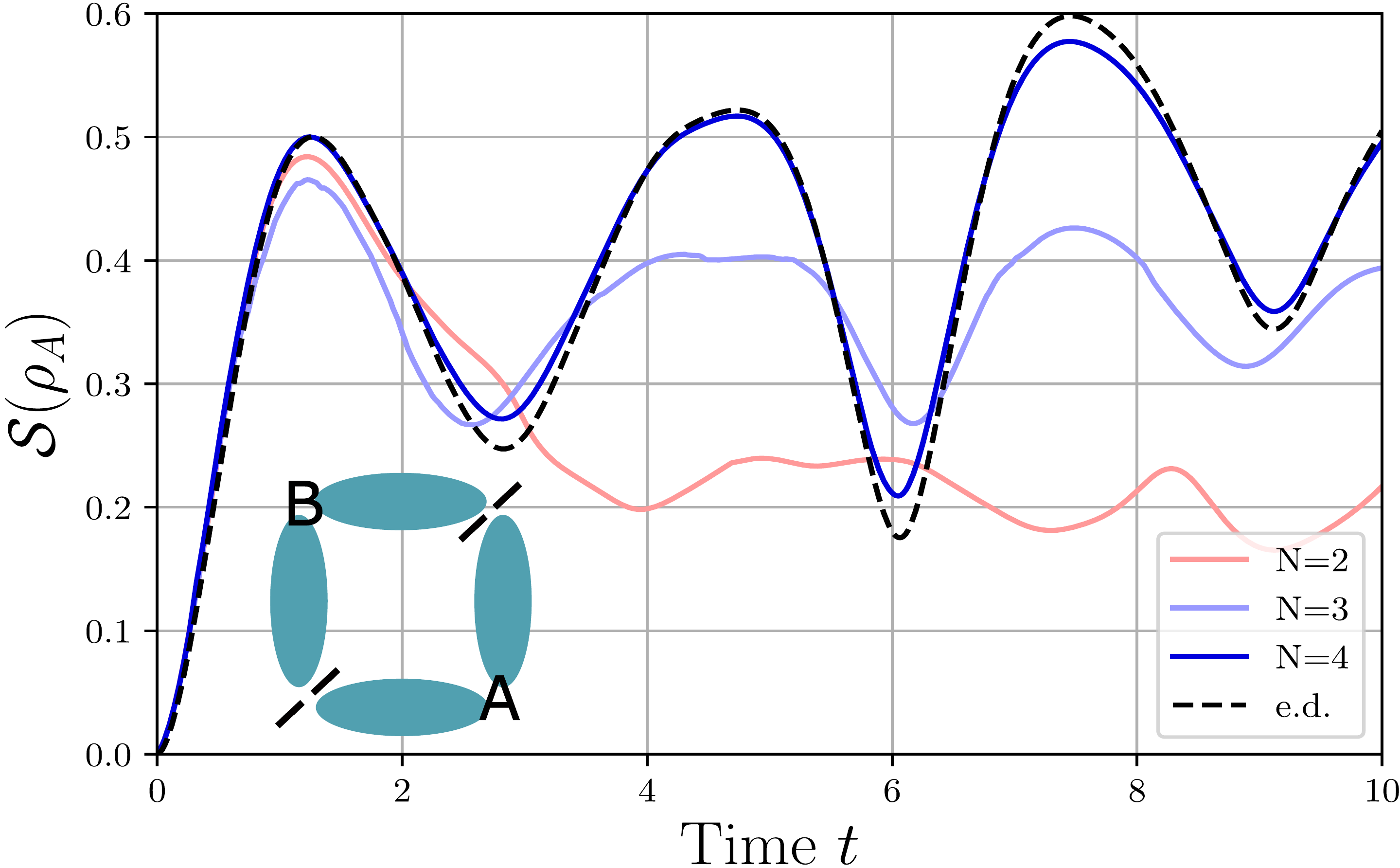}
    \caption{\textbf{Evolution of the entanglement entropy of subsystem A after a quench in a system of 1 plaquette in (2+1) dimensions:} The initial growth of the entropy is captured very accurately for $N = 2,3,4$ layers. Furthermore, for $N = 4$ layers, the time evolution of the entropy is very well approximated over the whole simulation time.}
    \label{fig:rte_1plaquette_plaquette_entropy}
\end{figure}
\section{How to measure $V_{\mu}$ for imaginary time evolution}
\label{app:HowToMeasureVImag} 
\begin{figure*}[t]
    \centering
    \includegraphics[width=\textwidth]{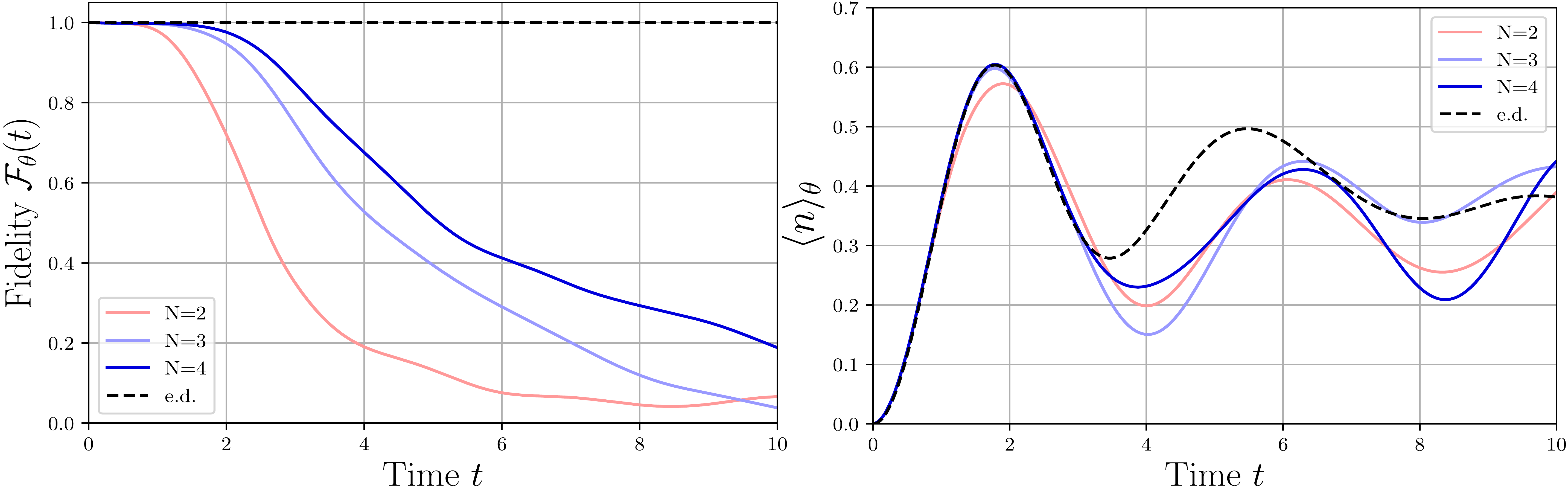}
    \caption{\textbf{Variational real-time evolution of a system of 7 qudits in (1+1) dimensions :} \textbf{(a)} Fidelity of the variational state $\ket{\psi(\theta(t))}$ with respect to the exactly time-evolved state $\ket{\psi(t)}$. \textbf{(b)} Real-time dynamics after a quench of the gauge invariant fermionic numbers in the middle of the one dimensional system. The dashed line corresponds to the exact time evolution of the fermion numbers and the colorful lines correspond to the approximate time evolution with different number of layers $N$.} 
    \label{fig:rte_7qudits_joint}
\end{figure*}
Since the components of the vector $V^I_{\mu}$ from the main text are connected to the derivatives of the function $v^I_{\mu}(a)$ (see Eq.~\eqref{eq:function_for_V}), we show here how to calculate these derivatives by measuring the function itself. Similar to $f_{\mu\nu}(a)$, the function $v^I_{\mu}(a)$ is of the form 
\begin{align}
    v^I_{\mu}(a) = \braket{\psi_{\mu}|Q^{\dagger}_{\mu}(a) R^{\dagger} H  R Q_{\mu}(a) |\psi_{\mu}} \,,  
\end{align}
with unitary operators $R$ and $Q_{\mu}(a)$. The latter is of the form  
\begin{align}
    Q_{\mu}(a) &= e^{ia \, q_{\mu} } \, 
\end{align}
with hermitian $q_{\mu}$. Using the eigenvalue equation for 
\begin{align}
    Q_{\mu}(a) \ket{q_{\mu,m}} = e^{iq_{\mu,m}a} \ket{q_{\mu,m}} \,,
\label{eq:ev_fctn_v}
\end{align}
we can rewrite the function $v^I_{\mu}(a)$ as
\begin{align}
    v^I_{\mu}(a) = \sum_{m,n} v^I_{\mu,mn} e^{ia(q_{\mu,m} - q_{\mu,n})} \,. \label{eq:SimplifiedFormOfV}
\end{align}
There are $R_{\mu,v}$ different values of $q_{\mu,m} - q_{\mu,n}$,
which we call $\chi_{\mu,r_v}$, allowing us to rewrite the sum in Eq.~\eqref{eq:SimplifiedFormOfV} as 
\begin{align}
    v^I_{\mu}(a) = \sum_{r_v} v^I_{\mu,r_v} e^{ia\chi_{\mu,r_v}}\,.
\end{align}
In order to determine the Fourier coefficients $v_{\mu,r_v}$ we
evaluate the function $v_{\mu}(a)$ at $R_{\mu,v}$ different values
of $a$ and determine the Fourier coefficients $v_{\mu,r_v}$
by solving the linear system
\begin{align}
    v^I_{\mu}(a_{\mu,r}) = \sum_{r_v} v^I_{\mu,r_v} e^{ia_{\mu,r}\chi_{\mu,r_v}} \,.
\end{align}
Given the analytical form of $v^I_{\mu}(a)$ the derivative is given by
\begin{align}
    \frac{\partial}{\partial a} v^I_{\mu}(a) = i \sum_{r_v}  v^I_{\mu,r_v} \chi_{\mu,r_v} e^{ia\chi_{\mu,r_v}}    
\end{align}
Having this derivative, the component of the vector $V^I_{\mu}$ are obtained by Eq.~\eqref{eq:v_der}.

\section{Representing a Hermitian matrix as a sum of two unitaries}
\label{app:sum_of_unitaries}
A simple calculation shows that every hermitian matrix $S$ can be represented as a sum of two unitary matrices. We construct those unitaries as follows:
\begin{itemize}
    \item Diagonalise $S \rightarrow D_S = V^{\dagger}SV.$
    \item Define $||S|| := \underset{i}{\max} |\lambda_i|$, where $\lambda_i$ are the diagonal elements of $D_S$.
    \item Define the normalised $\tilde{D}_S := \frac{1}{||S||}D_S$.
    \item Define the unitary matrix $U_D := \tilde{D}_S + i\sqrt{\mathbb{1}- \tilde{D}_S}.$
    \item Define $U_S := VU_DV^{\dagger}$.
\end{itemize}
Then, the relation holds 
\begin{align}
    S = \frac{1}{2}||S||(U_S+U^{\dagger}_S), 
\end{align}
where we have an explicit construction of the matrix $U_S$. 
\section{Variational real-time evolution with a four-body entangling gate}
\label{app:four_body_gate}

Here, we propose how to minimally enhance the variational quantum circuit from FIG.~\ref{fig:rte_quantum_circuit_2d}, so that we obtain significantly better results for the real-time evolution in (2+1) dimensions. The Hamiltonian of the system includes a four-body interaction, given by the plaquette term. If we include such a gate and assign a variational parameter to it
\begin{align}
    U_{\text{plaq}}(\theta) = \exp\left(-i(\tilde{U}_P+\tilde{U}^{\dagger}_P)\theta\right),
\end{align}
we can extend each layer of the parametrized quantum circuit by one such unitary. Then, performing the variational real-time evolution, we obtain the results shown in FIG.~\ref{fig:rte_1plaquette_plaquette_joint} and FIG.~\ref{fig:rte_1plaquette_plaquette_entropy}.
\begin{figure}
    \centering
    \includegraphics[width = \columnwidth]{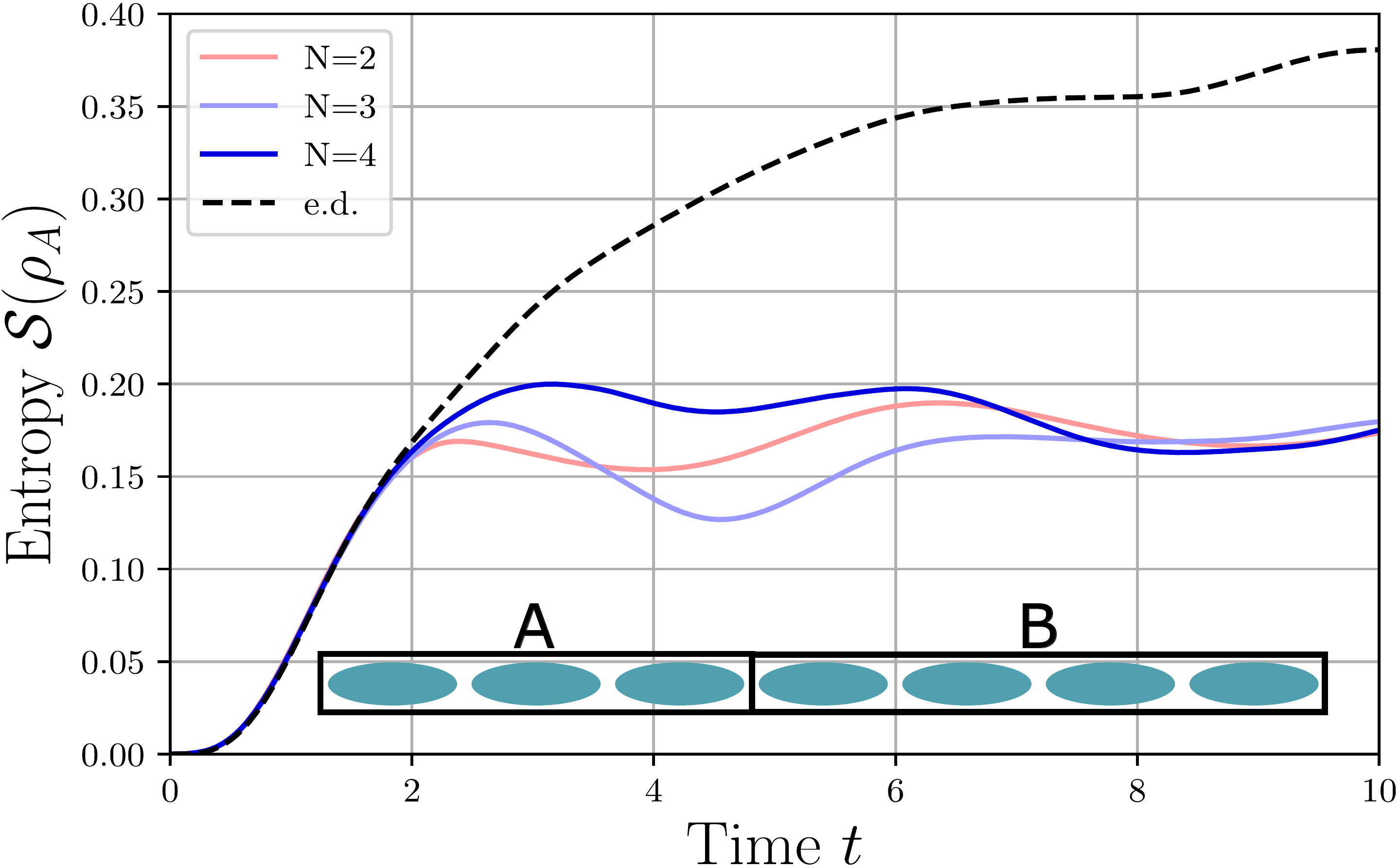}
    \caption{\textbf{Evolution of the entanglement entropy of subsystem A after a quench in a system of 7 qudits in (1+1) dimensions:} After initial linear growth, the entropy $\mathcal{S}(\rho_A)$ saturates to a finite constant up to fluctuations due to finite size effects. As we increase the number of layers $N$ in the variational circuits, we observe better and better approximation to this behaviour of the entropy in the trial state.}
    \label{fig:rte_7qudits_entropy}
\end{figure}
\section{Variational real-time evolution for 7 qudits in (1+1) dimensions}
\label{app:rte_7qudits}

In Section~\ref{sec:results}, we studied the variational real-time evolution of 5 qudits in (1+1) dimensions and observed qualitative agreement for the period of two oscillations (see FIG.~\ref{fig:rte_1d_joint}). Here, we complement the study of the variational real-time for a system of 7 qudits. We calculate the fidelity of the variational state at each time step concerning the exact state, see FIG.~\ref{fig:rte_7qudits_joint}. In addition, we calculate the dynamics of the fermion numbers in the center of the one-dimensional system. We observe that the variational circuit for $N = 3$ and $N=4$ layers can capture the correct behavior of the fermion number for up to one oscillation period. However, the fidelity drops significantly at later times, and the time evolution is no longer accurate. To understand this limitation, we investigate the entanglement entropy in the system of 7 qudits, shown in FIG.~\ref{fig:rte_7qudits_entropy} and observe a restriction in the growth of the entanglement entropy. The limitation of entanglement entropy suggests that the circuits cannot generate the correct form of entanglement to approximate the exact real-time evolution for times later than one oscillation period. In future studies, we will improve this circuit by going to deeper circuits but also by choosing different entangling gates.

\bibliography{ref}

\begin{thebibliography}{81}%
\makeatletter
\providecommand \@ifxundefined [1]{%
 \@ifx{#1\undefined}
}%
\providecommand \@ifnum [1]{%
 \ifnum #1\expandafter \@firstoftwo
 \else \expandafter \@secondoftwo
 \fi
}%
\providecommand \@ifx [1]{%
 \ifx #1\expandafter \@firstoftwo
 \else \expandafter \@secondoftwo
 \fi
}%
\providecommand \natexlab [1]{#1}%
\providecommand \enquote  [1]{``#1''}%
\providecommand \bibnamefont  [1]{#1}%
\providecommand \bibfnamefont [1]{#1}%
\providecommand \citenamefont [1]{#1}%
\providecommand \href@noop [0]{\@secondoftwo}%
\providecommand \href [0]{\begingroup \@sanitize@url \@href}%
\providecommand \@href[1]{\@@startlink{#1}\@@href}%
\providecommand \@@href[1]{\endgroup#1\@@endlink}%
\providecommand \@sanitize@url [0]{\catcode `\\12\catcode `\$12\catcode
  `\&12\catcode `\#12\catcode `\^12\catcode `\_12\catcode `\%12\relax}%
\providecommand \@@startlink[1]{}%
\providecommand \@@endlink[0]{}%
\providecommand \url  [0]{\begingroup\@sanitize@url \@url }%
\providecommand \@url [1]{\endgroup\@href {#1}{\urlprefix }}%
\providecommand \urlprefix  [0]{URL }%
\providecommand \Eprint [0]{\href }%
\providecommand \doibase [0]{http://dx.doi.org/}%
\providecommand \selectlanguage [0]{\@gobble}%
\providecommand \bibinfo  [0]{\@secondoftwo}%
\providecommand \bibfield  [0]{\@secondoftwo}%
\providecommand \translation [1]{[#1]}%
\providecommand \BibitemOpen [0]{}%
\providecommand \bibitemStop [0]{}%
\providecommand \bibitemNoStop [0]{.\EOS\space}%
\providecommand \EOS [0]{\spacefactor3000\relax}%
\providecommand \BibitemShut  [1]{\csname bibitem#1\endcsname}%
\let\auto@bib@innerbib\@empty
\bibitem [{\citenamefont {Wang}\ \emph {et~al.}(2020)\citenamefont {Wang},
  \citenamefont {Hu}, \citenamefont {Sanders},\ and\ \citenamefont
  {Kais}}]{Wang2020}%
  \BibitemOpen
  \bibfield  {author} {\bibinfo {author} {\bibfnamefont {Y.}~\bibnamefont
  {Wang}}, \bibinfo {author} {\bibfnamefont {Z.}~\bibnamefont {Hu}}, \bibinfo
  {author} {\bibfnamefont {B.~C.}\ \bibnamefont {Sanders}}, \ and\ \bibinfo
  {author} {\bibfnamefont {S.}~\bibnamefont {Kais}},\ }\href {\doibase
  10.3389/fphy.2020.589504} {\bibfield  {journal} {\bibinfo  {journal}
  {Frontiers in Physics}\ }\textbf {\bibinfo {volume} {8}} (\bibinfo {year}
  {2020}),\ 10.3389/fphy.2020.589504}\BibitemShut {NoStop}%
\bibitem [{\citenamefont {Lanyon}\ \emph {et~al.}(2008)\citenamefont {Lanyon},
  \citenamefont {Barbieri}, \citenamefont {Almeida}, \citenamefont {Jennewein},
  \citenamefont {Ralph}, \citenamefont {Resch}, \citenamefont {Pryde},
  \citenamefont {O’Brien}, \citenamefont {Gilchrist},\ and\ \citenamefont
  {White}}]{Lanyon2009}%
  \BibitemOpen
  \bibfield  {author} {\bibinfo {author} {\bibfnamefont {B.~P.}\ \bibnamefont
  {Lanyon}}, \bibinfo {author} {\bibfnamefont {M.}~\bibnamefont {Barbieri}},
  \bibinfo {author} {\bibfnamefont {M.~P.}\ \bibnamefont {Almeida}}, \bibinfo
  {author} {\bibfnamefont {T.}~\bibnamefont {Jennewein}}, \bibinfo {author}
  {\bibfnamefont {T.~C.}\ \bibnamefont {Ralph}}, \bibinfo {author}
  {\bibfnamefont {K.~J.}\ \bibnamefont {Resch}}, \bibinfo {author}
  {\bibfnamefont {G.~J.}\ \bibnamefont {Pryde}}, \bibinfo {author}
  {\bibfnamefont {J.~L.}\ \bibnamefont {O’Brien}}, \bibinfo {author}
  {\bibfnamefont {A.}~\bibnamefont {Gilchrist}}, \ and\ \bibinfo {author}
  {\bibfnamefont {A.~G.}\ \bibnamefont {White}},\ }\href {\doibase
  10.1038/nphys1150} {\bibfield  {journal} {\bibinfo  {journal} {Nature
  Physics}\ }\textbf {\bibinfo {volume} {5}},\ \bibinfo {pages} {134} (\bibinfo
  {year} {2008})}\BibitemShut {NoStop}%
\bibitem [{\citenamefont {Stricker}\ \emph {et~al.}(2022)\citenamefont
  {Stricker}, \citenamefont {Meth}, \citenamefont {Postler}, \citenamefont
  {Edmunds}, \citenamefont {Ferrie}, \citenamefont {Blatt}, \citenamefont
  {Schindler}, \citenamefont {Monz}, \citenamefont {Kueng},\ and\ \citenamefont
  {Ringbauer}}]{Stricker2022}%
  \BibitemOpen
  \bibfield  {author} {\bibinfo {author} {\bibfnamefont {R.}~\bibnamefont
  {Stricker}}, \bibinfo {author} {\bibfnamefont {M.}~\bibnamefont {Meth}},
  \bibinfo {author} {\bibfnamefont {L.}~\bibnamefont {Postler}}, \bibinfo
  {author} {\bibfnamefont {C.}~\bibnamefont {Edmunds}}, \bibinfo {author}
  {\bibfnamefont {C.}~\bibnamefont {Ferrie}}, \bibinfo {author} {\bibfnamefont
  {R.}~\bibnamefont {Blatt}}, \bibinfo {author} {\bibfnamefont
  {P.}~\bibnamefont {Schindler}}, \bibinfo {author} {\bibfnamefont
  {T.}~\bibnamefont {Monz}}, \bibinfo {author} {\bibfnamefont {R.}~\bibnamefont
  {Kueng}}, \ and\ \bibinfo {author} {\bibfnamefont {M.}~\bibnamefont
  {Ringbauer}},\ }\href {\doibase 10.1103/PRXQuantum.3.040310} {\bibfield
  {journal} {\bibinfo  {journal} {PRX Quantum}\ }\textbf {\bibinfo {volume}
  {3}},\ \bibinfo {pages} {040310} (\bibinfo {year} {2022})}\BibitemShut
  {NoStop}%
\bibitem [{\citenamefont {MacDonell}\ \emph {et~al.}(2021)\citenamefont
  {MacDonell}, \citenamefont {Dickerson}, \citenamefont {Birch}, \citenamefont
  {Kumar}, \citenamefont {Edmunds}, \citenamefont {Biercuk}, \citenamefont
  {Hempel},\ and\ \citenamefont {Kassal}}]{MacDonell2020}%
  \BibitemOpen
  \bibfield  {author} {\bibinfo {author} {\bibfnamefont {R.~J.}\ \bibnamefont
  {MacDonell}}, \bibinfo {author} {\bibfnamefont {C.~E.}\ \bibnamefont
  {Dickerson}}, \bibinfo {author} {\bibfnamefont {C.~J.~T.}\ \bibnamefont
  {Birch}}, \bibinfo {author} {\bibfnamefont {A.}~\bibnamefont {Kumar}},
  \bibinfo {author} {\bibfnamefont {C.~L.}\ \bibnamefont {Edmunds}}, \bibinfo
  {author} {\bibfnamefont {M.~J.}\ \bibnamefont {Biercuk}}, \bibinfo {author}
  {\bibfnamefont {C.}~\bibnamefont {Hempel}}, \ and\ \bibinfo {author}
  {\bibfnamefont {I.}~\bibnamefont {Kassal}},\ }\href {\doibase
  10.1039/D1SC02142G} {\bibfield  {journal} {\bibinfo  {journal} {Chemical
  Science}\ }\textbf {\bibinfo {volume} {12}},\ \bibinfo {pages} {9794}
  (\bibinfo {year} {2021})}\BibitemShut {NoStop}%
\bibitem [{\citenamefont {Ringbauer}\ \emph {et~al.}(2018)\citenamefont
  {Ringbauer}, \citenamefont {Bromley}, \citenamefont {Cianciaruso},
  \citenamefont {Lami}, \citenamefont {Lau}, \citenamefont {Adesso},
  \citenamefont {White}, \citenamefont {Fedrizzi},\ and\ \citenamefont
  {Piani}}]{Ringbauer2018}%
  \BibitemOpen
  \bibfield  {author} {\bibinfo {author} {\bibfnamefont {M.}~\bibnamefont
  {Ringbauer}}, \bibinfo {author} {\bibfnamefont {T.~R.}\ \bibnamefont
  {Bromley}}, \bibinfo {author} {\bibfnamefont {M.}~\bibnamefont
  {Cianciaruso}}, \bibinfo {author} {\bibfnamefont {L.}~\bibnamefont {Lami}},
  \bibinfo {author} {\bibfnamefont {W.~S.}\ \bibnamefont {Lau}}, \bibinfo
  {author} {\bibfnamefont {G.}~\bibnamefont {Adesso}}, \bibinfo {author}
  {\bibfnamefont {A.~G.}\ \bibnamefont {White}}, \bibinfo {author}
  {\bibfnamefont {A.}~\bibnamefont {Fedrizzi}}, \ and\ \bibinfo {author}
  {\bibfnamefont {M.}~\bibnamefont {Piani}},\ }\href@noop {} {\bibfield
  {journal} {\bibinfo  {journal} {Physical Review X}\ }\textbf {\bibinfo
  {volume} {8}},\ \bibinfo {pages} {041007} (\bibinfo {year}
  {2018})}\BibitemShut {NoStop}%
\bibitem [{\citenamefont {Kraft}\ \emph {et~al.}(2018)\citenamefont {Kraft},
  \citenamefont {Ritz}, \citenamefont {Brunner}, \citenamefont {Huber},\ and\
  \citenamefont {G{\"u}hne}}]{Kraft2018}%
  \BibitemOpen
  \bibfield  {author} {\bibinfo {author} {\bibfnamefont {T.}~\bibnamefont
  {Kraft}}, \bibinfo {author} {\bibfnamefont {C.}~\bibnamefont {Ritz}},
  \bibinfo {author} {\bibfnamefont {N.}~\bibnamefont {Brunner}}, \bibinfo
  {author} {\bibfnamefont {M.}~\bibnamefont {Huber}}, \ and\ \bibinfo {author}
  {\bibfnamefont {O.}~\bibnamefont {G{\"u}hne}},\ }\href@noop {} {\bibfield
  {journal} {\bibinfo  {journal} {Physical review letters}\ }\textbf {\bibinfo
  {volume} {120}},\ \bibinfo {pages} {060502} (\bibinfo {year}
  {2018})}\BibitemShut {NoStop}%
\bibitem [{\citenamefont {Cozzolino}\ \emph {et~al.}(2019)\citenamefont
  {Cozzolino}, \citenamefont {Da~Lio}, \citenamefont {Bacco},\ and\
  \citenamefont {Oxenl{\o}we}}]{Cozzolino2019}%
  \BibitemOpen
  \bibfield  {author} {\bibinfo {author} {\bibfnamefont {D.}~\bibnamefont
  {Cozzolino}}, \bibinfo {author} {\bibfnamefont {B.}~\bibnamefont {Da~Lio}},
  \bibinfo {author} {\bibfnamefont {D.}~\bibnamefont {Bacco}}, \ and\ \bibinfo
  {author} {\bibfnamefont {L.~K.}\ \bibnamefont {Oxenl{\o}we}},\ }\href@noop {}
  {\bibfield  {journal} {\bibinfo  {journal} {Advanced Quantum Technologies}\
  }\textbf {\bibinfo {volume} {2}},\ \bibinfo {pages} {1900038} (\bibinfo
  {year} {2019})}\BibitemShut {NoStop}%
\bibitem [{\citenamefont {Campbell}(2014)}]{Campbell2014}%
  \BibitemOpen
  \bibfield  {author} {\bibinfo {author} {\bibfnamefont {E.~T.}\ \bibnamefont
  {Campbell}},\ }\href {\doibase 10.1103/PhysRevLett.113.230501} {\bibfield
  {journal} {\bibinfo  {journal} {Phys. Rev. Lett.}\ }\textbf {\bibinfo
  {volume} {113}},\ \bibinfo {pages} {230501} (\bibinfo {year}
  {2014})}\BibitemShut {NoStop}%
\bibitem [{\citenamefont {Bru\ss{}}(1998)}]{Bruss1998}%
  \BibitemOpen
  \bibfield  {author} {\bibinfo {author} {\bibfnamefont {D.}~\bibnamefont
  {Bru\ss{}}},\ }\href {\doibase 10.1103/PhysRevLett.81.3018} {\bibfield
  {journal} {\bibinfo  {journal} {Phys. Rev. Lett.}\ }\textbf {\bibinfo
  {volume} {81}},\ \bibinfo {pages} {3018} (\bibinfo {year}
  {1998})}\BibitemShut {NoStop}%
\bibitem [{\citenamefont {Thew}\ \emph {et~al.}(2004)\citenamefont {Thew},
  \citenamefont {Acin}, \citenamefont {Zbinden},\ and\ \citenamefont
  {Gisin}}]{Thew2004}%
  \BibitemOpen
  \bibfield  {author} {\bibinfo {author} {\bibfnamefont {R.~T.}\ \bibnamefont
  {Thew}}, \bibinfo {author} {\bibfnamefont {A.}~\bibnamefont {Acin}}, \bibinfo
  {author} {\bibfnamefont {H.}~\bibnamefont {Zbinden}}, \ and\ \bibinfo
  {author} {\bibfnamefont {N.}~\bibnamefont {Gisin}},\ }\href@noop {}
  {\bibfield  {journal} {\bibinfo  {journal} {Physical review letters}\
  }\textbf {\bibinfo {volume} {93}},\ \bibinfo {pages} {010503} (\bibinfo
  {year} {2004})}\BibitemShut {NoStop}%
\bibitem [{\citenamefont {Morvan}\ \emph {et~al.}(2021)\citenamefont {Morvan},
  \citenamefont {Ramasesh}, \citenamefont {Blok}, \citenamefont {Kreikebaum},
  \citenamefont {O’Brien}, \citenamefont {Chen}, \citenamefont {Mitchell},
  \citenamefont {Naik}, \citenamefont {Santiago},\ and\ \citenamefont
  {Siddiqi}}]{Morvan2021}%
  \BibitemOpen
  \bibfield  {author} {\bibinfo {author} {\bibfnamefont {A.}~\bibnamefont
  {Morvan}}, \bibinfo {author} {\bibfnamefont {V.}~\bibnamefont {Ramasesh}},
  \bibinfo {author} {\bibfnamefont {M.}~\bibnamefont {Blok}}, \bibinfo {author}
  {\bibfnamefont {J.}~\bibnamefont {Kreikebaum}}, \bibinfo {author}
  {\bibfnamefont {K.}~\bibnamefont {O’Brien}}, \bibinfo {author}
  {\bibfnamefont {L.}~\bibnamefont {Chen}}, \bibinfo {author} {\bibfnamefont
  {B.}~\bibnamefont {Mitchell}}, \bibinfo {author} {\bibfnamefont
  {R.}~\bibnamefont {Naik}}, \bibinfo {author} {\bibfnamefont {D.}~\bibnamefont
  {Santiago}}, \ and\ \bibinfo {author} {\bibfnamefont {I.}~\bibnamefont
  {Siddiqi}},\ }\href@noop {} {\bibfield  {journal} {\bibinfo  {journal}
  {Physical review letters}\ }\textbf {\bibinfo {volume} {126}},\ \bibinfo
  {pages} {210504} (\bibinfo {year} {2021})}\BibitemShut {NoStop}%
\bibitem [{\citenamefont {{Chi}}\ \emph {et~al.}(2022)\citenamefont {{Chi}},
  \citenamefont {{Huang}}, \citenamefont {{Zhang}}, \citenamefont {{Mao}},
  \citenamefont {{Zhou}}, \citenamefont {{Chen}}, \citenamefont {{Zhai}},
  \citenamefont {{Bao}}, \citenamefont {{Dai}}, \citenamefont {{Yuan}},
  \citenamefont {{Zhang}}, \citenamefont {{Dai}}, \citenamefont {{Tang}},
  \citenamefont {{Yang}}, \citenamefont {{Li}}, \citenamefont {{Ding}},
  \citenamefont {{Oxenl{\o}we}}, \citenamefont {{Thompson}}, \citenamefont
  {{O'Brien}}, \citenamefont {{Li}}, \citenamefont {{Gong}},\ and\
  \citenamefont {{Wang}}}]{Chi2022}%
  \BibitemOpen
  \bibfield  {author} {\bibinfo {author} {\bibfnamefont {Y.}~\bibnamefont
  {{Chi}}}, \bibinfo {author} {\bibfnamefont {J.}~\bibnamefont {{Huang}}},
  \bibinfo {author} {\bibfnamefont {Z.}~\bibnamefont {{Zhang}}}, \bibinfo
  {author} {\bibfnamefont {J.}~\bibnamefont {{Mao}}}, \bibinfo {author}
  {\bibfnamefont {Z.}~\bibnamefont {{Zhou}}}, \bibinfo {author} {\bibfnamefont
  {X.}~\bibnamefont {{Chen}}}, \bibinfo {author} {\bibfnamefont
  {C.}~\bibnamefont {{Zhai}}}, \bibinfo {author} {\bibfnamefont
  {J.}~\bibnamefont {{Bao}}}, \bibinfo {author} {\bibfnamefont
  {T.}~\bibnamefont {{Dai}}}, \bibinfo {author} {\bibfnamefont
  {H.}~\bibnamefont {{Yuan}}}, \bibinfo {author} {\bibfnamefont
  {M.}~\bibnamefont {{Zhang}}}, \bibinfo {author} {\bibfnamefont
  {D.}~\bibnamefont {{Dai}}}, \bibinfo {author} {\bibfnamefont
  {B.}~\bibnamefont {{Tang}}}, \bibinfo {author} {\bibfnamefont
  {Y.}~\bibnamefont {{Yang}}}, \bibinfo {author} {\bibfnamefont
  {Z.}~\bibnamefont {{Li}}}, \bibinfo {author} {\bibfnamefont {Y.}~\bibnamefont
  {{Ding}}}, \bibinfo {author} {\bibfnamefont {L.~K.}\ \bibnamefont
  {{Oxenl{\o}we}}}, \bibinfo {author} {\bibfnamefont {M.~G.}\ \bibnamefont
  {{Thompson}}}, \bibinfo {author} {\bibfnamefont {J.~L.}\ \bibnamefont
  {{O'Brien}}}, \bibinfo {author} {\bibfnamefont {Y.}~\bibnamefont {{Li}}},
  \bibinfo {author} {\bibfnamefont {Q.}~\bibnamefont {{Gong}}}, \ and\ \bibinfo
  {author} {\bibfnamefont {J.}~\bibnamefont {{Wang}}},\ }\href {\doibase
  10.1038/s41467-022-28767-x} {\bibfield  {journal} {\bibinfo  {journal}
  {Nature Communications}\ }\textbf {\bibinfo {volume} {13}},\ \bibinfo {eid}
  {1166} (\bibinfo {year} {2022})}\BibitemShut {NoStop}%
\bibitem [{\citenamefont {Ringbauer}\ \emph {et~al.}(2022)\citenamefont
  {Ringbauer}, \citenamefont {Meth}, \citenamefont {Postler}, \citenamefont
  {Stricker}, \citenamefont {Blatt}, \citenamefont {Schindler},\ and\
  \citenamefont {Monz}}]{Ringbauer2022}%
  \BibitemOpen
  \bibfield  {author} {\bibinfo {author} {\bibfnamefont {M.}~\bibnamefont
  {Ringbauer}}, \bibinfo {author} {\bibfnamefont {M.}~\bibnamefont {Meth}},
  \bibinfo {author} {\bibfnamefont {L.}~\bibnamefont {Postler}}, \bibinfo
  {author} {\bibfnamefont {R.}~\bibnamefont {Stricker}}, \bibinfo {author}
  {\bibfnamefont {R.}~\bibnamefont {Blatt}}, \bibinfo {author} {\bibfnamefont
  {P.}~\bibnamefont {Schindler}}, \ and\ \bibinfo {author} {\bibfnamefont
  {T.}~\bibnamefont {Monz}},\ }\href {\doibase 10.1038/s41567-022-01658-0}
  {\bibfield  {journal} {\bibinfo  {journal} {Nature Physics}\ }\textbf
  {\bibinfo {volume} {18}},\ \bibinfo {pages} {1053} (\bibinfo {year}
  {2022})}\BibitemShut {NoStop}%
\bibitem [{\citenamefont {Hauke}\ \emph {et~al.}(2012)\citenamefont {Hauke},
  \citenamefont {Cucchietti}, \citenamefont {Tagliacozzo}, \citenamefont
  {Deutsch},\ and\ \citenamefont {Lewenstein}}]{Hauke2012_rev}%
  \BibitemOpen
  \bibfield  {author} {\bibinfo {author} {\bibfnamefont {P.}~\bibnamefont
  {Hauke}}, \bibinfo {author} {\bibfnamefont {F.~M.}\ \bibnamefont
  {Cucchietti}}, \bibinfo {author} {\bibfnamefont {L.}~\bibnamefont
  {Tagliacozzo}}, \bibinfo {author} {\bibfnamefont {I.}~\bibnamefont
  {Deutsch}}, \ and\ \bibinfo {author} {\bibfnamefont {M.}~\bibnamefont
  {Lewenstein}},\ }\href@noop {} {\bibfield  {journal} {\bibinfo  {journal}
  {Reports on Progress in Physics}\ }\textbf {\bibinfo {volume} {75}},\
  \bibinfo {pages} {082401} (\bibinfo {year} {2012})}\BibitemShut {NoStop}%
\bibitem [{\citenamefont {Preskill}(2018)}]{Preskill2018}%
  \BibitemOpen
  \bibfield  {author} {\bibinfo {author} {\bibfnamefont {J.}~\bibnamefont
  {Preskill}},\ }\href {\doibase 10.22331/q-2018-08-06-79} {\bibfield
  {journal} {\bibinfo  {journal} {Quantum}\ }\textbf {\bibinfo {volume} {2}},\
  \bibinfo {pages} {79} (\bibinfo {year} {2018})}\BibitemShut {NoStop}%
\bibitem [{\citenamefont {Bharti}\ \emph {et~al.}(2022)\citenamefont {Bharti},
  \citenamefont {Cervera-Lierta}, \citenamefont {Kyaw}, \citenamefont {Haug},
  \citenamefont {Alperin-Lea}, \citenamefont {Anand}, \citenamefont {Degroote},
  \citenamefont {Heimonen}, \citenamefont {Kottmann}, \citenamefont {Menke},
  \citenamefont {Mok}, \citenamefont {Sim}, \citenamefont {Kwek},\ and\
  \citenamefont {Aspuru-Guzik}}]{Bharti2022}%
  \BibitemOpen
  \bibfield  {author} {\bibinfo {author} {\bibfnamefont {K.}~\bibnamefont
  {Bharti}}, \bibinfo {author} {\bibfnamefont {A.}~\bibnamefont
  {Cervera-Lierta}}, \bibinfo {author} {\bibfnamefont {T.~H.}\ \bibnamefont
  {Kyaw}}, \bibinfo {author} {\bibfnamefont {T.}~\bibnamefont {Haug}}, \bibinfo
  {author} {\bibfnamefont {S.}~\bibnamefont {Alperin-Lea}}, \bibinfo {author}
  {\bibfnamefont {A.}~\bibnamefont {Anand}}, \bibinfo {author} {\bibfnamefont
  {M.}~\bibnamefont {Degroote}}, \bibinfo {author} {\bibfnamefont
  {H.}~\bibnamefont {Heimonen}}, \bibinfo {author} {\bibfnamefont {J.~S.}\
  \bibnamefont {Kottmann}}, \bibinfo {author} {\bibfnamefont {T.}~\bibnamefont
  {Menke}}, \bibinfo {author} {\bibfnamefont {W.-K.}\ \bibnamefont {Mok}},
  \bibinfo {author} {\bibfnamefont {S.}~\bibnamefont {Sim}}, \bibinfo {author}
  {\bibfnamefont {L.-C.}\ \bibnamefont {Kwek}}, \ and\ \bibinfo {author}
  {\bibfnamefont {A.}~\bibnamefont {Aspuru-Guzik}},\ }\href {\doibase
  10.1103/RevModPhys.94.015004} {\bibfield  {journal} {\bibinfo  {journal}
  {Rev. Mod. Phys.}\ }\textbf {\bibinfo {volume} {94}},\ \bibinfo {pages}
  {015004} (\bibinfo {year} {2022})}\BibitemShut {NoStop}%
\bibitem [{\citenamefont {Wiese}(2013)}]{Wiese2013}%
  \BibitemOpen
  \bibfield  {author} {\bibinfo {author} {\bibfnamefont {U.-J.}\ \bibnamefont
  {Wiese}},\ }\href@noop {} {\bibfield  {journal} {\bibinfo  {journal} {Annalen
  der Physik}\ }\textbf {\bibinfo {volume} {525}},\ \bibinfo {pages} {777}
  (\bibinfo {year} {2013})}\BibitemShut {NoStop}%
\bibitem [{\citenamefont {Zohar}\ \emph {et~al.}(2015)\citenamefont {Zohar},
  \citenamefont {Cirac},\ and\ \citenamefont {Reznik}}]{Zohar2015}%
  \BibitemOpen
  \bibfield  {author} {\bibinfo {author} {\bibfnamefont {E.}~\bibnamefont
  {Zohar}}, \bibinfo {author} {\bibfnamefont {J.~I.}\ \bibnamefont {Cirac}}, \
  and\ \bibinfo {author} {\bibfnamefont {B.}~\bibnamefont {Reznik}},\ }\href
  {https://link.aps.org/doi/10.1088/0034-4885/79/1/014401} {\bibfield
  {journal} {\bibinfo  {journal} {Rep. Prog. Phys.}\ }\textbf {\bibinfo
  {volume} {79}} (\bibinfo {year} {2015})}\BibitemShut {NoStop}%
\bibitem [{\citenamefont {Dalmonte}\ and\ \citenamefont
  {Montangero}(2016)}]{Dalmonte2016}%
  \BibitemOpen
  \bibfield  {author} {\bibinfo {author} {\bibfnamefont {M.}~\bibnamefont
  {Dalmonte}}\ and\ \bibinfo {author} {\bibfnamefont {S.}~\bibnamefont
  {Montangero}},\ }\href {\doibase 10.1080/00107514.2016.1151199} {\bibfield
  {journal} {\bibinfo  {journal} {Contemp. Phys.}\ }\textbf {\bibinfo {volume}
  {57}},\ \bibinfo {pages} {388} (\bibinfo {year} {2016})}\BibitemShut
  {NoStop}%
\bibitem [{\citenamefont {Ba{\~{n}}uls}\ \emph {et~al.}(2020)\citenamefont
  {Ba{\~{n}}uls} \emph {et~al.}}]{Banuls2020}%
  \BibitemOpen
  \bibfield  {author} {\bibinfo {author} {\bibfnamefont {M.~C.}\ \bibnamefont
  {Ba{\~{n}}uls}} \emph {et~al.},\ }\href
  {https://link.aps.org/doi/10.1140/epjd/e2020-100571-8} {\bibfield  {journal}
  {\bibinfo  {journal} {Eur. Phys. J. D}\ }\textbf {\bibinfo {volume} {74}}
  (\bibinfo {year} {2020})}\BibitemShut {NoStop}%
\bibitem [{\citenamefont {Aidelsburger}\ \emph {et~al.}(2022)\citenamefont
  {Aidelsburger}, \citenamefont {Barbiero}, \citenamefont {Bermudez},
  \citenamefont {Chanda}, \citenamefont {Dauphin}, \citenamefont
  {Gonz{\'a}lez-Cuadra}, \citenamefont {Grzybowski}, \citenamefont {Hands},
  \citenamefont {Jendrzejewski}, \citenamefont {J{\"u}nemann} \emph
  {et~al.}}]{Aidelsburger2022}%
  \BibitemOpen
  \bibfield  {author} {\bibinfo {author} {\bibfnamefont {M.}~\bibnamefont
  {Aidelsburger}}, \bibinfo {author} {\bibfnamefont {L.}~\bibnamefont
  {Barbiero}}, \bibinfo {author} {\bibfnamefont {A.}~\bibnamefont {Bermudez}},
  \bibinfo {author} {\bibfnamefont {T.}~\bibnamefont {Chanda}}, \bibinfo
  {author} {\bibfnamefont {A.}~\bibnamefont {Dauphin}}, \bibinfo {author}
  {\bibfnamefont {D.}~\bibnamefont {Gonz{\'a}lez-Cuadra}}, \bibinfo {author}
  {\bibfnamefont {P.~R.}\ \bibnamefont {Grzybowski}}, \bibinfo {author}
  {\bibfnamefont {S.}~\bibnamefont {Hands}}, \bibinfo {author} {\bibfnamefont
  {F.}~\bibnamefont {Jendrzejewski}}, \bibinfo {author} {\bibfnamefont
  {J.}~\bibnamefont {J{\"u}nemann}},  \emph {et~al.},\ }\href@noop {}
  {\bibfield  {journal} {\bibinfo  {journal} {Philosophical Transactions of the
  Royal Society A}\ }\textbf {\bibinfo {volume} {380}},\ \bibinfo {pages}
  {20210064} (\bibinfo {year} {2022})}\BibitemShut {NoStop}%
\bibitem [{\citenamefont {Zohar}(2022)}]{Zohar2022}%
  \BibitemOpen
  \bibfield  {author} {\bibinfo {author} {\bibfnamefont {E.}~\bibnamefont
  {Zohar}},\ }\href@noop {} {\bibfield  {journal} {\bibinfo  {journal}
  {Philosophical Transactions of the Royal Society A}\ }\textbf {\bibinfo
  {volume} {380}},\ \bibinfo {pages} {20210069} (\bibinfo {year}
  {2022})}\BibitemShut {NoStop}%
\bibitem [{\citenamefont {Bauer}\ \emph {et~al.}(2023)\citenamefont {Bauer},
  \citenamefont {Davoudi}, \citenamefont {Balantekin}, \citenamefont
  {Bhattacharya}, \citenamefont {Carena}, \citenamefont {de~Jong},
  \citenamefont {Draper}, \citenamefont {El-Khadra}, \citenamefont {Gemelke},
  \citenamefont {Hanada}, \citenamefont {Kharzeev}, \citenamefont {Lamm},
  \citenamefont {Li}, \citenamefont {Liu}, \citenamefont {Lukin}, \citenamefont
  {Meurice}, \citenamefont {Monroe}, \citenamefont {Nachman}, \citenamefont
  {Pagano}, \citenamefont {Preskill}, \citenamefont {Rinaldi}, \citenamefont
  {Roggero}, \citenamefont {Santiago}, \citenamefont {Savage}, \citenamefont
  {Siddiqi}, \citenamefont {Siopsis}, \citenamefont {Van~Zanten}, \citenamefont
  {Wiebe}, \citenamefont {Yamauchi}, \citenamefont {Yeter-Aydeniz},\ and\
  \citenamefont {Zorzetti}}]{Bauer2023}%
  \BibitemOpen
  \bibfield  {author} {\bibinfo {author} {\bibfnamefont {C.~W.}\ \bibnamefont
  {Bauer}}, \bibinfo {author} {\bibfnamefont {Z.}~\bibnamefont {Davoudi}},
  \bibinfo {author} {\bibfnamefont {A.~B.}\ \bibnamefont {Balantekin}},
  \bibinfo {author} {\bibfnamefont {T.}~\bibnamefont {Bhattacharya}}, \bibinfo
  {author} {\bibfnamefont {M.}~\bibnamefont {Carena}}, \bibinfo {author}
  {\bibfnamefont {W.~A.}\ \bibnamefont {de~Jong}}, \bibinfo {author}
  {\bibfnamefont {P.}~\bibnamefont {Draper}}, \bibinfo {author} {\bibfnamefont
  {A.}~\bibnamefont {El-Khadra}}, \bibinfo {author} {\bibfnamefont
  {N.}~\bibnamefont {Gemelke}}, \bibinfo {author} {\bibfnamefont
  {M.}~\bibnamefont {Hanada}}, \bibinfo {author} {\bibfnamefont
  {D.}~\bibnamefont {Kharzeev}}, \bibinfo {author} {\bibfnamefont
  {H.}~\bibnamefont {Lamm}}, \bibinfo {author} {\bibfnamefont {Y.-Y.}\
  \bibnamefont {Li}}, \bibinfo {author} {\bibfnamefont {J.}~\bibnamefont
  {Liu}}, \bibinfo {author} {\bibfnamefont {M.}~\bibnamefont {Lukin}}, \bibinfo
  {author} {\bibfnamefont {Y.}~\bibnamefont {Meurice}}, \bibinfo {author}
  {\bibfnamefont {C.}~\bibnamefont {Monroe}}, \bibinfo {author} {\bibfnamefont
  {B.}~\bibnamefont {Nachman}}, \bibinfo {author} {\bibfnamefont
  {G.}~\bibnamefont {Pagano}}, \bibinfo {author} {\bibfnamefont
  {J.}~\bibnamefont {Preskill}}, \bibinfo {author} {\bibfnamefont
  {E.}~\bibnamefont {Rinaldi}}, \bibinfo {author} {\bibfnamefont
  {A.}~\bibnamefont {Roggero}}, \bibinfo {author} {\bibfnamefont {D.~I.}\
  \bibnamefont {Santiago}}, \bibinfo {author} {\bibfnamefont {M.~J.}\
  \bibnamefont {Savage}}, \bibinfo {author} {\bibfnamefont {I.}~\bibnamefont
  {Siddiqi}}, \bibinfo {author} {\bibfnamefont {G.}~\bibnamefont {Siopsis}},
  \bibinfo {author} {\bibfnamefont {D.}~\bibnamefont {Van~Zanten}}, \bibinfo
  {author} {\bibfnamefont {N.}~\bibnamefont {Wiebe}}, \bibinfo {author}
  {\bibfnamefont {Y.}~\bibnamefont {Yamauchi}}, \bibinfo {author}
  {\bibfnamefont {K.}~\bibnamefont {Yeter-Aydeniz}}, \ and\ \bibinfo {author}
  {\bibfnamefont {S.}~\bibnamefont {Zorzetti}},\ }\href {\doibase
  10.1103/PRXQuantum.4.027001} {\bibfield  {journal} {\bibinfo  {journal} {PRX
  Quantum}\ }\textbf {\bibinfo {volume} {4}},\ \bibinfo {pages} {027001}
  (\bibinfo {year} {2023})}\BibitemShut {NoStop}%
\bibitem [{\citenamefont {Weinberg}(1996)}]{Weinberg1996}%
  \BibitemOpen
  \bibfield  {author} {\bibinfo {author} {\bibfnamefont {S.}~\bibnamefont
  {Weinberg}},\ }\href
  {http://www.slac.stanford.edu/spires/find/hep/www?key=3763846} {\emph
  {\bibinfo {title} {{The quantum theory of fields. Vol. 2: Modern
  applications}}}}\ (\bibinfo {year} {1996})\ \bibinfo {note} {cambridge, UK:
  Univ. Pr. (1996) 489 p}\BibitemShut {NoStop}%
\bibitem [{\citenamefont {Kitaev}(2003)}]{Kitaev1997}%
  \BibitemOpen
  \bibfield  {author} {\bibinfo {author} {\bibfnamefont {A.~Y.}\ \bibnamefont
  {Kitaev}},\ }\href@noop {} {\bibfield  {journal} {\bibinfo  {journal} {Annals
  of physics}\ }\textbf {\bibinfo {volume} {303}},\ \bibinfo {pages} {2}
  (\bibinfo {year} {2003})}\BibitemShut {NoStop}%
\bibitem [{\citenamefont {Tagliacozzo}\ \emph
  {et~al.}(2013{\natexlab{a}})\citenamefont {Tagliacozzo}, \citenamefont
  {Celi}, \citenamefont {Orland}, \citenamefont {Mitchell},\ and\ \citenamefont
  {Lewenstein}}]{Tagliacozzo2013}%
  \BibitemOpen
  \bibfield  {author} {\bibinfo {author} {\bibfnamefont {L.}~\bibnamefont
  {Tagliacozzo}}, \bibinfo {author} {\bibfnamefont {A.}~\bibnamefont {Celi}},
  \bibinfo {author} {\bibfnamefont {P.}~\bibnamefont {Orland}}, \bibinfo
  {author} {\bibfnamefont {M.}~\bibnamefont {Mitchell}}, \ and\ \bibinfo
  {author} {\bibfnamefont {M.}~\bibnamefont {Lewenstein}},\ }\href@noop {}
  {\bibfield  {journal} {\bibinfo  {journal} {Nature communications}\ }\textbf
  {\bibinfo {volume} {4}},\ \bibinfo {pages} {2615} (\bibinfo {year}
  {2013}{\natexlab{a}})}\BibitemShut {NoStop}%
\bibitem [{\citenamefont {Tagliacozzo}\ \emph
  {et~al.}(2013{\natexlab{b}})\citenamefont {Tagliacozzo}, \citenamefont
  {Celi}, \citenamefont {Zamora},\ and\ \citenamefont
  {Lewenstein}}]{Tagliacozzo2013_2}%
  \BibitemOpen
  \bibfield  {author} {\bibinfo {author} {\bibfnamefont {L.}~\bibnamefont
  {Tagliacozzo}}, \bibinfo {author} {\bibfnamefont {A.}~\bibnamefont {Celi}},
  \bibinfo {author} {\bibfnamefont {A.}~\bibnamefont {Zamora}}, \ and\ \bibinfo
  {author} {\bibfnamefont {M.}~\bibnamefont {Lewenstein}},\ }\href@noop {}
  {\bibfield  {journal} {\bibinfo  {journal} {Annals of Physics}\ }\textbf
  {\bibinfo {volume} {330}},\ \bibinfo {pages} {160} (\bibinfo {year}
  {2013}{\natexlab{b}})}\BibitemShut {NoStop}%
\bibitem [{\citenamefont {Zohar}\ \emph {et~al.}(2013)\citenamefont {Zohar},
  \citenamefont {Cirac},\ and\ \citenamefont {Reznik}}]{Zohar2013}%
  \BibitemOpen
  \bibfield  {author} {\bibinfo {author} {\bibfnamefont {E.}~\bibnamefont
  {Zohar}}, \bibinfo {author} {\bibfnamefont {J.~I.}\ \bibnamefont {Cirac}}, \
  and\ \bibinfo {author} {\bibfnamefont {B.}~\bibnamefont {Reznik}},\ }\href
  {https://link.aps.org/doi/10.1103/PhysRevLett.110.125304} {\bibfield
  {journal} {\bibinfo  {journal} {Phys. Rev. Lett.}\ }\textbf {\bibinfo
  {volume} {110}} (\bibinfo {year} {2013})}\BibitemShut {NoStop}%
\bibitem [{\citenamefont {Banerjee}\ \emph {et~al.}(2013)\citenamefont
  {Banerjee}, \citenamefont {B{\"{o}}gli}, \citenamefont {Dalmonte},
  \citenamefont {Rico}, \citenamefont {Stebler}, \citenamefont {Wiese},\ and\
  \citenamefont {Zoller}}]{Banerjee2013}%
  \BibitemOpen
  \bibfield  {author} {\bibinfo {author} {\bibfnamefont {D.}~\bibnamefont
  {Banerjee}}, \bibinfo {author} {\bibfnamefont {M.}~\bibnamefont
  {B{\"{o}}gli}}, \bibinfo {author} {\bibfnamefont {M.}~\bibnamefont
  {Dalmonte}}, \bibinfo {author} {\bibfnamefont {E.}~\bibnamefont {Rico}},
  \bibinfo {author} {\bibfnamefont {P.}~\bibnamefont {Stebler}}, \bibinfo
  {author} {\bibfnamefont {U.-J.}\ \bibnamefont {Wiese}}, \ and\ \bibinfo
  {author} {\bibfnamefont {P.}~\bibnamefont {Zoller}},\ }\href {\doibase
  10.1103/PhysRevLett.110.125303} {\bibfield  {journal} {\bibinfo  {journal}
  {Phys. Rev. Lett.}\ }\textbf {\bibinfo {volume} {110}},\ \bibinfo {pages}
  {125303} (\bibinfo {year} {2013})}\BibitemShut {NoStop}%
\bibitem [{\citenamefont {Stannigel}\ \emph {et~al.}(2014)\citenamefont
  {Stannigel}, \citenamefont {Hauke}, \citenamefont {Marcos}, \citenamefont
  {Hafezi}, \citenamefont {Diehl}, \citenamefont {Dalmonte},\ and\
  \citenamefont {Zoller}}]{Stannigel2014}%
  \BibitemOpen
  \bibfield  {author} {\bibinfo {author} {\bibfnamefont {K.}~\bibnamefont
  {Stannigel}}, \bibinfo {author} {\bibfnamefont {P.}~\bibnamefont {Hauke}},
  \bibinfo {author} {\bibfnamefont {D.}~\bibnamefont {Marcos}}, \bibinfo
  {author} {\bibfnamefont {M.}~\bibnamefont {Hafezi}}, \bibinfo {author}
  {\bibfnamefont {S.}~\bibnamefont {Diehl}}, \bibinfo {author} {\bibfnamefont
  {M.}~\bibnamefont {Dalmonte}}, \ and\ \bibinfo {author} {\bibfnamefont
  {P.}~\bibnamefont {Zoller}},\ }\href {\doibase
  10.1103/PhysRevLett.112.120406} {\bibfield  {journal} {\bibinfo  {journal}
  {Phys. Rev. Lett.}\ }\textbf {\bibinfo {volume} {112}},\ \bibinfo {pages}
  {120406} (\bibinfo {year} {2014})}\BibitemShut {NoStop}%
\bibitem [{\citenamefont {Kasper}\ \emph {et~al.}(2017)\citenamefont {Kasper},
  \citenamefont {Hebenstreit}, \citenamefont {Jendrzejewski}, \citenamefont
  {Oberthaler},\ and\ \citenamefont {Berges}}]{Kasper2017}%
  \BibitemOpen
  \bibfield  {author} {\bibinfo {author} {\bibfnamefont {V.}~\bibnamefont
  {Kasper}}, \bibinfo {author} {\bibfnamefont {F.}~\bibnamefont {Hebenstreit}},
  \bibinfo {author} {\bibfnamefont {F.}~\bibnamefont {Jendrzejewski}}, \bibinfo
  {author} {\bibfnamefont {M.~K.}\ \bibnamefont {Oberthaler}}, \ and\ \bibinfo
  {author} {\bibfnamefont {J.}~\bibnamefont {Berges}},\ }\href {\doibase
  10.1088/1367-2630/aa54e0} {\bibfield  {journal} {\bibinfo  {journal} {New J.
  Phys.}\ }\textbf {\bibinfo {volume} {19}},\ \bibinfo {pages} {023030}
  (\bibinfo {year} {2017})}\BibitemShut {NoStop}%
\bibitem [{\citenamefont {Zohar}\ \emph {et~al.}(2017)\citenamefont {Zohar},
  \citenamefont {Farace}, \citenamefont {Reznik},\ and\ \citenamefont
  {Cirac}}]{Zohar2017}%
  \BibitemOpen
  \bibfield  {author} {\bibinfo {author} {\bibfnamefont {E.}~\bibnamefont
  {Zohar}}, \bibinfo {author} {\bibfnamefont {A.}~\bibnamefont {Farace}},
  \bibinfo {author} {\bibfnamefont {B.}~\bibnamefont {Reznik}}, \ and\ \bibinfo
  {author} {\bibfnamefont {J.~I.}\ \bibnamefont {Cirac}},\ }\href {\doibase
  10.1103/PhysRevLett.118.070501} {\bibfield  {journal} {\bibinfo  {journal}
  {Phys. Rev. Lett.}\ }\textbf {\bibinfo {volume} {118}},\ \bibinfo {pages}
  {070501} (\bibinfo {year} {2017})}\BibitemShut {NoStop}%
\bibitem [{\citenamefont {Bender}\ \emph {et~al.}(2018)\citenamefont {Bender},
  \citenamefont {Zohar}, \citenamefont {Farace},\ and\ \citenamefont
  {Cirac}}]{Bender2018}%
  \BibitemOpen
  \bibfield  {author} {\bibinfo {author} {\bibfnamefont {J.}~\bibnamefont
  {Bender}}, \bibinfo {author} {\bibfnamefont {E.}~\bibnamefont {Zohar}},
  \bibinfo {author} {\bibfnamefont {A.}~\bibnamefont {Farace}}, \ and\ \bibinfo
  {author} {\bibfnamefont {J.~I.}\ \bibnamefont {Cirac}},\ }\href@noop {}
  {\bibfield  {journal} {\bibinfo  {journal} {New Journal of Physics}\ }\textbf
  {\bibinfo {volume} {20}},\ \bibinfo {pages} {093001} (\bibinfo {year}
  {2018})}\BibitemShut {NoStop}%
\bibitem [{\citenamefont {Kasper}\ \emph {et~al.}(2023)\citenamefont {Kasper},
  \citenamefont {Zache}, \citenamefont {Jendrzejewski}, \citenamefont
  {Lewenstein},\ and\ \citenamefont {Zohar}}]{Kasper2023}%
  \BibitemOpen
  \bibfield  {author} {\bibinfo {author} {\bibfnamefont {V.}~\bibnamefont
  {Kasper}}, \bibinfo {author} {\bibfnamefont {T.~V.}\ \bibnamefont {Zache}},
  \bibinfo {author} {\bibfnamefont {F.}~\bibnamefont {Jendrzejewski}}, \bibinfo
  {author} {\bibfnamefont {M.}~\bibnamefont {Lewenstein}}, \ and\ \bibinfo
  {author} {\bibfnamefont {E.}~\bibnamefont {Zohar}},\ }\href@noop {}
  {\bibfield  {journal} {\bibinfo  {journal} {Physical Review D}\ }\textbf
  {\bibinfo {volume} {107}},\ \bibinfo {pages} {014506} (\bibinfo {year}
  {2023})}\BibitemShut {NoStop}%
\bibitem [{\citenamefont {Gonz{\'a}lez-Cuadra}\ \emph
  {et~al.}(2022)\citenamefont {Gonz{\'a}lez-Cuadra}, \citenamefont {Zache},
  \citenamefont {Carrasco}, \citenamefont {Kraus},\ and\ \citenamefont
  {Zoller}}]{Gonzalez-Cuadra2022}%
  \BibitemOpen
  \bibfield  {author} {\bibinfo {author} {\bibfnamefont {D.}~\bibnamefont
  {Gonz{\'a}lez-Cuadra}}, \bibinfo {author} {\bibfnamefont {T.~V.}\
  \bibnamefont {Zache}}, \bibinfo {author} {\bibfnamefont {J.}~\bibnamefont
  {Carrasco}}, \bibinfo {author} {\bibfnamefont {B.}~\bibnamefont {Kraus}}, \
  and\ \bibinfo {author} {\bibfnamefont {P.}~\bibnamefont {Zoller}},\
  }\href@noop {} {\bibfield  {journal} {\bibinfo  {journal} {Physical Review
  Letters}\ }\textbf {\bibinfo {volume} {129}},\ \bibinfo {pages} {160501}
  (\bibinfo {year} {2022})}\BibitemShut {NoStop}%
\bibitem [{\citenamefont {Osborne}\ \emph {et~al.}(2022)\citenamefont
  {Osborne}, \citenamefont {McCulloch}, \citenamefont {Yang}, \citenamefont
  {Hauke},\ and\ \citenamefont {Halimeh}}]{Osborne2022}%
  \BibitemOpen
  \bibfield  {author} {\bibinfo {author} {\bibfnamefont {J.}~\bibnamefont
  {Osborne}}, \bibinfo {author} {\bibfnamefont {I.~P.}\ \bibnamefont
  {McCulloch}}, \bibinfo {author} {\bibfnamefont {B.}~\bibnamefont {Yang}},
  \bibinfo {author} {\bibfnamefont {P.}~\bibnamefont {Hauke}}, \ and\ \bibinfo
  {author} {\bibfnamefont {J.~C.}\ \bibnamefont {Halimeh}},\ }\href@noop {} {\
  (\bibinfo {year} {2022})}\BibitemShut {NoStop}%
\bibitem [{\citenamefont {Zache}\ \emph {et~al.}(2023)\citenamefont {Zache},
  \citenamefont {Gonz{\'a}lez-Cuadra},\ and\ \citenamefont
  {Zoller}}]{Zache2023}%
  \BibitemOpen
  \bibfield  {author} {\bibinfo {author} {\bibfnamefont {T.~V.}\ \bibnamefont
  {Zache}}, \bibinfo {author} {\bibfnamefont {D.}~\bibnamefont
  {Gonz{\'a}lez-Cuadra}}, \ and\ \bibinfo {author} {\bibfnamefont
  {P.}~\bibnamefont {Zoller}},\ }\href@noop {} {\bibfield  {journal} {\bibinfo
  {journal} {arXiv preprint arXiv:2303.08683}\ } (\bibinfo {year}
  {2023})}\BibitemShut {NoStop}%
\bibitem [{\citenamefont {Hauke}\ \emph {et~al.}(2013)\citenamefont {Hauke},
  \citenamefont {Marcos}, \citenamefont {Dalmonte},\ and\ \citenamefont
  {Zoller}}]{Hauke2013}%
  \BibitemOpen
  \bibfield  {author} {\bibinfo {author} {\bibfnamefont {P.}~\bibnamefont
  {Hauke}}, \bibinfo {author} {\bibfnamefont {D.}~\bibnamefont {Marcos}},
  \bibinfo {author} {\bibfnamefont {M.}~\bibnamefont {Dalmonte}}, \ and\
  \bibinfo {author} {\bibfnamefont {P.}~\bibnamefont {Zoller}},\ }\href
  {\doibase 10.1103/PhysRevX.3.041018} {\bibfield  {journal} {\bibinfo
  {journal} {Phys. Rev. X}\ }\textbf {\bibinfo {volume} {3}},\ \bibinfo {pages}
  {041018} (\bibinfo {year} {2013})}\BibitemShut {NoStop}%
\bibitem [{\citenamefont {Yang}\ \emph {et~al.}(2016)\citenamefont {Yang},
  \citenamefont {Giri}, \citenamefont {Johanning}, \citenamefont {Wunderlich},
  \citenamefont {Zoller},\ and\ \citenamefont {Hauke}}]{Yang2016}%
  \BibitemOpen
  \bibfield  {author} {\bibinfo {author} {\bibfnamefont {D.}~\bibnamefont
  {Yang}}, \bibinfo {author} {\bibfnamefont {G.~S.}\ \bibnamefont {Giri}},
  \bibinfo {author} {\bibfnamefont {M.}~\bibnamefont {Johanning}}, \bibinfo
  {author} {\bibfnamefont {C.}~\bibnamefont {Wunderlich}}, \bibinfo {author}
  {\bibfnamefont {P.}~\bibnamefont {Zoller}}, \ and\ \bibinfo {author}
  {\bibfnamefont {P.}~\bibnamefont {Hauke}},\ }\href {\doibase
  10.1103/PhysRevA.94.052321} {\bibfield  {journal} {\bibinfo  {journal} {Phys.
  Rev. A}\ }\textbf {\bibinfo {volume} {94}},\ \bibinfo {pages} {052321}
  (\bibinfo {year} {2016})}\BibitemShut {NoStop}%
\bibitem [{\citenamefont {Muschik}\ \emph {et~al.}(2017)\citenamefont
  {Muschik}, \citenamefont {Heyl}, \citenamefont {Martinez}, \citenamefont
  {Monz}, \citenamefont {Schindler}, \citenamefont {Vogell}, \citenamefont
  {Dalmonte}, \citenamefont {Hauke}, \citenamefont {Blatt},\ and\ \citenamefont
  {Zoller}}]{Muschik2017}%
  \BibitemOpen
  \bibfield  {author} {\bibinfo {author} {\bibfnamefont {C.}~\bibnamefont
  {Muschik}}, \bibinfo {author} {\bibfnamefont {M.}~\bibnamefont {Heyl}},
  \bibinfo {author} {\bibfnamefont {E.}~\bibnamefont {Martinez}}, \bibinfo
  {author} {\bibfnamefont {T.}~\bibnamefont {Monz}}, \bibinfo {author}
  {\bibfnamefont {P.}~\bibnamefont {Schindler}}, \bibinfo {author}
  {\bibfnamefont {B.}~\bibnamefont {Vogell}}, \bibinfo {author} {\bibfnamefont
  {M.}~\bibnamefont {Dalmonte}}, \bibinfo {author} {\bibfnamefont
  {P.}~\bibnamefont {Hauke}}, \bibinfo {author} {\bibfnamefont
  {R.}~\bibnamefont {Blatt}}, \ and\ \bibinfo {author} {\bibfnamefont
  {P.}~\bibnamefont {Zoller}},\ }\href@noop {} {\bibfield  {journal} {\bibinfo
  {journal} {New Journal of Physics}\ }\textbf {\bibinfo {volume} {19}},\
  \bibinfo {pages} {103020} (\bibinfo {year} {2017})}\BibitemShut {NoStop}%
\bibitem [{\citenamefont {Paulson}\ \emph {et~al.}(2021)\citenamefont
  {Paulson}, \citenamefont {Dellantonio}, \citenamefont {Haase}, \citenamefont
  {Celi}, \citenamefont {Kan}, \citenamefont {Jena}, \citenamefont {Kokail},
  \citenamefont {van Bijnen}, \citenamefont {Jansen}, \citenamefont {Zoller},\
  and\ \citenamefont {Muschik}}]{Paulson2021}%
  \BibitemOpen
  \bibfield  {author} {\bibinfo {author} {\bibfnamefont {D.}~\bibnamefont
  {Paulson}}, \bibinfo {author} {\bibfnamefont {L.}~\bibnamefont
  {Dellantonio}}, \bibinfo {author} {\bibfnamefont {J.~F.}\ \bibnamefont
  {Haase}}, \bibinfo {author} {\bibfnamefont {A.}~\bibnamefont {Celi}},
  \bibinfo {author} {\bibfnamefont {A.}~\bibnamefont {Kan}}, \bibinfo {author}
  {\bibfnamefont {A.}~\bibnamefont {Jena}}, \bibinfo {author} {\bibfnamefont
  {C.}~\bibnamefont {Kokail}}, \bibinfo {author} {\bibfnamefont
  {R.}~\bibnamefont {van Bijnen}}, \bibinfo {author} {\bibfnamefont
  {K.}~\bibnamefont {Jansen}}, \bibinfo {author} {\bibfnamefont
  {P.}~\bibnamefont {Zoller}}, \ and\ \bibinfo {author} {\bibfnamefont {C.~A.}\
  \bibnamefont {Muschik}},\ }\href {\doibase 10.1103/PRXQuantum.2.030334}
  {\bibfield  {journal} {\bibinfo  {journal} {PRX Quantum}\ }\textbf {\bibinfo
  {volume} {2}},\ \bibinfo {pages} {030334} (\bibinfo {year}
  {2021})}\BibitemShut {NoStop}%
\bibitem [{\citenamefont {Davoudi}\ \emph {et~al.}(2021)\citenamefont
  {Davoudi}, \citenamefont {Linke},\ and\ \citenamefont
  {Pagano}}]{Davoudi2021}%
  \BibitemOpen
  \bibfield  {author} {\bibinfo {author} {\bibfnamefont {Z.}~\bibnamefont
  {Davoudi}}, \bibinfo {author} {\bibfnamefont {N.~M.}\ \bibnamefont {Linke}},
  \ and\ \bibinfo {author} {\bibfnamefont {G.}~\bibnamefont {Pagano}},\ }\href
  {\doibase 10.1103/PhysRevResearch.3.043072} {\bibfield  {journal} {\bibinfo
  {journal} {Phys. Rev. Res.}\ }\textbf {\bibinfo {volume} {3}},\ \bibinfo
  {pages} {043072} (\bibinfo {year} {2021})}\BibitemShut {NoStop}%
\bibitem [{\citenamefont {Mezzacapo}\ \emph {et~al.}(2015)\citenamefont
  {Mezzacapo}, \citenamefont {Rico}, \citenamefont {Sab\'{\i}n}, \citenamefont
  {Egusquiza}, \citenamefont {Lamata},\ and\ \citenamefont
  {Solano}}]{Mezzacapo2015}%
  \BibitemOpen
  \bibfield  {author} {\bibinfo {author} {\bibfnamefont {A.}~\bibnamefont
  {Mezzacapo}}, \bibinfo {author} {\bibfnamefont {E.}~\bibnamefont {Rico}},
  \bibinfo {author} {\bibfnamefont {C.}~\bibnamefont {Sab\'{\i}n}}, \bibinfo
  {author} {\bibfnamefont {I.~L.}\ \bibnamefont {Egusquiza}}, \bibinfo {author}
  {\bibfnamefont {L.}~\bibnamefont {Lamata}}, \ and\ \bibinfo {author}
  {\bibfnamefont {E.}~\bibnamefont {Solano}},\ }\href {\doibase
  10.1103/PhysRevLett.115.240502} {\bibfield  {journal} {\bibinfo  {journal}
  {Phys. Rev. Lett.}\ }\textbf {\bibinfo {volume} {115}},\ \bibinfo {pages}
  {240502} (\bibinfo {year} {2015})}\BibitemShut {NoStop}%
\bibitem [{\citenamefont {Klco}\ \emph {et~al.}(2018)\citenamefont {Klco},
  \citenamefont {Dumitrescu}, \citenamefont {McCaskey}, \citenamefont {Morris},
  \citenamefont {Pooser}, \citenamefont {Sanz}, \citenamefont {Solano},
  \citenamefont {Lougovski},\ and\ \citenamefont {Savage}}]{Klco2018}%
  \BibitemOpen
  \bibfield  {author} {\bibinfo {author} {\bibfnamefont {N.}~\bibnamefont
  {Klco}}, \bibinfo {author} {\bibfnamefont {E.~F.}\ \bibnamefont
  {Dumitrescu}}, \bibinfo {author} {\bibfnamefont {A.~J.}\ \bibnamefont
  {McCaskey}}, \bibinfo {author} {\bibfnamefont {T.~D.}\ \bibnamefont
  {Morris}}, \bibinfo {author} {\bibfnamefont {R.~C.}\ \bibnamefont {Pooser}},
  \bibinfo {author} {\bibfnamefont {M.}~\bibnamefont {Sanz}}, \bibinfo {author}
  {\bibfnamefont {E.}~\bibnamefont {Solano}}, \bibinfo {author} {\bibfnamefont
  {P.}~\bibnamefont {Lougovski}}, \ and\ \bibinfo {author} {\bibfnamefont
  {M.~J.}\ \bibnamefont {Savage}},\ }\href {\doibase
  10.1103/PhysRevA.98.032331} {\bibfield  {journal} {\bibinfo  {journal} {Phys.
  Rev. A}\ }\textbf {\bibinfo {volume} {98}},\ \bibinfo {pages} {032331}
  (\bibinfo {year} {2018})}\BibitemShut {NoStop}%
\bibitem [{\citenamefont {Atas}\ \emph {et~al.}(2021)\citenamefont {Atas},
  \citenamefont {Zhang}, \citenamefont {Lewis}, \citenamefont {Jahanpour},
  \citenamefont {Haase},\ and\ \citenamefont {Muschik}}]{Atas2021}%
  \BibitemOpen
  \bibfield  {author} {\bibinfo {author} {\bibfnamefont {Y.~Y.}\ \bibnamefont
  {Atas}}, \bibinfo {author} {\bibfnamefont {J.}~\bibnamefont {Zhang}},
  \bibinfo {author} {\bibfnamefont {R.}~\bibnamefont {Lewis}}, \bibinfo
  {author} {\bibfnamefont {A.}~\bibnamefont {Jahanpour}}, \bibinfo {author}
  {\bibfnamefont {J.~F.}\ \bibnamefont {Haase}}, \ and\ \bibinfo {author}
  {\bibfnamefont {C.~A.}\ \bibnamefont {Muschik}},\ }\href@noop {} {\bibfield
  {journal} {\bibinfo  {journal} {Nature communications}\ }\textbf {\bibinfo
  {volume} {12}},\ \bibinfo {pages} {6499} (\bibinfo {year}
  {2021})}\BibitemShut {NoStop}%
\bibitem [{\citenamefont {Ciavarella}\ \emph {et~al.}(2021)\citenamefont
  {Ciavarella}, \citenamefont {Klco},\ and\ \citenamefont
  {Savage}}]{Ciavarella2021}%
  \BibitemOpen
  \bibfield  {author} {\bibinfo {author} {\bibfnamefont {A.}~\bibnamefont
  {Ciavarella}}, \bibinfo {author} {\bibfnamefont {N.}~\bibnamefont {Klco}}, \
  and\ \bibinfo {author} {\bibfnamefont {M.~J.}\ \bibnamefont {Savage}},\
  }\href {\doibase 10.1103/PhysRevD.103.094501} {\bibfield  {journal} {\bibinfo
   {journal} {Phys. Rev. D}\ }\textbf {\bibinfo {volume} {103}},\ \bibinfo
  {pages} {094501} (\bibinfo {year} {2021})}\BibitemShut {NoStop}%
\bibitem [{\citenamefont {Klco}\ \emph {et~al.}(2020)\citenamefont {Klco},
  \citenamefont {Savage},\ and\ \citenamefont {Stryker}}]{Klco2020}%
  \BibitemOpen
  \bibfield  {author} {\bibinfo {author} {\bibfnamefont {N.}~\bibnamefont
  {Klco}}, \bibinfo {author} {\bibfnamefont {M.~J.}\ \bibnamefont {Savage}}, \
  and\ \bibinfo {author} {\bibfnamefont {J.~R.}\ \bibnamefont {Stryker}},\
  }\href {\doibase 10.1103/PhysRevD.101.074512} {\bibfield  {journal} {\bibinfo
   {journal} {Phys. Rev. D}\ }\textbf {\bibinfo {volume} {101}},\ \bibinfo
  {pages} {74512} (\bibinfo {year} {2020})}\BibitemShut {NoStop}%
\bibitem [{\citenamefont {Mathis}\ \emph {et~al.}(2020)\citenamefont {Mathis},
  \citenamefont {Mazzola},\ and\ \citenamefont {Tavernelli}}]{Mathis2020}%
  \BibitemOpen
  \bibfield  {author} {\bibinfo {author} {\bibfnamefont {S.~V.}\ \bibnamefont
  {Mathis}}, \bibinfo {author} {\bibfnamefont {G.}~\bibnamefont {Mazzola}}, \
  and\ \bibinfo {author} {\bibfnamefont {I.}~\bibnamefont {Tavernelli}},\
  }\href {\doibase 10.1103/PhysRevD.102.094501} {\bibfield  {journal} {\bibinfo
   {journal} {Phys. Rev. D}\ }\textbf {\bibinfo {volume} {102}},\ \bibinfo
  {pages} {094501} (\bibinfo {year} {2020})}\BibitemShut {NoStop}%
\bibitem [{\citenamefont {Armon}\ \emph {et~al.}(2021)\citenamefont {Armon},
  \citenamefont {Ashkenazi}, \citenamefont {Garc\'{\i}a-Moreno}, \citenamefont
  {Gonz\'alez-Tudela},\ and\ \citenamefont {Zohar}}]{Armon2021}%
  \BibitemOpen
  \bibfield  {author} {\bibinfo {author} {\bibfnamefont {T.}~\bibnamefont
  {Armon}}, \bibinfo {author} {\bibfnamefont {S.}~\bibnamefont {Ashkenazi}},
  \bibinfo {author} {\bibfnamefont {G.}~\bibnamefont {Garc\'{\i}a-Moreno}},
  \bibinfo {author} {\bibfnamefont {A.}~\bibnamefont {Gonz\'alez-Tudela}}, \
  and\ \bibinfo {author} {\bibfnamefont {E.}~\bibnamefont {Zohar}},\ }\href
  {\doibase 10.1103/PhysRevLett.127.250501} {\bibfield  {journal} {\bibinfo
  {journal} {Phys. Rev. Lett.}\ }\textbf {\bibinfo {volume} {127}},\ \bibinfo
  {pages} {250501} (\bibinfo {year} {2021})}\BibitemShut {NoStop}%
\bibitem [{\citenamefont {Martinez}\ \emph {et~al.}(2016)\citenamefont
  {Martinez} \emph {et~al.}}]{Martinez2016}%
  \BibitemOpen
  \bibfield  {author} {\bibinfo {author} {\bibfnamefont {E.~A.}\ \bibnamefont
  {Martinez}} \emph {et~al.},\ }\href {\doibase 10.1038/nature18318} {\bibfield
   {journal} {\bibinfo  {journal} {Nature}\ }\textbf {\bibinfo {volume}
  {534}},\ \bibinfo {pages} {516} (\bibinfo {year} {2016})}\BibitemShut
  {NoStop}%
\bibitem [{\citenamefont {Schweizer}\ \emph {et~al.}(2019)\citenamefont
  {Schweizer}, \citenamefont {Grusdt}, \citenamefont {Berngruber},
  \citenamefont {Barbiero}, \citenamefont {Demler}, \citenamefont {Goldman},
  \citenamefont {Bloch},\ and\ \citenamefont {Aidelsburger}}]{Schweizer2019}%
  \BibitemOpen
  \bibfield  {author} {\bibinfo {author} {\bibfnamefont {C.}~\bibnamefont
  {Schweizer}}, \bibinfo {author} {\bibfnamefont {F.}~\bibnamefont {Grusdt}},
  \bibinfo {author} {\bibfnamefont {M.}~\bibnamefont {Berngruber}}, \bibinfo
  {author} {\bibfnamefont {L.}~\bibnamefont {Barbiero}}, \bibinfo {author}
  {\bibfnamefont {E.}~\bibnamefont {Demler}}, \bibinfo {author} {\bibfnamefont
  {N.}~\bibnamefont {Goldman}}, \bibinfo {author} {\bibfnamefont
  {I.}~\bibnamefont {Bloch}}, \ and\ \bibinfo {author} {\bibfnamefont
  {M.}~\bibnamefont {Aidelsburger}},\ }\href {\doibase
  10.1038/s41567-019-0649-7} {\bibfield  {journal} {\bibinfo  {journal} {Nat.
  Phys.}\ }\textbf {\bibinfo {volume} {15}},\ \bibinfo {pages} {1168} (\bibinfo
  {year} {2019})}\BibitemShut {NoStop}%
\bibitem [{\citenamefont {Kokail}\ \emph {et~al.}(2019)\citenamefont {Kokail},
  \citenamefont {Maier}, \citenamefont {van Bijnen}, \citenamefont {Brydges},
  \citenamefont {Joshi}, \citenamefont {Jurcevic}, \citenamefont {Muschik},
  \citenamefont {Silvi}, \citenamefont {Blatt}, \citenamefont {Roos},\ and\
  \citenamefont {Zoller}}]{Kokail2019}%
  \BibitemOpen
  \bibfield  {author} {\bibinfo {author} {\bibfnamefont {C.}~\bibnamefont
  {Kokail}}, \bibinfo {author} {\bibfnamefont {C.}~\bibnamefont {Maier}},
  \bibinfo {author} {\bibfnamefont {R.}~\bibnamefont {van Bijnen}}, \bibinfo
  {author} {\bibfnamefont {T.}~\bibnamefont {Brydges}}, \bibinfo {author}
  {\bibfnamefont {M.~K.}\ \bibnamefont {Joshi}}, \bibinfo {author}
  {\bibfnamefont {P.}~\bibnamefont {Jurcevic}}, \bibinfo {author}
  {\bibfnamefont {C.~A.}\ \bibnamefont {Muschik}}, \bibinfo {author}
  {\bibfnamefont {P.}~\bibnamefont {Silvi}}, \bibinfo {author} {\bibfnamefont
  {R.}~\bibnamefont {Blatt}}, \bibinfo {author} {\bibfnamefont {C.~F.}\
  \bibnamefont {Roos}}, \ and\ \bibinfo {author} {\bibfnamefont
  {P.}~\bibnamefont {Zoller}},\ }\href {\doibase 10.1038/s41586-019-1177-4}
  {\bibfield  {journal} {\bibinfo  {journal} {Nature}\ }\textbf {\bibinfo
  {volume} {569}},\ \bibinfo {pages} {355} (\bibinfo {year}
  {2019})}\BibitemShut {NoStop}%
\bibitem [{\citenamefont {Mil}\ \emph {et~al.}(2020)\citenamefont {Mil},
  \citenamefont {Zache}, \citenamefont {Hegde}, \citenamefont {Xia},
  \citenamefont {Bhatt}, \citenamefont {Oberthaler}, \citenamefont {Hauke},
  \citenamefont {Berges},\ and\ \citenamefont {Jendrzejewski}}]{Mil2020}%
  \BibitemOpen
  \bibfield  {author} {\bibinfo {author} {\bibfnamefont {A.}~\bibnamefont
  {Mil}}, \bibinfo {author} {\bibfnamefont {T.~V.}\ \bibnamefont {Zache}},
  \bibinfo {author} {\bibfnamefont {A.}~\bibnamefont {Hegde}}, \bibinfo
  {author} {\bibfnamefont {A.}~\bibnamefont {Xia}}, \bibinfo {author}
  {\bibfnamefont {R.~P.}\ \bibnamefont {Bhatt}}, \bibinfo {author}
  {\bibfnamefont {M.~K.}\ \bibnamefont {Oberthaler}}, \bibinfo {author}
  {\bibfnamefont {P.}~\bibnamefont {Hauke}}, \bibinfo {author} {\bibfnamefont
  {J.}~\bibnamefont {Berges}}, \ and\ \bibinfo {author} {\bibfnamefont
  {F.}~\bibnamefont {Jendrzejewski}},\ }\href {\doibase
  10.1126/science.aaz5312} {\bibfield  {journal} {\bibinfo  {journal}
  {Science}\ }\textbf {\bibinfo {volume} {367}},\ \bibinfo {pages} {1128 LP }
  (\bibinfo {year} {2020})}\BibitemShut {NoStop}%
\bibitem [{\citenamefont {Yang}\ \emph {et~al.}(2020)\citenamefont {Yang},
  \citenamefont {Sun}, \citenamefont {Ott}, \citenamefont {Wang}, \citenamefont
  {Zache}, \citenamefont {Halimeh}, \citenamefont {Yuan}, \citenamefont
  {Hauke},\ and\ \citenamefont {Pan}}]{Yang2020}%
  \BibitemOpen
  \bibfield  {author} {\bibinfo {author} {\bibfnamefont {B.}~\bibnamefont
  {Yang}}, \bibinfo {author} {\bibfnamefont {H.}~\bibnamefont {Sun}}, \bibinfo
  {author} {\bibfnamefont {R.}~\bibnamefont {Ott}}, \bibinfo {author}
  {\bibfnamefont {H.-Y.}\ \bibnamefont {Wang}}, \bibinfo {author}
  {\bibfnamefont {T.~V.}\ \bibnamefont {Zache}}, \bibinfo {author}
  {\bibfnamefont {J.~C.}\ \bibnamefont {Halimeh}}, \bibinfo {author}
  {\bibfnamefont {Z.-S.}\ \bibnamefont {Yuan}}, \bibinfo {author}
  {\bibfnamefont {P.}~\bibnamefont {Hauke}}, \ and\ \bibinfo {author}
  {\bibfnamefont {J.-W.}\ \bibnamefont {Pan}},\ }\href {\doibase
  10.1038/s41586-020-2910-8} {\bibfield  {journal} {\bibinfo  {journal}
  {Nature}\ }\textbf {\bibinfo {volume} {587}},\ \bibinfo {pages} {392}
  (\bibinfo {year} {2020})}\BibitemShut {NoStop}%
\bibitem [{\citenamefont {Zhou}\ \emph {et~al.}(2022)\citenamefont {Zhou},
  \citenamefont {Su}, \citenamefont {Halimeh}, \citenamefont {Ott},
  \citenamefont {Sun}, \citenamefont {Hauke}, \citenamefont {Yang},
  \citenamefont {Yuan}, \citenamefont {Berges},\ and\ \citenamefont
  {Pan}}]{Zhou2022}%
  \BibitemOpen
  \bibfield  {author} {\bibinfo {author} {\bibfnamefont {Z.-Y.}\ \bibnamefont
  {Zhou}}, \bibinfo {author} {\bibfnamefont {G.-X.}\ \bibnamefont {Su}},
  \bibinfo {author} {\bibfnamefont {J.~C.}\ \bibnamefont {Halimeh}}, \bibinfo
  {author} {\bibfnamefont {R.}~\bibnamefont {Ott}}, \bibinfo {author}
  {\bibfnamefont {H.}~\bibnamefont {Sun}}, \bibinfo {author} {\bibfnamefont
  {P.}~\bibnamefont {Hauke}}, \bibinfo {author} {\bibfnamefont
  {B.}~\bibnamefont {Yang}}, \bibinfo {author} {\bibfnamefont {Z.-S.}\
  \bibnamefont {Yuan}}, \bibinfo {author} {\bibfnamefont {J.}~\bibnamefont
  {Berges}}, \ and\ \bibinfo {author} {\bibfnamefont {J.-W.}\ \bibnamefont
  {Pan}},\ }\href@noop {} {\bibfield  {journal} {\bibinfo  {journal} {Science}\
  }\textbf {\bibinfo {volume} {377}},\ \bibinfo {pages} {311} (\bibinfo {year}
  {2022})}\BibitemShut {NoStop}%
\bibitem [{\citenamefont {Nguyen}\ \emph {et~al.}(2022)\citenamefont {Nguyen},
  \citenamefont {Tran}, \citenamefont {Zhu}, \citenamefont {Green},
  \citenamefont {Alderete}, \citenamefont {Davoudi},\ and\ \citenamefont
  {Linke}}]{Nguyen2022}%
  \BibitemOpen
  \bibfield  {author} {\bibinfo {author} {\bibfnamefont {N.~H.}\ \bibnamefont
  {Nguyen}}, \bibinfo {author} {\bibfnamefont {M.~C.}\ \bibnamefont {Tran}},
  \bibinfo {author} {\bibfnamefont {Y.}~\bibnamefont {Zhu}}, \bibinfo {author}
  {\bibfnamefont {A.~M.}\ \bibnamefont {Green}}, \bibinfo {author}
  {\bibfnamefont {C.~H.}\ \bibnamefont {Alderete}}, \bibinfo {author}
  {\bibfnamefont {Z.}~\bibnamefont {Davoudi}}, \ and\ \bibinfo {author}
  {\bibfnamefont {N.~M.}\ \bibnamefont {Linke}},\ }\href {\doibase
  10.1103/PRXQuantum.3.020324} {\bibfield  {journal} {\bibinfo  {journal} {PRX
  Quantum}\ }\textbf {\bibinfo {volume} {3}},\ \bibinfo {pages} {020324}
  (\bibinfo {year} {2022})}\BibitemShut {NoStop}%
\bibitem [{\citenamefont {Mildenberger}\ \emph {et~al.}(2022)\citenamefont
  {Mildenberger}, \citenamefont {Mruczkiewicz}, \citenamefont {Halimeh},
  \citenamefont {Jiang},\ and\ \citenamefont {Hauke}}]{Mildenberger2022}%
  \BibitemOpen
  \bibfield  {author} {\bibinfo {author} {\bibfnamefont {J.}~\bibnamefont
  {Mildenberger}}, \bibinfo {author} {\bibfnamefont {W.}~\bibnamefont
  {Mruczkiewicz}}, \bibinfo {author} {\bibfnamefont {J.~C.}\ \bibnamefont
  {Halimeh}}, \bibinfo {author} {\bibfnamefont {Z.}~\bibnamefont {Jiang}}, \
  and\ \bibinfo {author} {\bibfnamefont {P.}~\bibnamefont {Hauke}},\
  }\href@noop {} {\bibfield  {journal} {\bibinfo  {journal} {arXiv preprint
  arXiv:2203.08905}\ } (\bibinfo {year} {2022})}\BibitemShut {NoStop}%
\bibitem [{\citenamefont {Zohar}\ and\ \citenamefont
  {Cirac}(2018)}]{Zohar2018}%
  \BibitemOpen
  \bibfield  {author} {\bibinfo {author} {\bibfnamefont {E.}~\bibnamefont
  {Zohar}}\ and\ \bibinfo {author} {\bibfnamefont {J.~I.}\ \bibnamefont
  {Cirac}},\ }\href {\doibase 10.1103/PhysRevB.98.075119} {\bibfield  {journal}
  {\bibinfo  {journal} {Phys. Rev. B}\ }\textbf {\bibinfo {volume} {98}},\
  \bibinfo {pages} {075119} (\bibinfo {year} {2018})}\BibitemShut {NoStop}%
\bibitem [{\citenamefont {Zohar}\ and\ \citenamefont
  {Cirac}(2019)}]{Zohar2019}%
  \BibitemOpen
  \bibfield  {author} {\bibinfo {author} {\bibfnamefont {E.}~\bibnamefont
  {Zohar}}\ and\ \bibinfo {author} {\bibfnamefont {J.~I.}\ \bibnamefont
  {Cirac}},\ }\href {\doibase 10.1103/PhysRevD.99.114511} {\bibfield  {journal}
  {\bibinfo  {journal} {Phys. Rev. D}\ }\textbf {\bibinfo {volume} {99}},\
  \bibinfo {pages} {114511} (\bibinfo {year} {2019})}\BibitemShut {NoStop}%
\bibitem [{\citenamefont {Pardo}\ \emph {et~al.}(2023)\citenamefont {Pardo},
  \citenamefont {Greenberg}, \citenamefont {Fortinsky}, \citenamefont {Katz},\
  and\ \citenamefont {Zohar}}]{Pardo2023}%
  \BibitemOpen
  \bibfield  {author} {\bibinfo {author} {\bibfnamefont {G.}~\bibnamefont
  {Pardo}}, \bibinfo {author} {\bibfnamefont {T.}~\bibnamefont {Greenberg}},
  \bibinfo {author} {\bibfnamefont {A.}~\bibnamefont {Fortinsky}}, \bibinfo
  {author} {\bibfnamefont {N.}~\bibnamefont {Katz}}, \ and\ \bibinfo {author}
  {\bibfnamefont {E.}~\bibnamefont {Zohar}},\ }\href {\doibase
  10.1103/PhysRevResearch.5.023077} {\bibfield  {journal} {\bibinfo  {journal}
  {Phys. Rev. Res.}\ }\textbf {\bibinfo {volume} {5}},\ \bibinfo {pages}
  {023077} (\bibinfo {year} {2023})}\BibitemShut {NoStop}%
\bibitem [{\citenamefont {Irmejs}\ \emph {et~al.}(2022)\citenamefont {Irmejs},
  \citenamefont {Banuls},\ and\ \citenamefont {Cirac}}]{Irmejs2022}%
  \BibitemOpen
  \bibfield  {author} {\bibinfo {author} {\bibfnamefont {R.}~\bibnamefont
  {Irmejs}}, \bibinfo {author} {\bibfnamefont {M.~C.}\ \bibnamefont {Banuls}},
  \ and\ \bibinfo {author} {\bibfnamefont {J.~I.}\ \bibnamefont {Cirac}},\
  }\href@noop {} {\bibfield  {journal} {\bibinfo  {journal} {arXiv preprint
  arXiv:2206.08909}\ } (\bibinfo {year} {2022})}\BibitemShut {NoStop}%
\bibitem [{\citenamefont {Tilly}\ \emph {et~al.}(2022)\citenamefont {Tilly},
  \citenamefont {Chen}, \citenamefont {Cao}, \citenamefont {Picozzi},
  \citenamefont {Setia}, \citenamefont {Li}, \citenamefont {Grant},
  \citenamefont {Wossnig}, \citenamefont {Rungger}, \citenamefont {Booth},\
  and\ \citenamefont {Tennyson}}]{Tilly2022}%
  \BibitemOpen
  \bibfield  {author} {\bibinfo {author} {\bibfnamefont {J.}~\bibnamefont
  {Tilly}}, \bibinfo {author} {\bibfnamefont {H.}~\bibnamefont {Chen}},
  \bibinfo {author} {\bibfnamefont {S.}~\bibnamefont {Cao}}, \bibinfo {author}
  {\bibfnamefont {D.}~\bibnamefont {Picozzi}}, \bibinfo {author} {\bibfnamefont
  {K.}~\bibnamefont {Setia}}, \bibinfo {author} {\bibfnamefont
  {Y.}~\bibnamefont {Li}}, \bibinfo {author} {\bibfnamefont {E.}~\bibnamefont
  {Grant}}, \bibinfo {author} {\bibfnamefont {L.}~\bibnamefont {Wossnig}},
  \bibinfo {author} {\bibfnamefont {I.}~\bibnamefont {Rungger}}, \bibinfo
  {author} {\bibfnamefont {G.~H.}\ \bibnamefont {Booth}}, \ and\ \bibinfo
  {author} {\bibfnamefont {J.}~\bibnamefont {Tennyson}},\ }\href {\doibase
  10.1016/j.physrep.2022.08.003} {\bibfield  {journal} {\bibinfo  {journal}
  {Physics Reports}\ }\textbf {\bibinfo {volume} {986}},\ \bibinfo {pages} {1}
  (\bibinfo {year} {2022})}\BibitemShut {NoStop}%
\bibitem [{\citenamefont {Ferguson}\ \emph {et~al.}(2021)\citenamefont
  {Ferguson}, \citenamefont {Dellantonio}, \citenamefont {Balushi},
  \citenamefont {Jansen}, \citenamefont {Dür},\ and\ \citenamefont
  {Muschik}}]{Ferguson2021}%
  \BibitemOpen
  \bibfield  {author} {\bibinfo {author} {\bibfnamefont {R.}~\bibnamefont
  {Ferguson}}, \bibinfo {author} {\bibfnamefont {L.}~\bibnamefont
  {Dellantonio}}, \bibinfo {author} {\bibfnamefont {A.~A.}\ \bibnamefont
  {Balushi}}, \bibinfo {author} {\bibfnamefont {K.}~\bibnamefont {Jansen}},
  \bibinfo {author} {\bibfnamefont {W.}~\bibnamefont {Dür}}, \ and\ \bibinfo
  {author} {\bibfnamefont {C.}~\bibnamefont {Muschik}},\ }\href {\doibase
  10.1103/physrevlett.126.220501} {\bibfield  {journal} {\bibinfo  {journal}
  {Physical Review Letters}\ }\textbf {\bibinfo {volume} {126}} (\bibinfo
  {year} {2021}),\ 10.1103/physrevlett.126.220501}\BibitemShut {NoStop}%
\bibitem [{\citenamefont {Atas}\ \emph {et~al.}(2023)\citenamefont {Atas},
  \citenamefont {Haase}, \citenamefont {Zhang}, \citenamefont {Wei},
  \citenamefont {Pfaendler}, \citenamefont {Lewis},\ and\ \citenamefont
  {Muschik}}]{Atas2023}%
  \BibitemOpen
  \bibfield  {author} {\bibinfo {author} {\bibfnamefont {Y.~Y.}\ \bibnamefont
  {Atas}}, \bibinfo {author} {\bibfnamefont {J.~F.}\ \bibnamefont {Haase}},
  \bibinfo {author} {\bibfnamefont {J.}~\bibnamefont {Zhang}}, \bibinfo
  {author} {\bibfnamefont {V.}~\bibnamefont {Wei}}, \bibinfo {author}
  {\bibfnamefont {S.~M.~L.}\ \bibnamefont {Pfaendler}}, \bibinfo {author}
  {\bibfnamefont {R.}~\bibnamefont {Lewis}}, \ and\ \bibinfo {author}
  {\bibfnamefont {C.~A.}\ \bibnamefont {Muschik}},\ }\href@noop {} {\enquote
  {\bibinfo {title} {Simulating one-dimensional quantum chromodynamics on a
  quantum computer: Real-time evolutions of tetra- and pentaquarks},}\ }
  (\bibinfo {year} {2023}),\ \Eprint {http://arxiv.org/abs/2207.03473}
  {arXiv:2207.03473 [quant-ph]} \BibitemShut {NoStop}%
\bibitem [{\citenamefont {Chan}\ \emph {et~al.}(2023)\citenamefont {Chan},
  \citenamefont {Shi}, \citenamefont {Dellantonio}, \citenamefont {D{\"u}r},\
  and\ \citenamefont {Muschik}}]{Chan2023}%
  \BibitemOpen
  \bibfield  {author} {\bibinfo {author} {\bibfnamefont {A.}~\bibnamefont
  {Chan}}, \bibinfo {author} {\bibfnamefont {Z.}~\bibnamefont {Shi}}, \bibinfo
  {author} {\bibfnamefont {L.}~\bibnamefont {Dellantonio}}, \bibinfo {author}
  {\bibfnamefont {W.}~\bibnamefont {D{\"u}r}}, \ and\ \bibinfo {author}
  {\bibfnamefont {C.~A.}\ \bibnamefont {Muschik}},\ }\href@noop {} {\bibfield
  {journal} {\bibinfo  {journal} {arXiv preprint arXiv:2305.19200}\ } (\bibinfo
  {year} {2023})}\BibitemShut {NoStop}%
\bibitem [{\citenamefont {Peruzzo}\ \emph {et~al.}(2014)\citenamefont
  {Peruzzo}, \citenamefont {McClean}, \citenamefont {Shadbolt}, \citenamefont
  {Yung}, \citenamefont {Zhou}, \citenamefont {Love}, \citenamefont
  {Aspuru-Guzik},\ and\ \citenamefont {O'Brien}}]{Peruzzo2014}%
  \BibitemOpen
  \bibfield  {author} {\bibinfo {author} {\bibfnamefont {A.}~\bibnamefont
  {Peruzzo}}, \bibinfo {author} {\bibfnamefont {J.}~\bibnamefont {McClean}},
  \bibinfo {author} {\bibfnamefont {P.}~\bibnamefont {Shadbolt}}, \bibinfo
  {author} {\bibfnamefont {M.-H.}\ \bibnamefont {Yung}}, \bibinfo {author}
  {\bibfnamefont {X.-Q.}\ \bibnamefont {Zhou}}, \bibinfo {author}
  {\bibfnamefont {P.~J.}\ \bibnamefont {Love}}, \bibinfo {author}
  {\bibfnamefont {A.}~\bibnamefont {Aspuru-Guzik}}, \ and\ \bibinfo {author}
  {\bibfnamefont {J.~L.}\ \bibnamefont {O'Brien}},\ }\href {\doibase
  10.1038/ncomms5213} {\bibfield  {journal} {\bibinfo  {journal} {Nature
  Communications}\ }\textbf {\bibinfo {volume} {5}} (\bibinfo {year} {2014}),\
  10.1038/ncomms5213}\BibitemShut {NoStop}%
\bibitem [{\citenamefont {McArdle}\ \emph {et~al.}(2019)\citenamefont
  {McArdle}, \citenamefont {Jones}, \citenamefont {Endo}, \citenamefont {Li},
  \citenamefont {Benjamin},\ and\ \citenamefont {Yuan}}]{McArdle2019}%
  \BibitemOpen
  \bibfield  {author} {\bibinfo {author} {\bibfnamefont {S.}~\bibnamefont
  {McArdle}}, \bibinfo {author} {\bibfnamefont {T.}~\bibnamefont {Jones}},
  \bibinfo {author} {\bibfnamefont {S.}~\bibnamefont {Endo}}, \bibinfo {author}
  {\bibfnamefont {Y.}~\bibnamefont {Li}}, \bibinfo {author} {\bibfnamefont
  {S.~C.}\ \bibnamefont {Benjamin}}, \ and\ \bibinfo {author} {\bibfnamefont
  {X.}~\bibnamefont {Yuan}},\ }\href {\doibase 10.1038/s41534-019-0187-2}
  {\bibfield  {journal} {\bibinfo  {journal} {npj Quantum Information}\
  }\textbf {\bibinfo {volume} {5}} (\bibinfo {year} {2019}),\
  10.1038/s41534-019-0187-2}\BibitemShut {NoStop}%
\bibitem [{\citenamefont {Verdon}\ \emph {et~al.}(2019)\citenamefont {Verdon},
  \citenamefont {Marks}, \citenamefont {Nanda}, \citenamefont {Leichenauer},\
  and\ \citenamefont {Hidary}}]{Verdon2019}%
  \BibitemOpen
  \bibfield  {author} {\bibinfo {author} {\bibfnamefont {G.}~\bibnamefont
  {Verdon}}, \bibinfo {author} {\bibfnamefont {J.}~\bibnamefont {Marks}},
  \bibinfo {author} {\bibfnamefont {S.}~\bibnamefont {Nanda}}, \bibinfo
  {author} {\bibfnamefont {S.}~\bibnamefont {Leichenauer}}, \ and\ \bibinfo
  {author} {\bibfnamefont {J.}~\bibnamefont {Hidary}},\ }\href@noop {}
  {\enquote {\bibinfo {title} {Quantum hamiltonian-based models and the
  variational quantum thermalizer algorithm},}\ } (\bibinfo {year} {2019}),\
  \Eprint {http://arxiv.org/abs/1910.02071} {arXiv:1910.02071 [quant-ph]}
  \BibitemShut {NoStop}%
\bibitem [{\citenamefont {Higgott}\ \emph {et~al.}(2019)\citenamefont
  {Higgott}, \citenamefont {Wang},\ and\ \citenamefont
  {Brierley}}]{Higgott2019}%
  \BibitemOpen
  \bibfield  {author} {\bibinfo {author} {\bibfnamefont {O.}~\bibnamefont
  {Higgott}}, \bibinfo {author} {\bibfnamefont {D.}~\bibnamefont {Wang}}, \
  and\ \bibinfo {author} {\bibfnamefont {S.}~\bibnamefont {Brierley}},\ }\href
  {\doibase 10.22331/q-2019-07-01-156} {\bibfield  {journal} {\bibinfo
  {journal} {Quantum}\ }\textbf {\bibinfo {volume} {3}},\ \bibinfo {pages}
  {156} (\bibinfo {year} {2019})}\BibitemShut {NoStop}%
\bibitem [{\citenamefont {Yuan}\ \emph {et~al.}(2019)\citenamefont {Yuan},
  \citenamefont {Endo}, \citenamefont {Zhao}, \citenamefont {Li},\ and\
  \citenamefont {Benjamin}}]{Yuan2019}%
  \BibitemOpen
  \bibfield  {author} {\bibinfo {author} {\bibfnamefont {X.}~\bibnamefont
  {Yuan}}, \bibinfo {author} {\bibfnamefont {S.}~\bibnamefont {Endo}}, \bibinfo
  {author} {\bibfnamefont {Q.}~\bibnamefont {Zhao}}, \bibinfo {author}
  {\bibfnamefont {Y.}~\bibnamefont {Li}}, \ and\ \bibinfo {author}
  {\bibfnamefont {S.~C.}\ \bibnamefont {Benjamin}},\ }\href {\doibase
  10.22331/q-2019-10-07-191} {\bibfield  {journal} {\bibinfo  {journal}
  {Quantum}\ }\textbf {\bibinfo {volume} {3}},\ \bibinfo {pages} {191}
  (\bibinfo {year} {2019})}\BibitemShut {NoStop}%
\bibitem [{\citenamefont {Barison}\ \emph {et~al.}(2021)\citenamefont
  {Barison}, \citenamefont {Vicentini},\ and\ \citenamefont
  {Carleo}}]{Barison2021}%
  \BibitemOpen
  \bibfield  {author} {\bibinfo {author} {\bibfnamefont {S.}~\bibnamefont
  {Barison}}, \bibinfo {author} {\bibfnamefont {F.}~\bibnamefont {Vicentini}},
  \ and\ \bibinfo {author} {\bibfnamefont {G.}~\bibnamefont {Carleo}},\ }\href
  {\doibase 10.22331/q-2021-07-28-512} {\bibfield  {journal} {\bibinfo
  {journal} {Quantum}\ }\textbf {\bibinfo {volume} {5}},\ \bibinfo {pages}
  {512} (\bibinfo {year} {2021})}\BibitemShut {NoStop}%
\bibitem [{\citenamefont {Wierichs}\ \emph {et~al.}(2022)\citenamefont
  {Wierichs}, \citenamefont {Izaac}, \citenamefont {Wang},\ and\ \citenamefont
  {Lin}}]{Wierichs2022}%
  \BibitemOpen
  \bibfield  {author} {\bibinfo {author} {\bibfnamefont {D.}~\bibnamefont
  {Wierichs}}, \bibinfo {author} {\bibfnamefont {J.}~\bibnamefont {Izaac}},
  \bibinfo {author} {\bibfnamefont {C.}~\bibnamefont {Wang}}, \ and\ \bibinfo
  {author} {\bibfnamefont {C.~Y.-Y.}\ \bibnamefont {Lin}},\ }\href {\doibase
  10.22331/q-2022-03-30-677} {\bibfield  {journal} {\bibinfo  {journal}
  {Quantum}\ }\textbf {\bibinfo {volume} {6}},\ \bibinfo {pages} {677}
  (\bibinfo {year} {2022})}\BibitemShut {NoStop}%
\bibitem [{\citenamefont {Vermersch}\ \emph {et~al.}(2019)\citenamefont
  {Vermersch}, \citenamefont {Elben}, \citenamefont {Sieberer}, \citenamefont
  {Yao},\ and\ \citenamefont {Zoller}}]{Vermersch2019}%
  \BibitemOpen
  \bibfield  {author} {\bibinfo {author} {\bibfnamefont {B.}~\bibnamefont
  {Vermersch}}, \bibinfo {author} {\bibfnamefont {A.}~\bibnamefont {Elben}},
  \bibinfo {author} {\bibfnamefont {L.~M.}\ \bibnamefont {Sieberer}}, \bibinfo
  {author} {\bibfnamefont {N.~Y.}\ \bibnamefont {Yao}}, \ and\ \bibinfo
  {author} {\bibfnamefont {P.}~\bibnamefont {Zoller}},\ }\href {\doibase
  10.1103/PhysRevX.9.021061} {\bibfield  {journal} {\bibinfo  {journal} {Phys.
  Rev. X}\ }\textbf {\bibinfo {volume} {9}},\ \bibinfo {pages} {021061}
  (\bibinfo {year} {2019})}\BibitemShut {NoStop}%
\bibitem [{\citenamefont {Schuckert}\ and\ \citenamefont
  {Knap}(2020)}]{Schuckert2020}%
  \BibitemOpen
  \bibfield  {author} {\bibinfo {author} {\bibfnamefont {A.}~\bibnamefont
  {Schuckert}}\ and\ \bibinfo {author} {\bibfnamefont {M.}~\bibnamefont
  {Knap}},\ }\href {\doibase 10.1103/PhysRevResearch.2.043315} {\bibfield
  {journal} {\bibinfo  {journal} {Phys. Rev. Res.}\ }\textbf {\bibinfo {volume}
  {2}},\ \bibinfo {pages} {043315} (\bibinfo {year} {2020})}\BibitemShut
  {NoStop}%
\bibitem [{\citenamefont {Popov}\ \emph {et~al.}()\citenamefont {Popov},
  \citenamefont {Geier}, \citenamefont {Lewenstein}, \citenamefont {Kasper},\
  and\ \citenamefont {Hauke}}]{Popov2023}%
  \BibitemOpen
  \bibfield  {author} {\bibinfo {author} {\bibfnamefont {P.}~\bibnamefont
  {Popov}}, \bibinfo {author} {\bibfnamefont {K.}~\bibnamefont {Geier}},
  \bibinfo {author} {\bibfnamefont {M.}~\bibnamefont {Lewenstein}}, \bibinfo
  {author} {\bibfnamefont {V.}~\bibnamefont {Kasper}}, \ and\ \bibinfo {author}
  {\bibfnamefont {P.}~\bibnamefont {Hauke}},\ }\href@noop {} {}\bibinfo {note}
  {In preparation}\BibitemShut {NoStop}%
\bibitem [{\citenamefont {Katz}\ \emph {et~al.}(2022)\citenamefont {Katz},
  \citenamefont {Cetina},\ and\ \citenamefont {Monroe}}]{Katz2022}%
  \BibitemOpen
  \bibfield  {author} {\bibinfo {author} {\bibfnamefont {O.}~\bibnamefont
  {Katz}}, \bibinfo {author} {\bibfnamefont {M.}~\bibnamefont {Cetina}}, \ and\
  \bibinfo {author} {\bibfnamefont {C.}~\bibnamefont {Monroe}},\ }\href
  {\doibase 10.1103/PhysRevLett.129.063603} {\bibfield  {journal} {\bibinfo
  {journal} {Phys. Rev. Lett.}\ }\textbf {\bibinfo {volume} {129}},\ \bibinfo
  {pages} {063603} (\bibinfo {year} {2022})}\BibitemShut {NoStop}%
\bibitem [{\citenamefont {S\o{}rensen}\ and\ \citenamefont
  {M\o{}lmer}(1999)}]{Sorensen1999}%
  \BibitemOpen
  \bibfield  {author} {\bibinfo {author} {\bibfnamefont {A.}~\bibnamefont
  {S\o{}rensen}}\ and\ \bibinfo {author} {\bibfnamefont {K.}~\bibnamefont
  {M\o{}lmer}},\ }\href {\doibase 10.1103/PhysRevLett.82.1971} {\bibfield
  {journal} {\bibinfo  {journal} {Phys. Rev. Lett.}\ }\textbf {\bibinfo
  {volume} {82}},\ \bibinfo {pages} {1971} (\bibinfo {year}
  {1999})}\BibitemShut {NoStop}%
\bibitem [{\citenamefont {M\o{}lmer}\ and\ \citenamefont
  {S\o{}rensen}(1999)}]{Molmer1999}%
  \BibitemOpen
  \bibfield  {author} {\bibinfo {author} {\bibfnamefont {K.}~\bibnamefont
  {M\o{}lmer}}\ and\ \bibinfo {author} {\bibfnamefont {A.}~\bibnamefont
  {S\o{}rensen}},\ }\href {\doibase 10.1103/PhysRevLett.82.1835} {\bibfield
  {journal} {\bibinfo  {journal} {Phys. Rev. Lett.}\ }\textbf {\bibinfo
  {volume} {82}},\ \bibinfo {pages} {1835} (\bibinfo {year}
  {1999})}\BibitemShut {NoStop}%
\bibitem [{\citenamefont {Calabrese}\ and\ \citenamefont
  {Cardy}(2004)}]{Calabrese2004}%
  \BibitemOpen
  \bibfield  {author} {\bibinfo {author} {\bibfnamefont {P.}~\bibnamefont
  {Calabrese}}\ and\ \bibinfo {author} {\bibfnamefont {J.}~\bibnamefont
  {Cardy}},\ }\href@noop {} {\bibfield  {journal} {\bibinfo  {journal} {Journal
  of statistical mechanics: theory and experiment}\ }\textbf {\bibinfo {volume}
  {2004}},\ \bibinfo {pages} {P06002} (\bibinfo {year} {2004})}\BibitemShut
  {NoStop}%
\bibitem [{\citenamefont {Meth}\ \emph {et~al.}(2022)\citenamefont {Meth},
  \citenamefont {Kuzmin}, \citenamefont {van Bijnen}, \citenamefont {Postler},
  \citenamefont {Stricker}, \citenamefont {Blatt}, \citenamefont {Ringbauer},
  \citenamefont {Monz}, \citenamefont {Silvi},\ and\ \citenamefont
  {Schindler}}]{Meth2022}%
  \BibitemOpen
  \bibfield  {author} {\bibinfo {author} {\bibfnamefont {M.}~\bibnamefont
  {Meth}}, \bibinfo {author} {\bibfnamefont {V.}~\bibnamefont {Kuzmin}},
  \bibinfo {author} {\bibfnamefont {R.}~\bibnamefont {van Bijnen}}, \bibinfo
  {author} {\bibfnamefont {L.}~\bibnamefont {Postler}}, \bibinfo {author}
  {\bibfnamefont {R.}~\bibnamefont {Stricker}}, \bibinfo {author}
  {\bibfnamefont {R.}~\bibnamefont {Blatt}}, \bibinfo {author} {\bibfnamefont
  {M.}~\bibnamefont {Ringbauer}}, \bibinfo {author} {\bibfnamefont
  {T.}~\bibnamefont {Monz}}, \bibinfo {author} {\bibfnamefont {P.}~\bibnamefont
  {Silvi}}, \ and\ \bibinfo {author} {\bibfnamefont {P.}~\bibnamefont
  {Schindler}},\ }\href {\doibase 10.1103/PhysRevX.12.041035} {\bibfield
  {journal} {\bibinfo  {journal} {Phys. Rev. X}\ }\textbf {\bibinfo {volume}
  {12}},\ \bibinfo {pages} {041035} (\bibinfo {year} {2022})}\BibitemShut
  {NoStop}%
\bibitem [{\citenamefont {Andrade}\ \emph {et~al.}(2022)\citenamefont
  {Andrade}, \citenamefont {Davoudi}, \citenamefont {Gra{\ss}}, \citenamefont
  {Hafezi}, \citenamefont {Pagano},\ and\ \citenamefont {Seif}}]{Andrade2022}%
  \BibitemOpen
  \bibfield  {author} {\bibinfo {author} {\bibfnamefont {B.}~\bibnamefont
  {Andrade}}, \bibinfo {author} {\bibfnamefont {Z.}~\bibnamefont {Davoudi}},
  \bibinfo {author} {\bibfnamefont {T.}~\bibnamefont {Gra{\ss}}}, \bibinfo
  {author} {\bibfnamefont {M.}~\bibnamefont {Hafezi}}, \bibinfo {author}
  {\bibfnamefont {G.}~\bibnamefont {Pagano}}, \ and\ \bibinfo {author}
  {\bibfnamefont {A.}~\bibnamefont {Seif}},\ }\href {\doibase
  10.1088/2058-9565/ac5f5b} {\bibfield  {journal} {\bibinfo  {journal} {Quantum
  Science and Technology}\ }\textbf {\bibinfo {volume} {7}},\ \bibinfo {pages}
  {034001} (\bibinfo {year} {2022})}\BibitemShut {NoStop}%
\end{thebibliography}%

\end{document}